\documentclass[a4paper,11pt]{article}
\pdfoutput=1 

\usepackage{jheppub} 

\usepackage[T1]{fontenc} 

\usepackage{amssymb}
\usepackage{enumerate} 

\usepackage{multirow} 
\usepackage{booktabs} 
\usepackage{makecell} 

\def\dilep{{\it di-lepton} }
\def\jetlep{{\it lepton $+$ jet} }

\title{Associated production of a top-quark pair with two isolated 
photons at the LHC through NLO in QCD}

\author{Daniel Stremmer}
\author{and Malgorzata  Worek}
\affiliation{ Institute for Theoretical Particle Physics
and Cosmology, RWTH Aachen University, \\D-52056 Aachen, Germany}

\emailAdd{daniel.stremmer@rwth-aachen.de}
\emailAdd{worek@physik.rwth-aachen.de}

\abstract{
We report on the computation of NLO QCD corrections to top-quark pair production in association with two  photons at the LHC. Higher-order effects and photon bremsstrahlung are taken into account in the production and decays of the top-quark pair. Top-quark and $W$-boson decays are treated in the Narrow Width Approximation conserving spin correlations up to NLO in QCD. 
This is the first time that the complete set of NLO QCD corrections to the $pp \to t\bar{t}\gamma\gamma$ process including top-quark decays is calculated. We present results at the integrated and differential cross-section level in the \dilep and \jetlep channel.  In addition, we investigate the effect of  photon bremsstrahlung in $t\bar{t}$ production and top-quark decays, as well as the mixed contribution. The latter contribution, in which two photons occur simultaneously in the production and decay of the $t\bar{t}$ pair, proved to be significant at both the integrated and differential cross-section level.}

\dedicated{\rm P3H-23-040, TTK-23-12}


\begin{document} 
\maketitle
\flushbottom

%
\section{Introduction}
\label{sec:introduction}
%

The observation  of  the $pp \to t\bar{t}H$ process at the Large Hadron Collider (LHC) reported by the  CMS \cite{CMS:2018uxb} and ATLAS \cite{ATLAS:2018mme} collaborations has launched a new endeavour to investigate  the tree-level top quark Yukawa coupling $(Y_t)$ and the ${\cal CP}$ structure of the Higgs boson. One of the most sensitive Higgs-boson decay channels for probing the $pp \to t\bar{t}H$ process is $H \to \gamma\gamma$.  Despite the small branching ratio the Higgs-boson signal can be extracted in this channel thanks to the excellent photon reconstruction and identification efficiency of the ATLAS and CMS detectors. Even though by probing the interactions between the $H$ boson and electroweak $W/Z$ gauge bosons, CMS and ATLAS  have determined that the $H$ boson quantum numbers are consistent with the Standard Model (SM) \cite{ATLAS:2016ifi,CMS:2016tad,ATLAS:2017azn,ATLAS:2018hxb,
CMS:2019ekd,CMS:2019jdw}, the presence of a  pseudoscalar admixture, which introduces a second coupling to the top quark, has not yet been ruled out and is worth investigating.   The observation of a non-zero  ${\cal CP}$-odd coupling component would signal the existence of physics beyond the SM, and open up the possibility of ${\cal CP}$-violation in the Higgs-boson sector, see e.g.  \cite{Gunion:1996xu,Demartin:2014fia,Mileo:2016mxg,Gritsan:2016hjl,Demartin:2016axk,AmorDosSantos:2017ayi,Bernreuther:2018ynm,Goncalves:2018agy,Martini:2021uey,Hermann:2022vit,Barman:2021yfh,Bahl:2021dnc,Azevedo:2022jnd} and references therein. Such a new source of ${\cal CP}$ violation could play a fundamental role in explaining the matter–antimatter asymmetry of the universe. Therefore, studies of the $Y_t$ coupling  in the $H\to \gamma\gamma$ channel would provide an alternative and independent path for $\cal{CP}$ tests in the Higgs-boson  sector. Indeed, present analyses of the SM Higgs boson at the LHC in the $pp\to t\bar{t}H$ production mode focus on the Higgs boson decaying into two photons. In fact,  the first single-channel observation of the $pp \to t\bar{t}H$ process by both ATLAS and CMS \cite{CMS:2020cga,ATLAS:2020ior} has been reported in the $H\to \gamma\gamma$ channel, together with the measurement of the ${\cal CP}$ structure of the $Y_t$ coupling. Despite the fact that  the data disfavored the pure ${\cal CP}$-odd model of the $Ht\bar{t}$ coupling, still only rather weak constrains exist on the possible admixture between the ${\cal CP}$-even and ${\cal CP}$-odd component of $Y_t$. A close scrutiny of the backgrounds shows that the direct production of $t\bar{t}\gamma\gamma$ is the most relevant, often referred to as the irreducible $t\bar{t}\gamma\gamma$ background. Experimental analyses at the LHC rely on data-driven approaches to estimate this background, however, Monte Carlo simulations are also used for this purpose, but mainly with LO accuracy.

On the theory side, NLO predictions for the $pp \to t\bar{t} \gamma \gamma$ process with stable top quarks have already been known for some time and  have been further matched to parton shower programs \cite{Alwall:2014hca, Kardos:2014pba,Maltoni:2015ena,vanDeurzen:2015cga}. In all these studies NLO QCD corrections were calculated for the $pp \to t\bar{t} \gamma \gamma$ production stage only. On the one hand, parton shower programs contain dominant soft-collinear logarithmic corrections and can approximate radiative effects in top-quark decays. On the other hand, such effects  are described by matrix elements formally accurate only at LO. Moreover, photon radiation from the charged top-quark decay products is omitted in such theoretical predictions.  In addition to higher-order QCD effects also NLO EW corrections  for $t\bar{t}$ production in association with two photons have recently been presented in literature \cite{Pagani:2021iwa}, but again only for stable top quarks. The contribution of photons emitted after the decay of top quarks might be, however, significant and should be incorporated in theoretical predictions for this process.  Consequently, a complete study with  higher-order corrections for the following final state $W^+W^- b\bar{b}\,\gamma\gamma$ including $W$ decays  is required to deeper understand the dynamics of the $pp \to t\bar{t}\gamma\gamma$ process.  Similar studies for the simpler $pp\to t\bar{t}\gamma$ process with top-quark and $W$ decays included, either in the Narrow Width Approximation (NWA) or with full  off-shell  effects, are already available in literature \cite{Melnikov:2011ta,Bevilacqua:2018woc,Bevilacqua:2018dny,Bergner:2018lgm,Bevilacqua:2019quz}. They have shown, among others, that photon radiation is distributed evenly between the $t\bar{t}\gamma$ production and the two top-quark decays: $t\to b W^+ \, (\gamma) \to b \ell^+ \nu_\ell \, \gamma$ as well as  $\bar{t}\to \bar{b} W^- \, (\gamma)\to \bar{b} \ell^- \bar{\nu}_\ell\, \gamma$, where $\ell=e, \mu$ \cite{Bevilacqua:2019quz}.

The purpose of this paper is to mitigate the current situation and to calculate for the first time NLO QCD corrections to the $pp \to t\bar{t}\gamma\gamma$ process taking into account higher-order effects in both the $t\bar{t}$ production and the decay of the top-quark pair. In this calculation top-quark decays are treated in the NWA. Thus, NLO $t\bar{t}$ spin correlations are preserved throughout the calculation. Furthermore, effects of photon bremsstrahlung from the charged top-quark decay products are consistently included. In detail, we calculate NLO QCD corrections  to the following final states 
\begin{itemize}
\item $pp \to (t\to W^+ (\ell^+\nu_\ell) \, b)\times (\bar{t}\to W^-(\ell^- \bar{\nu}_\ell) \, \bar{b}) \, \gamma\gamma$,
\item $pp \to (t\to W^+ (q\bar{q}^{\, \prime}) \, b)\times (\bar{t}\to W^-(\ell^- \bar{\nu}_\ell) \, \bar{b}) \, \gamma\gamma$,
\end{itemize}
denoted as  the \dilep and \jetlep  channel, respectively. In the remainder of the paper we refer to these two processes as $pp\to \ell^+\nu_{\ell}\, \ell^-\bar{\nu}_{\ell} \, b\bar{b}\,\gamma\gamma +X$ and $pp\to \ell^-\bar{\nu}_{\ell} \, jj \, b\bar{b}\,\gamma\gamma +X$. In the case of the \dilep channel we consider the following leptonic combinations: $(\ell^+\ell^-) \in (e^+e^-, e^+\mu^-, \mu^+e^-,\mu^+\mu^-)$. We do not include $\tau^\pm$ as most of the experimental analyses at the LHC distinguish the electron and muon from the $\tau$ channel, which is more difficult to reconstruct. For the  \jetlep  decay channel we employ $\ell^-=e^-,\mu^-$ and $W^+$ gauge boson decays into two families of light quarks, i.e. $q\bar{q}^{\, \prime}= u\bar{d}, c\bar{s}$. For $W^+\to u\bar{d}$ and $W^+ \to c\bar{s}$ decays QCD radiative corrections are taken into account.  We examine the size of higher-order corrections and theoretical uncertainties  for both decay channels. We additionally address the choice of a judicious renormalisation and factorisation scale setting  and the size of parton distribution function (PDF) uncertainties.  Having included photon emissions from various stages, we can assess their distribution and impact on the integrated and differential fiducial cross sections for both \dilep and \jetlep channels.   Our results are obtained with the help of \textsc{Helac-1Loop}/\textsc{Recola} and \textsc{Helac-Dipoles}.  For this work, the \textsc{Helac-Nlo} MC framework, that comprises  \textsc{Helac-1Loop} and \textsc{Helac-Dipoles}, is used for the first time in NLO QCD calculations involving hadronically decaying top quarks.

The article is organized as follows:  in Section \ref{sec:description} we outline the framework of the calculation and discuss cross-checks that have been performed. Input parameters and cuts that have been used to simulate detector response are summarised in Section \ref{sec:setup}. Numerical results for the integrated and differential cross sections for the $pp\to \ell^+\nu_{\ell}\, \ell^-\bar{\nu}_{\ell} \, b\bar{b}\,\gamma\gamma +X$  process at the LHC with $\sqrt{s}=13$ TeV for two renormalisation ($\mu_R$) and factorisation ($\mu_F$) scale choices are presented in detail in Section  \ref{sec:ttaa-lep}. The theoretical uncertainties  that are associated with neglected higher order terms in the perturbative expansion and with the different parametrisation of PDFs, are also given there.  Our findings for the $pp\to \ell^-\bar{\nu}_{\ell} \, jj \, b\bar{b}\,\gamma\gamma +X$ process are provided in Section \ref{sec:ttaa-semi} following the same structure as for the \dilep  channel.  In addition, different parameter choices of the smooth photon isolation prescription are briefly discussed there as well. Lastly, in Section \ref{sec:sum} our results for the  $pp \to t\bar{t}\gamma\gamma$ production process are briefly summarised and conclusions are outlined.

%
\section{Description of the calculation}
\label{sec:description}
%

\begin{figure}[t!]
  \begin{center}
  \includegraphics[trim= 20 710 20 20, width=\textwidth]{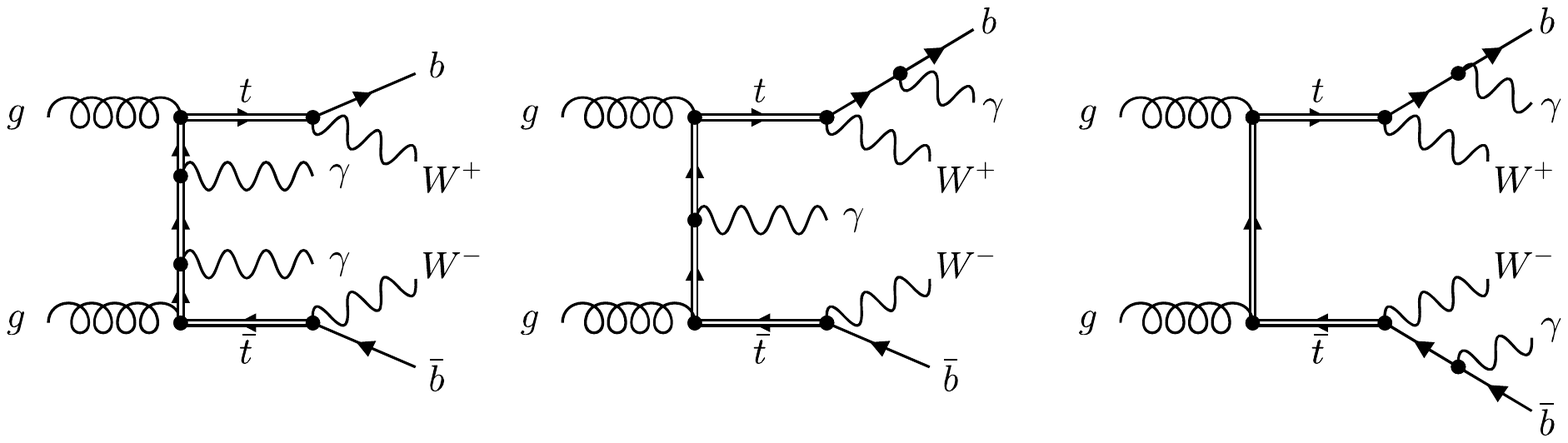}
\end{center}
  \caption{\label{fig:fd-LO} \it  Representative Feynman diagrams  for contributions:  {\it Prod.}, {\it Mixed} and {\it Decay} at LO with suppressed $W$ gauge boson decays. Feynman diagrams were produced with the help of the \textsc{FeynGame} program \cite{Harlander:2020cyh}.}
\end{figure}

We calculate NLO QCD corrections to the $pp\to t\bar{t}\gamma\gamma$ process  at the LHC. In particular, we evaluate $\alpha_s$ corrections to the Born-level process at $\mathcal{O}(\alpha_s^2\alpha^6)$. Unstable top quarks in the \dilep and \jetlep channels are considered. This leads to the following final states respectively
\begin{equation}
\begin{split}
pp &\to t\bar{t}(\gamma\gamma)\to W^+W^-\,b\bar{b}(\gamma\gamma) \to \ell^+\nu_{\ell}\, \ell^-\bar{\nu}_{\ell} \, b\bar{b}\,\gamma\gamma +X, \\[0.2cm]
pp &\to t\bar{t}(\gamma\gamma)\to W^+W^-\,b\bar{b}(\gamma\gamma) \to \ell^-\bar{\nu}_{\ell} \, jj \, b\bar{b}\,\gamma\gamma +X,
\end{split}
\end{equation}
where  $\ell^{\pm}=\mu^{\pm},e^{\pm}$. The decays of  top quarks and  $W$ bosons are performed in the NWA, i.e. in the limit  $\Gamma_t/m_t\to 0$. In this approach all terms less singular than $\Gamma^{-2}_t$ are consistently neglected and the Breit-Wigner propagators become delta-functions which force unstable particles to be on-shell, see e.g. Refs \cite{Denner:1999gp,Denner:2005fg,Melnikov:2009dn,Melnikov:2011qx,Campbell:2012uf,Behring:2019iiv,Bevilacqua:2019quz,Czakon:2020qbd}. Thus, the differential cross section can be factorised in the production of top quarks (and photons) and top-quark decays (with photons) according to
\begin{equation}
\label{eq_res}
\begin{split}
    d\sigma_{\rm Full}&=\overbrace{d\sigma_{t\bar{t}\gamma\gamma}\times\frac{d\Gamma_t}{\Gamma_t}\times\frac{d\Gamma_{\bar{t}}}{\Gamma_t}}^{\sigma_{\rm Prod.}} + \overbrace{d\sigma_{t\bar{t}\gamma}\times\left( \frac{d\Gamma_{t\gamma}}{\Gamma_t}\times\frac{d\Gamma_{\bar{t}}}{\Gamma_t} + \frac{d\Gamma_t}{\Gamma_t}\times\frac{d\Gamma_{\bar{t}\gamma}}{\Gamma_t} \right)}^{\sigma_{\rm Mixed}}
    \\&
    +\underbrace{d\sigma_{t\bar{t}}\times\left( \frac{d\Gamma_{t\gamma\gamma}}{\Gamma_t}\times\frac{d\Gamma_{\bar{t}}}{\Gamma_t} + \frac{d\Gamma_t}{\Gamma_t}\times\frac{d\Gamma_{\bar{t}\gamma\gamma}}{\Gamma_t} + \frac{d\Gamma_{t\gamma}}{\Gamma_t}\times\frac{d\Gamma_{\bar{t}\gamma}}{\Gamma_t} \right)}_{\sigma_{\rm Decay}}\,.
\end{split}
\end{equation}
In addition, treating  the $W$ gauge boson in the NWA, the differential top-quark decay rate can be further expanded as 
\begin{equation}
    d\Gamma_{t+n\gamma}=\sum_{i=0}^{n} d\Gamma_{t\to bW^+ + i\gamma}\frac{d\Gamma_{W^+ +(n-i)\gamma } }{\Gamma_{W}}\,,
\end{equation}
where $W^+ \to \ell^+ \nu_\ell$ or $ W^+ \to q\bar{q}^{\, \prime}$. This leads to a total of $15$ possibilities (resonant histories) from which photons can be radiated in the decay chain at LO. Furthermore, we defined in Eq. \eqref{eq_res} the following three contributions: {\it Prod.}, {\it Mixed} and {\it Decay}, based on the number of photons in the $t\bar{t}$ production process, in order to study the origin of photon radiation in more detail. In Figure \ref{fig:fd-LO} we depict a few examples of Feynman diagrams for the three contributions. At NLO, however, the number of resonant histories increases up to $45$ in the \dilep and $60$ in the \jetlep channel due to additional QCD radiation. At the Born level in the 
\dilep and \jetlep channels, we encounter the following subprocesses, respectively
\begin{equation}
\begin{array}{clllcll}
gg&\to& \ell^+\nu_{\ell}\, \ell^-\bar{\nu}_{\ell} \, 
b\bar{b}\,\gamma\gamma \,,
&\quad\quad\quad\quad\quad\quad\quad\quad\quad\quad& 
gg&\to& \ell^-\bar{\nu}_{\ell} \, q\bar{q}^{\, \prime} 
\, b\bar{b}\,\gamma\gamma \,,     \\[0.2cm]
q\bar{q}/\bar{q}q&\to& \ell^+\nu_{\ell}\, 
\ell^{-}\bar{\nu}_{\ell} \, b\bar{b}\,\gamma\gamma \,,
&\quad\quad\quad\quad\quad\quad\quad\quad\quad\quad
& q\bar{q}/\bar{q}q&\to& \ell^-\bar{\nu}_{\ell} 
\, q\bar{q}^{\, \prime} \, b\bar{b}\,\gamma\gamma \,, \\[0.2cm]
\end{array}
\end{equation}
where $q=u,d,c,s,b$ and $\ell^\pm = e^\pm,\mu^\pm$ as well as $q\bar{q}^{\, \prime} = 
u\bar{d},c\bar{s}$. Since we work in the NWA there is no cross-talk between the $t\bar{t}$ 
production and top-quark decays or between the $t$ and $\bar{t}$ decay. Thus, no additional
contributions arise when both leptons in the \dilep channel or the quarks in the initial and
final state in the \jetlep decay channel are coming from the same generation.  
Although the contribution of bottom quarks in the initial state is only $0.1\%$ of the integrated
fiducial LO cross section in both decay channels,  and therefore numerical  insignificant, we
still include it in our calculations. At NLO in QCD additional subprocesses, that can be
constructed from Born-level subprocesses by gluon radiation and crossing, must be  included. This
leads to the following set of subprocesses contributing to the real emission part of the NLO QCD
calculation 
\begin{equation}
\begin{array}{clllcll}
gg&\to& \ell^+\nu_{\ell}\, \ell^-\bar{\nu}_{\ell} 
\, b\bar{b}\,\gamma\gamma\, g \,,
&\quad\quad\quad\quad\quad\quad\quad\quad\quad\quad& gg&\to& \ell^-\bar{\nu}_{\ell} \, q\bar{q}^{\, \prime} \, b\bar{b}\,\gamma\gamma \, g \,,    \\[0.2cm]
q\bar{q}/\bar{q}q&\to& \ell^+\nu_{\ell}\, \ell^-\bar{\nu}_{\ell} 
\, b\bar{b}\,\gamma\gamma  \, g \,,
&\quad\quad\quad\quad\quad\quad\quad\quad\quad\quad& q\bar{q}/\bar{q}q
&\to& \ell^-\bar{\nu}_{\ell} \, q\bar{q}^{\, \prime} 
\, b\bar{b}\,\gamma\gamma   \,g \,,\\[0.2cm]
gq/qg &\to& \ell^+\nu_{\ell}\, \ell^-\bar{\nu}_{\ell} \, b\bar{b}\,\gamma\gamma \, q \,,
&\quad\quad\quad\quad\quad\quad\quad\quad\quad\quad
& gq/qg&\to& \ell^-\bar{\nu}_{\ell} \, q\bar{q}^{\, \prime} 
\, b\bar{b}\,\gamma\gamma \, q\,,\\[0.2cm]
g\bar{q}/\bar{q}g&\to& \ell^+\nu_{\ell}\, \ell^-\bar{\nu}_{\ell} 
\, b\bar{b}\,\gamma\gamma  \, \bar{q}\,,
&\quad\quad\quad\quad\quad\quad\quad\quad\quad\quad
& g\bar{q}/\bar{q}g&\to& \ell^-\bar{\nu}_{\ell} \, q\bar{q}^{\, \prime} 
\, b\bar{b}\,\gamma\gamma  \,\bar{q} \,,
\end{array}
\end{equation}
We note that the contribution from bottom initiated subprocesses also at the NLO level is  phenomenologically negligible, but we nevertheless include all subprocesses for consistency reasons.  Furthermore, since subleading NLO corrections are not included in our calculation, the three contributions: {\it Prod.}, {\it Mixed} and {\it Decay} do not mix with each other. Only the presence of an additional photon from the real emission part of such corrections would introduce some ambiguity and  lead to different resonant structures
in singular limits. The same problem also occurs in processes such as $pp\to t\bar{t}j(j)$ production if higher-order QCD effects are included \cite{Melnikov:2011qx,Bevilacqua:2022ozv}.  Consequently, the calculation of NLO QCD corrections for the $pp \to t\bar{t}\gamma\gamma$ process can be performed  for each resonant history, that is present at LO, independently.
We work in the five-flavour scheme and keep the Cabibbo-Kobayashi-Maskawa (CKM) mixing matrix diagonal. For the calculation of tree-level and one-loop matrix elements the program \textsc{Recola} \cite{Actis:2016mpe,Actis:2012qn} is used, together with the library \textsc{Collier} \cite{Denner:2016kdg} that is employed for the numerical evaluation of one-loop scalar and tensor integrals. Because \textsc{Recola} is also able to provide matrix elements in the so-called double-pole approximation, see e.g. Refs. \cite{Denner:2000bj,Accomando:2004de,Denner:2016jyo}, it was straightforward  to interface it to the \textsc{Helac-NLO} MC program \cite{Bevilacqua:2011xh} and adopt it for use in the NWA \cite{Bevilacqua:2019quz}.  In addition, we have implemented in the \textsc{Recola} framework the random polarisation method as introduced in Refs. \cite{Draggiotis:1998gr,Draggiotis:2002hm,Bevilacqua:2013iha}. In that way, the polarisation state is replaced by a linear combination of helicity eigenstates while the spin summation is replaced by an integration over a phase parameter. This leads to a drastic speed improvement, especially for processes involving many final state particles. For example, for the polarisation state of a gluon, we can write 
\begin{equation}
    \epsilon_\mu(k, \phi) = e^{i\phi} \epsilon_\mu (k, +) + e^{-i\phi} \epsilon_\mu (k, -)\,,
\end{equation}
where $\epsilon_\mu (k, \pm)$ are helicity eigenstates. Then, the sum over the helicity of this gluon can be written as 
\begin{equation}
\label{sum_lambda}
   \sum_\lambda |{\cal M}_\lambda|^2 = \frac{1}{2\pi}\int_0^{2\pi} |{\cal M}_\phi|^2 \, d\phi\,.
\end{equation}
We do not have large differences in the values of $|{\cal M}_\phi|^2$ as function of $\phi$, since for every value of $\phi$ we have both helicities contributing. Consequently, a flat distribution in Monte Carlo sampling leads to satisfactory results.  Notice that the range of integration could have been reduced to $[0,\pi]$ with the same result. We keep $2\pi$   as the upper bound instead, to accommodate the third degree of freedom of massive gauge bosons. The latter degree of freedom can  be added to Eq. \eqref{sum_lambda} without any phase factor. For a few phase-space points the Born-level and one-loop matrix elements of all subprocesses have been cross-checked against \textsc{Helac-1Loop} \cite{vanHameren:2009dr}. In this framework the one-loop matrix elements are reduced to scalar integrals at the integrand level using the OPP reduction technique \cite{Ossola:2006us} as implemented in the \textsc{CutTools} program \cite{Ossola:2007ax}. Furthermore, for the evaluation of one-loop scalar functions the program \textsc{OneLOop} \cite{vanHameren:2010cp} is employed. The calculation of the real emission part is performed with the Nagy-Soper subtraction scheme \cite{Bevilacqua:2013iha}, which has recently been extended  to also handle calculations in the NWA \cite{Bevilacqua:2022ozv}. In addition, we have used the Catani-Seymour dipole formalism \cite{Catani:1996vz,Catani:2002hc} and its extension to top-quark decays \cite{Bevilacqua:2019quz,Campbell:2004ch} to cross-check our results for a few resonant histories. Both subtraction schemes, together with a phase-space restriction on the subtraction terms \cite{Nagy:1998bb,Nagy:2003tz,Bevilacqua:2009zn,Czakon:2015cla}, are implemented in the \textsc{Helac-Dipoles} MC program \cite{Czakon:2009ss}. The phase-space integration is performed with \textsc{Parni} \cite{vanHameren:2007pt} and \textsc{Kaleu} \cite{vanHameren:2010gg}. To improve the efficiency of the phase-space generation for the subtracted real emission part, additional channels  have been added for the subtraction terms contributing at the different decay stages, see e.g. Ref. \cite{Bevilacqua:2010qb} for more details. All results are stored in modified Les Houches Event files (LHEFs) \cite{Alwall:2006yp} as partially unweighted events \cite{Bevilacqua:2016jfk}. These files include supplementary matrix element and PDF information \cite{Bern:2013zja} needed for reweighting. This makes it possible to create new differential cross-section distributions, change their binning and/or ranges. Furthermore,  different PDF sets as well as  renormalisation/factorialisation scale settings can be used without having to perform new time-consuming calculations.

%
\section{Computational setup}
\label{sec:setup}
%

In this section let us define the setup for the calculations of the present work. As already mentioned we consider top-quark pair production in association with two photons at $\mathcal{O}(\alpha_s^3\alpha^6)$ in the \dilep and \jetlep decay channel. We shall provide the results for the LHC Run 2 energy of $\sqrt{s} = 13$ TeV. As recommended by the PDF4LHC working group we employ three sets of PDFs for the use at the LHC \cite{PDF4LHCWorkingGroup:2022cjn}. The NLO NNPDF3.1 PDF set \cite{NNPDF:2017mvq}, that we employ both at  LO and NLO,  is our default PDF set. The running of the strong coupling constant is, therefore, always performed with two-loop accuracy via the LHAPDF interface
\cite{Buckley:2014ana}. Furthermore, we present additional theoretical results for the NLO MSHT20 \cite{Bailey:2020ooq} and NLO CT18 \cite{Hou:2019efy} PDF sets to quantify the differences between various PDF sets. The $G_\mu$-scheme is used for the derivation of the electromagnetic coupling constant $\alpha$ according to
\begin{equation}
	\alpha_{G_\mu} =\frac{\sqrt{2}}{\pi} \,G_\mu \, m_W^2\,\left(1-\frac{m_W^2}{m_Z^2}\right)\,,
	~~~~~~~~~~~~~~~~~~~~~
	G_{ \mu}=1.1663787 \cdot 10^{-5} \textrm{ GeV}^{-2}\,,
\end{equation}
where $m_{W}= 80.379$ GeV and $m_{Z}=91.1876$ GeV. However, for the emission of the two external hard photons we use  the $\alpha(0)$-scheme with $\alpha^{-1}= \alpha^{-1}(0) =137.035999084$ \cite{Workman:2022ynf}. This choice corresponds to a mixed scheme where the total power of $\alpha$ is split into two parts according to the number of final state photons $\alpha^6\to\alpha^2\alpha_{G_\mu}^4$ and should be renormalised in different schemes when NLO EW corrections are included \cite{Denner:2019vbn,Pagani:2021iwa}. In practise, we use $\alpha_{G_\mu}$ as input value and rescal the final results by $\alpha^2/\alpha_{G_\mu}^2=0.93044...$.  Consequently, the prediction for the $pp \to t\bar{t}\gamma\gamma$ cross section is decreased by about $7 \%$. Furthermore, the following SM parameters are used $m_{t}=172.5$ GeV and $\Gamma_{W} = 2.0972$ GeV. All other particles are considered massless. The LO and NLO top-quark widths are calculated based on Refs. \cite{Jezabek:1988iv,Denner:2012yc} and their numerical values  are given by
\begin{equation}
 \Gamma_{t}^{\rm LO} = 1.4806842 ~{\rm GeV}\,, 
 \quad \quad\quad\quad 
 \Gamma_{t}^{\rm NLO} = 1.3535983  ~{\rm GeV}\,,
\end{equation}
where $\alpha_s(\mu_R=m_t)$ is used to compute $\Gamma_{t}^{\rm NLO}$. The width of the top quark is kept fixed during the estimation of scale uncertainties. However,  the error introduced by this treatment is small. In particular,  for the two scales $\mu_R=m_t/2$ and $\mu_R=2m_t$ is at the level of $1.5\%$ only. The $anti$-$k_T$ jet algorithm \cite{Cacciari:2008gp} with the radius parameter $R = 0.4$ is used to cluster final state partons with pseudo-rapidity $|\eta|<5$ into jets. In the \dilep decay channel we require two 
opposite-sign  charged leptons and exactly two $b$-jets. On the other hand in the \jetlep decay channel we require one negatively charged lepton, exactly two $b$-jets and at least two light jets. In order to avoid QED collinear  singularities in photon emission due to $q \to q \gamma$ splittings, a separation between quarks and photons is required. Since on the experimental side  quark and gluon jets are indistinguishable, a separation between photons and gluons is additionally induced. Consequently, for a given transverse momentum of the photon $(p_{T,\,\gamma})$ an angular restriction is introduced on the phase space of the soft gluon emission. The soft divergence in the real emission part is,  therefore,  different from that in the virtual correction impairing the cancellation of  infrared divergences. To ensure IR-safety at NLO QCD in presence of two isolated prompt photons, the smooth photon isolation prescription, as described in Ref. \cite{Frixione:1998jh}, is used. According to this prescription, the event is rejected unless the following condition is fulfilled before the jet clustering is performed 
\begin{equation}
 \label{eq_iso}
\sum_{i} E_{T\,i}  \, \Theta(R - R_{\gamma i})  \le \epsilon_\gamma \, E_{T\,\gamma} \left(
\frac{1-\cos(R)}{1-\cos(R_{\gamma j})}
\right)^n \,,
\end{equation}
for all $R\le R_{\gamma j}$. We use $R_{\gamma j}=0.4$ and set $\epsilon_{\gamma}=n=1$. Furthermore, $E_{T\,i}$ is the transverse energy of the parton $i$, $E_{T\,\gamma}$ is the transverse energy of the photon and $R_{\gamma i}$ is given by
\begin{equation}
R_{\gamma i}=\sqrt{(y_\gamma-y_i)^2+(\phi_\gamma-\phi_i)^2}\,.
\end{equation}
At the cost of no longer reproducing the form of cone isolation applied in experimental analyses, the smooth photon isolation prescription ensures that arbitrarily soft radiation will always pass the condition, but (hard) collinear $(R \to 0)$ radiation is forbidden. It is important to quantify the differences between prediction obtained with alternative choices of  $\epsilon_{\gamma}$,  $n$ and $E_{T\,\gamma}$, which we also do later in this paper. Of course, the ultimate aim is to compare the results from this work with ATLAS and CMS data when they become available. In addition to the prompt photon requirements described above, the two photons must satisfy 
\begin{equation}
\begin{array}{lll}
p_{T,\,\gamma}>25 ~{\rm GeV}\,,  
&\quad \quad\quad\quad\quad |y_\gamma|<2.5 \,, 
 &\quad \quad\quad \quad \quad
\Delta R_{\gamma\gamma}>0.4\,.
\end{array}
\end{equation}
On the other hand,  for the two $b$-flavoured jets as well as light jets we require 
\begin{equation}
\begin{array}{lll}
p_{T,\,b}>25 ~{\rm GeV}\,,  
&\quad \quad\quad\quad\quad |y_b|<2.5 \,, 
 &\quad \quad\quad \quad \quad
\Delta R_{bb}>0.4\,,
\\[0.2cm]
p_{T,\,j}>25 ~{\rm GeV}\,,  
&\quad \quad\quad\quad\quad |y_j|<2.5 \,, 
 &\quad \quad\quad \quad \quad
\Delta R_{jj}>0.4\,.
\end{array}
\end{equation}
Additionally, charged leptons have to be in the following fiducial volume
\begin{equation}
\begin{array}{lll}
 p_{T,\,\ell}>25 ~{\rm GeV}\,,    
 &\quad \quad \quad \quad\quad|y_\ell|<2.5\,,&
\quad \quad \quad \quad \quad
\Delta R_{\ell
 \ell} > 0.4\,.
\end{array}
\end{equation}
There are no restrictions on the kinematics of the extra  light jet (if resolved by the jet algorithm) and the missing transverse momentum defined as $p_T^{miss} = |\vec{p}_{T,\, \nu_\ell} + \vec{p}_{T,\, \bar{\nu}_\ell}|$. On the other hand, both prompt photons and all jets must be well separated from each other and from charged leptons
\begin{equation}
\begin{array}{lll}
\Delta R_{l\gamma}>0.4\,,  
&\quad \quad\quad\quad\quad \Delta R_{lb}>0.4 \,, 
 &\quad \quad\quad \quad \quad
\Delta R_{lj}>0.4\,,
\\[0.2cm]
\Delta R_{b\gamma}>0.4\,,  
&\quad \quad\quad\quad\quad \Delta R_{bj}>0.4 \,, 
 &\quad \quad\quad \quad \quad
\Delta R_{\gamma j}>0.4\,.
\end{array}
\end{equation}
Finally, in the \jetlep  channel we require that the invariant mass of at least one light-jet pair, denoted as $M_{jj}$, has to  be in the following $W$ boson mass window
\begin{equation}
\label{eq_qcut}
|m_{W}-M_{jj}|<15 ~{\rm GeV}\,,
\end{equation}
where $m_W =80.379$ GeV. Such a cut has already been applied in NLO QCD calculations involving hadronically decaying top quarks. Indeed, it has been used to suppress kinematical configurations from real radiation where the two light jets originating from the $W$ boson are recombined into a single jet, while the extra real radiation gives rise to the presence of a second resolved jet that passes all the cuts \cite{Denner:2017kzu}. It has also been utilised to reduce photon radiation from the hadronically decaying $W$ gauge boson \cite{Melnikov:2011ta}. We employ a dynamical factorisation/renormalisation scale choice
\begin{equation}
\mu_R=\mu_F=\mu_0=\frac{E_T}{4} \,, 
\end{equation}
where $E_T$ is defined according to
 \begin{equation}
 \label{eq_scale}
E_T=\sqrt{m^2_{t}+p_{T, \,t}^2}+\sqrt{m^2_{t}+p_{T,\, \bar{t}}^2 }  + p_{T,\,\gamma_1}+p_{T,\,\gamma_2}\,.
\end{equation} 
In this scale setting, that is defined on an event-by-event basis, $p_{T,\,t}/p_{T,\,\bar{t}}$ are the transverse momenta of the on-shell top quarks. We have checked that variations of that scale setting lead to very similar results. In detail, we have reconstructed top-quark momenta based on the MC truth information with and without photons. We have also used 
a more resonant aware version of this scale definition in which the transverse momentum of the photon has been  included or not in the definition of $E_T$ depending whether this photon is produced in the production or decay stage. As an alternative and for comparison purposes  we also employ a fixed scale setting defined as
\begin{equation}
\mu_R=\mu_F=\mu_0=m_t\,. 
\end{equation}
The scale uncertainties are estimated by a 7-point scale variation in which the factorisation and renormalisation scales are varied independently in the range
\begin{equation}
\frac{1}{2} \, \mu_0  \le \mu_R\,,\mu_F \le  2 \,  \mu_0\,, \quad \quad 
\quad \quad \quad \quad \quad \quad \frac{1}{2}  \le
\frac{\mu_R}{\mu_F} \le  2 \,,
\end{equation}
which leads to the following pairs
\begin{equation}
\label{scan}
\left(\frac{\mu_R}{\mu_0}\,,\frac{\mu_F}{\mu_0}\right) = \Big\{
\left(2,1\right),\left(0.5,1  
\right),\left(1,2\right), (1,1), (1,0.5), (2,2),(0.5,0.5)
\Big\} \,.
\end{equation}
The final scale uncertainties are obtained by finding the minimum and maximum of the resulting cross sections.

%
\section{Di-lepton  channel}
\label{sec:ttaa-lep}
%

%
\subsection{Integrated fiducial cross sections}
\label{sec:ttaa-lep-int}
%

%
\begin{table*}[t!]
    \centering
    \renewcommand{\arraystretch}{1.5}
    \begin{tabular}{l@{\hskip 10mm}ll@{\hskip 10mm}ll@{\hskip 10mm}l}
        \hline\noalign{\smallskip}
        $\mu_0$ &  &  & LO & NLO &${\cal K} = \sigma_{\text{NLO}} / \sigma_{\text{LO}}$ \\
        \noalign{\smallskip}\midrule[0.5mm]\noalign{\smallskip}
        \multirow{4}{*}{$E_T/4$}   
        & $\sigma_{\rm Full}$ & [fb] & $ 0.13868(3)^{+31.2\%}_{-22.1\%} $ & $ 0.1773(1)^{+1.8\%}_{-6.2\%} $ & $ 1.28 $ \\
        & $\sigma_{\rm Prod.}$ & [fb] & $ 0.05399(2)^{+30.6\%}_{-21.7\%} $ & $ 0.07130(6)^{+2.5\%}_{-7.2\%} $ & $ 1.32 $\\
        & $\sigma_{\rm Mixed}$ & [fb] & $ 0.06022(2)^{+31.9\%}_{-22.5\%} $ & $ 0.07733(8)^{+1.5\%}_{-6.2\%} $ & $ 1.28 $\\
        & $\sigma_{\rm Decay}$ & [fb] & $ 0.024473(7)^{+30.9\%}_{-22.1\%} $ & $ 0.02863(4)^{+0.9\%}_{-4.9\%} $ & $ 1.17 $\\

        \noalign{\smallskip}\midrule[0.5mm]\noalign{\smallskip}
        \multirow{4}{*}{$m_t$}   
        & $\sigma_{\rm Full}$ & [fb] & $ 0.13620(3)^{+31.3\%}_{-22.1\%} $ & $ 0.1758(1)^{+1.6\%}_{-6.3\%} $ & $ 1.29 $ \\
        & $\sigma_{\rm Prod.}$ & [fb] & $ 0.05484(2)^{+31.2\%}_{-21.9\%} $ & $ 0.07091(6)^{+2.2\%}_{-6.7\%} $ & $ 1.29 $\\
        & $\sigma_{\rm Mixed}$ & [fb] & $ 0.05847(2)^{+31.8\%}_{-22.4\%} $ & $ 0.07651(8)^{+1.4\%}_{-6.5\%} $ & $ 1.31 $\\
        & $\sigma_{\rm Decay}$ & [fb] & $ 0.022883(7)^{+30.5\%}_{-21.9\%} $ & $ 0.02840(3)^{+0.8\%}_{-4.7\%} $ & $ 1.24 $\\

        \noalign{\smallskip}\hline\noalign{\smallskip}
    \end{tabular}
    \caption{\label{tab:int-lep-scale} \it Integrated fiducial cross sections at LO and NLO QCD for the $pp\to \ell^+\nu_{\ell}\, \ell^-\bar{\nu}_{\ell} \, b\bar{b}\,\gamma\gamma +X$ process at the LHC with $\sqrt{s}=13~{\rm TeV}$. Results are presented for two scale settings $\mu_0=E_T/4$ and $\mu_0=m_t$ as well as for the three contributions: {\it Prod.}, {\it Mixed} and {\it Decay}. The NLO NNPDF3.1 PDF set is employed. The theoretical uncertainties from the 7-point  scale variation and MC integration errors (in parentheses) are also displayed. }
\end{table*}
We begin the presentation of our results with a discussion of the integrated fiducial cross section for the process $pp\to \ell^+\nu_{\ell}\, \ell^-\bar{\nu}_{\ell} \, b\bar{b}\,\gamma\gamma +X$. In Table \ref{tab:int-lep-scale} we show our findings in the \dilep  channel at LO and NLO QCD.  The results are shown for two scale settings $\mu_0=E_T/4$ and $\mu_0=m_t$ using the NLO NNPDF3.1 PDF set in both cases. The NLO QCD corrections to the full process are rather moderate, of the order of $30\%$ for both scale choices and thus of the same size as the LO scale uncertainties. The latter uncertainties are reduced at NLO in QCD to $6\%$, so  by a factor of $5$. At the integrated cross-section level, the difference in the theoretical results between the two scale settings is negligible and decreases from $2\%$ at LO to $1\%$ at NLO. In order to examine the radiation pattern of the prompt photons, the cross section is divided into the following three contributions: {\it Prod.}, {\it Mixed} and {\it Decay}, as defined in Eq. \eqref{eq_res}.  In particular, within our fiducial phase-space volume, independently of the scale choice as well as the perturbative order in QCD, the integrated cross section is dominated by the {\it Mixed} contribution $(43\%-44\%)$ rather than the  {\it Prod.} one. The latter contribution is at the level of $39\%-40\%$ only. On the other hand, the {\it Decay} contribution is about half the size $(16\%-18\%)$. Similar large effects from photon radiation  from top-quark decays have already been observed for the simpler process $pp \to t \bar{t}\gamma$ \cite{Melnikov:2011ta,Bevilacqua:2019quz}. In that case, due to the absence of the {\it Mixed} contribution, photon radiation is evenly distributed between {\it Prod.} and {\it Decay}. For the $pp \to t\bar{t}\gamma\gamma$ process NLO QCD corrections for the {\it Prod.} and {\it Mixed} contribution are of the order of $30\%$. Substantially smaller higher order effects are found for the {\it Decay} part. Indeed, they are at the level of $17\%$ and $24\%$ for $\mu_0=E_T/4$ and $\mu_0=m_t$ respectively. The overall behaviour is very similar for both scale settings. The largest differences between the two scales choices can be found for the {\it Decay} contribution at LO, which are at the level of $7\%$ while the differences at NLO are always below $1\%$.

In the following, we examine the decomposition into different resonance configurations, {\it Prod.}, {\it Mixed} and {\it Decay}, in more detail by dividing the NLO QCD cross section into the different production channels, namely: $gg$, $q\bar{q}$ and $qg/\bar{q}g$. Results are shown in Table \ref{tab:int-lep-ch}. They are obtained for the scale setting $\mu_0=E_T/4$ and  the NLO NNPDF3.1 PDF set.  We first note that at the central scale the $gg$ channel dominates the  integrated fiducial NLO QCD cross section by about $56.4\%$ followed by the $q\bar{q}$ channel with $24.3\%$ and at last the $qg/\bar{q}g$ channel with $19.3\%$. The relative size of the $gg$ channel decreases for the {\it Prod.} contribution alone and amounts to $36.3\%$ which is about the same size as the $q\bar{q}$ channel with $37.5\%$. In absolute size the {\it Mixed} contribution of the $q\bar{q}$ and $qg/\bar{q}g$ channels is smaller compared to the {\it Prod.} contribution as expected from simple phase-space arguments. Indeed, the finite mass of the top quark and $W$ boson limits the available phase space of the decay products. In contrast, in the $gg$ channel the {\it Mixed} contribution increases in absolute size and becomes almost twice as large as the {\it Prod.} part. Furthermore, it dominates the overall {\it Mixed} contribution for the whole process with about $64\%$.  We find that for all three production channels the {\it Decay} contribution is the least relevant. When comparing it to the {\it Mixed} contribution we observe that in the $gg$ channel its contributions is reduced by a factor of $2$, while  in the $q\bar{q}$ and $qg/\bar{q}g$ channels we receive a factor of $6$ and $9$ respectively. Thus, the {\it Decay} contribution is clearly dominated by the $gg$ channel with $86.5\%$. To conclude, on the one hand, we find that the $gg$ production channel is suppressed as the number of photons in the $t\bar{t}$ production stage increases, because photons can only be radiated from the top-quark line. With more photon emissions in top-quark decays, on the other hand, the available phase space is reduced and both $q\bar{q}$ and $qg$ channels become smaller in absolute size. In the $gg$ production channel these two effects  cancel each other out and the absolute cross section can even increase with a decreasing number of photons in the $t\bar{t}$ production. This ultimately leads to the observed large {\it Mixed} and {\it Decay} contributions  that we have found earlier.  We note that this is the opposite behaviour to that for the  $pp \to t\bar{t}jj$ process at NLO in QCD, in which jet radiation in $t\to Wb$ decays is much stronger suppressed,  as has been shown in Ref.  \cite{Bevilacqua:2022ozv}, and the two contributions, {\it Mixed} and {\it Decay}, are generally much smaller.
\begin{table*}[t!]
    \centering
    \renewcommand{\arraystretch}{1.5}
    \begin{tabular}{ll@{\hskip 3mm}cc@{\hskip 3mm}cc@{\hskip 3mm}cc}
        \hline\noalign{\smallskip}
        &&$gg$& $gg/pp$& $q\bar{q}$ &  $q\bar{q}/pp$ &$qg+\bar{q}g$ & $(qg+\bar{q}g)/pp$\\
        \noalign{\smallskip}\midrule[0.5mm]\noalign{\smallskip}
        $\sigma_{\rm Full}^{\textrm{NLO}}$ & [fb] & $ 0.0999(1) $ &  $56.4\%$& $ 0.04307(4) $ & $24.3\%$& $ 0.03428(4) $& $19.3\%$  \\
        $\sigma_{\rm Prod.}^{\textrm{NLO}}$ & [fb] & $ 0.02587(4) $ & $36.3\%$ & $ 0.02672(4) $ &$37.5\%$& $ 0.01871(3) $  & $26.2\%$ \\
        $\sigma_{\rm Mixed}^{\textrm{NLO}}$ & [fb] & $ 0.04928(8) $& $63.7\%$ & $ 0.01408(2) $ & $18.2\%$& $ 0.01398(2) $  & $18.1\%$ \\
        $\sigma_{\rm Decay}^{\textrm{NLO}}$ & [fb] & $ 0.02476(4) $ & $86.5\%$ & $ 0.002268(3) $ & $7.9\%$& $ 0.00160(2) $  & $5.6\%$\\
        \noalign{\smallskip}\hline\noalign{\smallskip}
    \end{tabular}
    \caption{\label{tab:int-lep-ch} \it Integrated fiducial cross sections at NLO QCD for the $pp\to \ell^+\nu_{\ell}\, \ell^-\bar{\nu}_{\ell} \, b\bar{b}\,\gamma\gamma +X$ process at the LHC with $\sqrt{s}=13~{\rm TeV}$. Results are divided into the different channels: $gg$, $q\bar{q}$ and $qg/\bar{q}g$ as well as given for the three contributions: {\it Prod.}, {\it Mixed} and {\it Decay}.  Relative contributions for these three categories are also provided. The $\mu_0=E_T/4$ scale setting and  NNPDF3.1 PDF set are employed. MC integration errors (in parentheses) are also displayed. }
\end{table*}
\begin{table*}[t!]
    \centering
    \renewcommand{\arraystretch}{1.5}
    \begin{tabular}{l@{\hskip 10mm}l@{\hskip 10mm}l@{\hskip 10mm}l}
        \hline\noalign{\smallskip}
        PDF set&$\sigma_{\text{\textrm{Full}}}^{\textrm{NLO}}$ [fb]&$\delta_{scale}$&$\delta_{\rm PDF}$ \\
        \noalign{\smallskip}\midrule[0.5mm]\noalign{\smallskip}
        NNPDF3.1 & $ 0.1773(1) $ & ${}^{+1.8\%}_{-6.2\%}$ & ${}^{+1.0\%}_{-1.0\%}$ \\
        CT18     & $ 0.1730(2) $ & ${}^{+1.8\%}_{-6.2\%}$ & ${}^{+1.9\%}_{-2.0\%}$ \\
        MSHT20   & $ 0.1742(2) $ & ${}^{+1.8\%}_{-6.2\%}$ & ${}^{+1.4\%}_{-1.3\%}$ \\
        \noalign{\smallskip}\hline\noalign{\smallskip}
    \end{tabular}
    \caption{\label{tab:int-lep-pdf} \it 
    Integrated fiducial cross sections at NLO QCD for the $pp\to \ell^+\nu_{\ell}\, \ell^-\bar{\nu}_{\ell} \, b\bar{b}\,\gamma\gamma +X$ process at the LHC with $\sqrt{s}=13~{\rm TeV}$. Results for three different PDF sets are shown using the $\mu_0=E_T/4$ scale setting. The theoretical uncertainties from the 7-point scale variation are denoted as $\delta_{scale}$, whereas the internal PDF uncertainties are labeled as $\delta_{\rm PDF}$. The former uncertainties are given for comparison purposes.}
\end{table*}

Finally, while employing the dynamical scale setting $\mu_0 = E_T /4$, we examine the second main source of theoretical uncertainties that comes from the choice of the PDF set. We use the corresponding prescriptions from each PDF fitting group to provide the $68\%$ confidence level  internal PDF uncertainties. Our findings are summarised in Table \ref{tab:int-lep-pdf}, where the integrated fiducial cross section at NLO in QCD for the three PDF sets NNPDF3.1, CT18 and MSHT20 is given. The internal PDF uncertainties amount to $1\%$ for NNPDF3.1, $2.0\%$ for CT18 and $1.4\%$ for MSHT20. When comparing NLO theoretical predictions  for NNPDF3.1, CT18 and MSHT20, the largest differences are found between the NNPDF3.1 and CT18 PDF sets of about $2.4\%$. They are therefore of a similar size as the internal PDF uncertainties. The theoretical uncertainties from the 7-point scale variation of about $6\%$ are larger by a factor of $3-6$ than the internal PDF uncertainties and remain the dominant source of theoretical uncertainties.

%
\subsection{Differential fiducial cross sections}
\label{sec:ttaa-lep-diff}
%

We turn our attention to the size of NLO QCD corrections at the differential cross-section level. Our goal here is to assess additional shape distortions on top of the NLO QCD  correction of about $30\%$, which are already present for the normalisation.
\begin{figure}[t!]
    \begin{center}
	\includegraphics[width=0.49\textwidth]{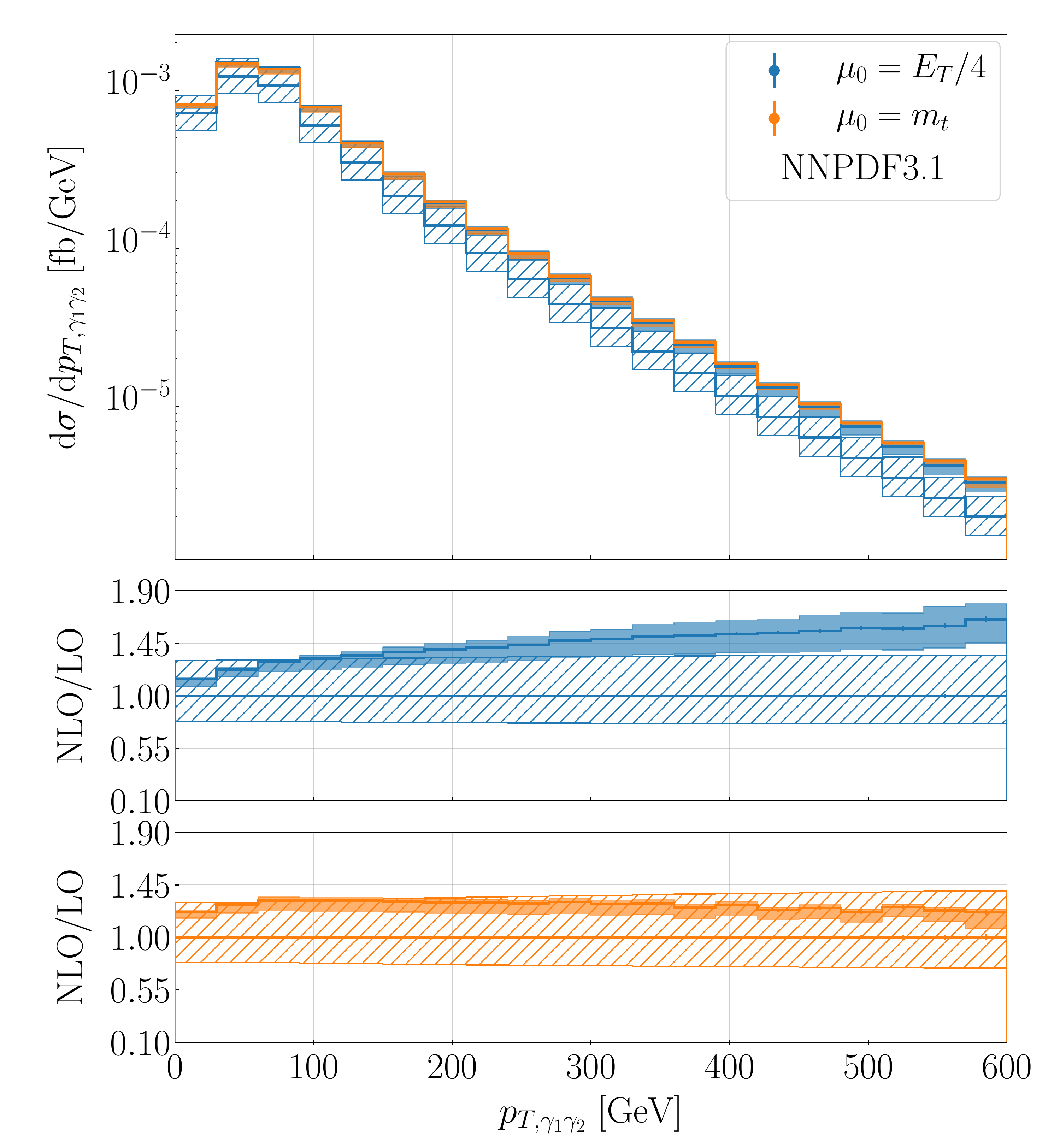}
	\includegraphics[width=0.49\textwidth]{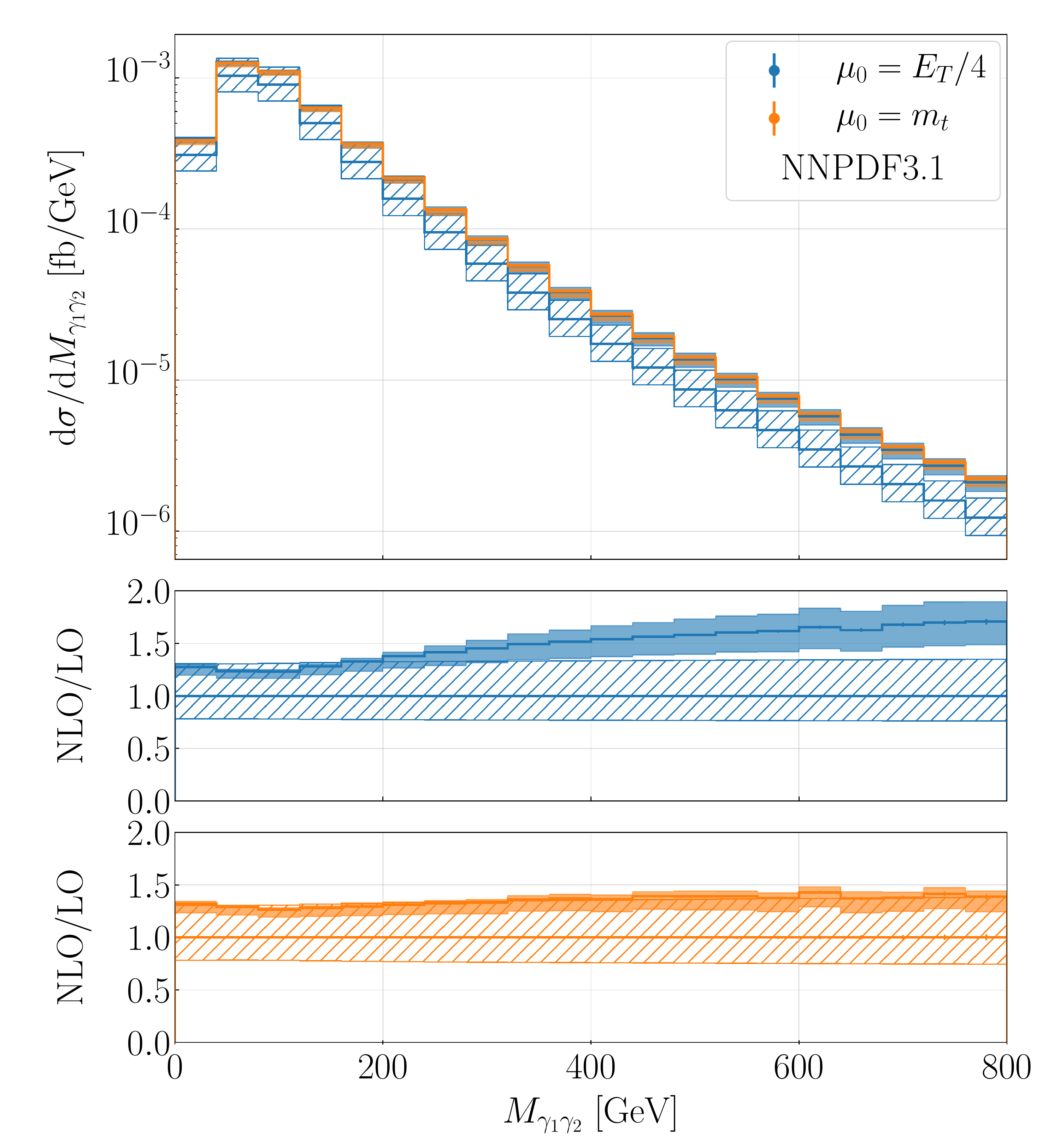}
	\includegraphics[width=0.49\textwidth]{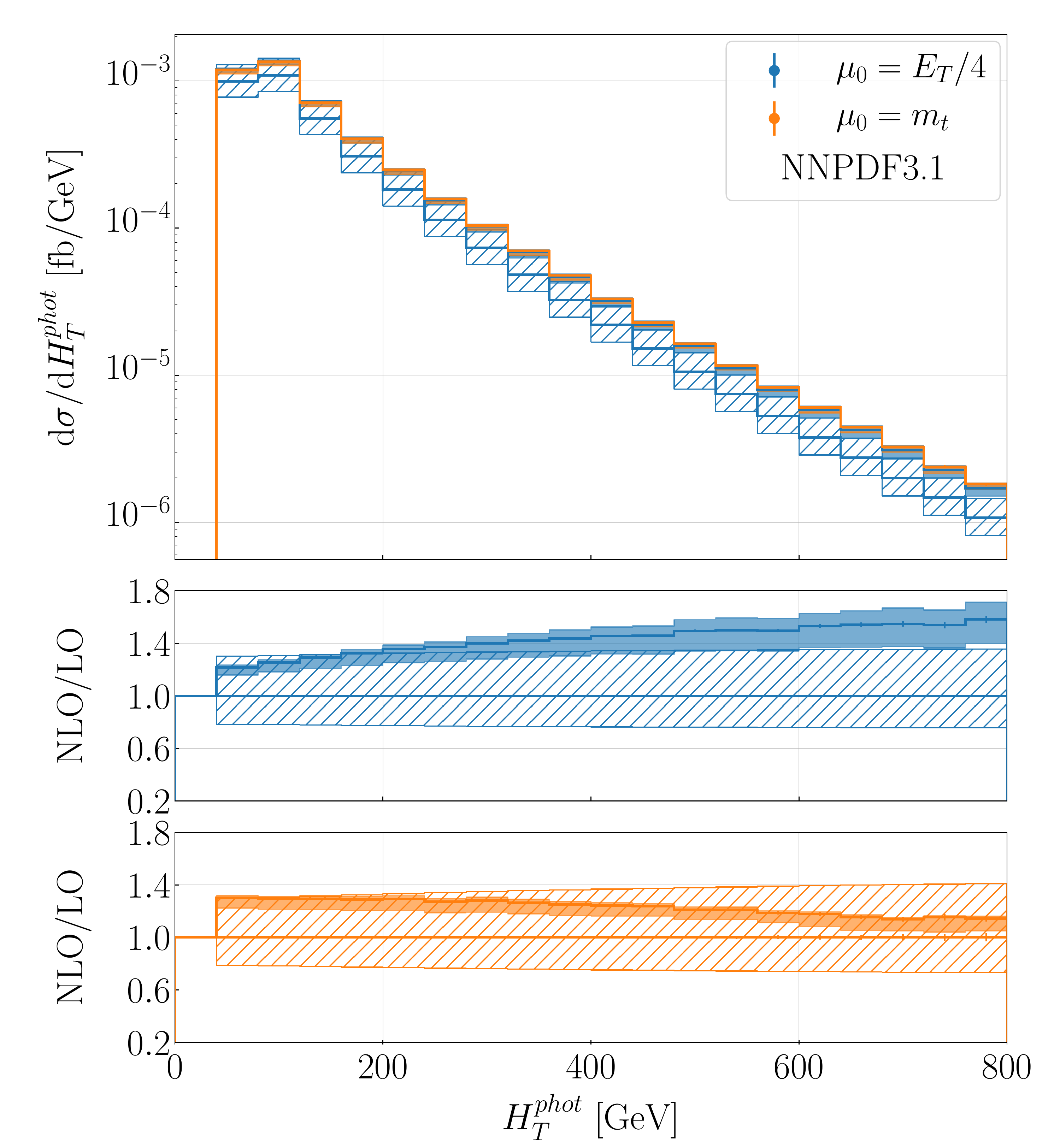}
	\includegraphics[width=0.49\textwidth]{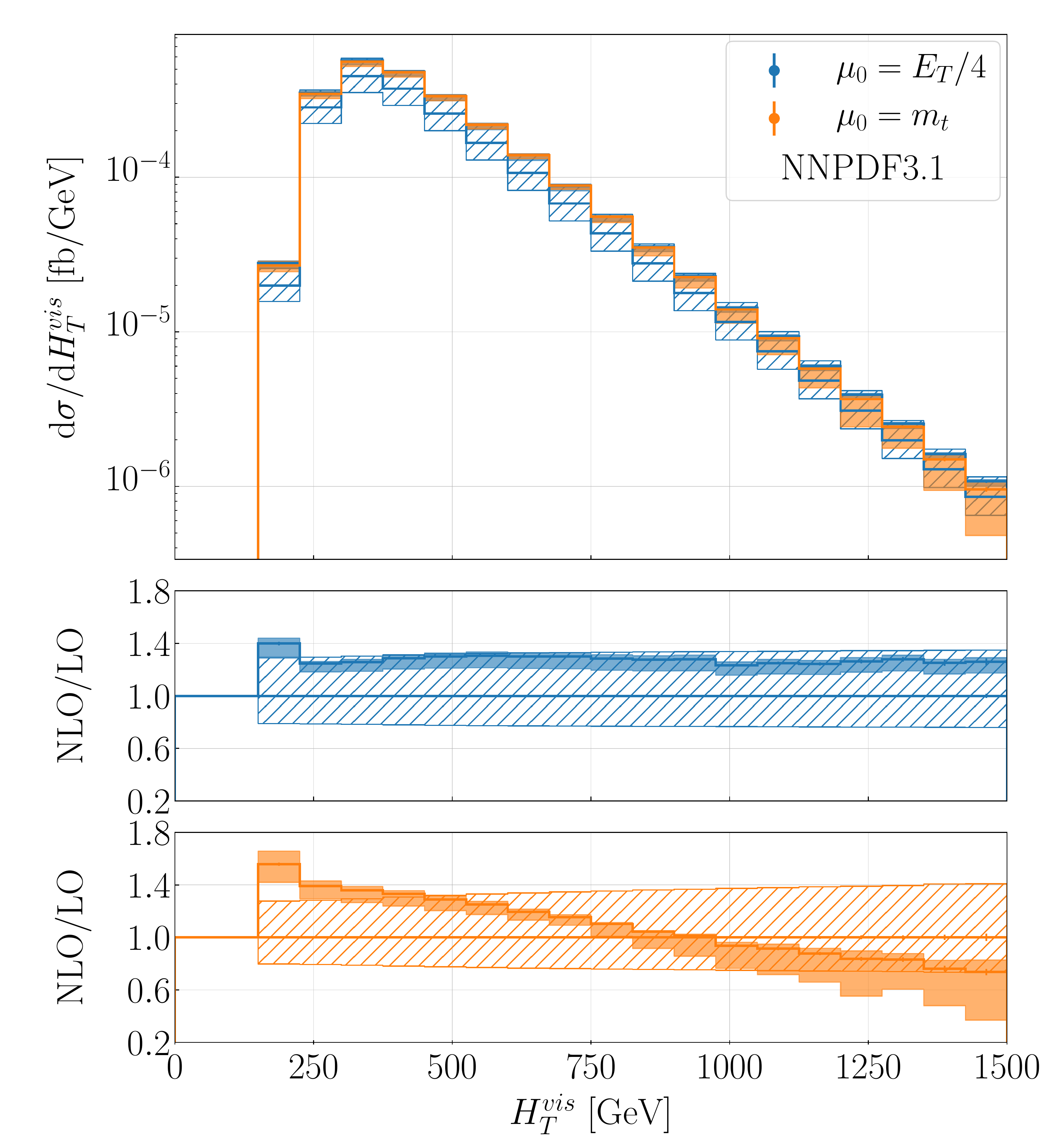}
    \end{center}
    \caption{\label{fig-lep:kfac1} \it Differential cross-section distributions for the observables $p_{T,\gamma_1\gamma_2}$, $M_{\gamma_1\gamma_2}$, $H_T^{phot}$ and $H_T^{vis}$ for the $pp\to \ell^+\nu_{\ell}\, \ell^-\bar{\nu}_{\ell} \, b\bar{b}\,\gamma\gamma +X$ process at the LHC with $\sqrt{s}=13$ TeV. The corresponding  theoretical uncertainties  from the 7-point scale variation and MC integration errors are also shown. Results are presented for  $\mu_0=E_T/4$ (blue) and $\mu_0=m_t$ (orange) at NLO (solid) and LO (dashed) using the NNPDF3.1 PDF set. The two lower panels display the differential ${\cal K}$-factor for both scale choices with the uncertainty band and the relative scale uncertainties of the LO cross section. }
\end{figure}
In Figure \ref{fig-lep:kfac1} we show differential cross section distributions for a few observables for the $pp\to \ell^+\nu_{\ell}\, \ell^-\bar{\nu}_{\ell} \, b\bar{b}\,\gamma\gamma +X$ process at the LHC with $\sqrt{s}=13$ TeV. In detail, we display the transverse momentum, $p_{T,\gamma_1\gamma_2}$, and the invariant mass, $M_{\gamma_1\gamma_2}$, of the $\gamma_1\gamma_2$ system, the scalar sum of transverse momenta of the two prompt photons, $H_T^{phot}$, defined according to 
\begin{equation}
    H_T^{phot}=p_{T,\, \gamma_1}+p_{T,\, \gamma_2}\,,
\end{equation}
and the scalar sum of transverse momenta of all visible particles in the final state, $H_T^{vis}$, given by
\begin{equation}
    H_T^{vis}=p_{T,\, \ell^+}+p_{T,\, \ell^-}+p_{T,\, b_1}+p_{T,\, b_2}+p_{T,\, \gamma_1}+p_{T,\, \gamma_2}.
\end{equation}
Results are presented for $\mu_0=E_T/4$ (blue) and $\mu_0=m_t$ (orange) at  NLO (solid) and LO (dashed). Also given are the corresponding uncertainty bands resulting from the 7-point scale variation. These uncertainties are assessed on a bin-by-bin basis. The two lower panels display the differential ${\cal K}$-factor for both scale choices with
the corresponding uncertainty bands of the LO and NLO predictions.  The first two observables, $p_{T,\gamma_1\gamma_2}$ and $M_{\gamma_1\gamma_2}$, are very interesting to study as they represent an important irreducible background to the transverse momentum and invariant mass of the Higgs boson in the $pp \to t\bar{t}H$ process with $H\to \gamma\gamma$. Large NLO QCD corrections up to  $65\%$ for $p_{T,\gamma_1\gamma_2}$ and $70\%$ for $M_{\gamma_1\gamma_2}$ are found in
the tail of the two distributions for the dynamical scale setting. Thus, higher-order corrections  exceed the LO scale uncertainties, that are in the range of  $31\%-35\%$, by a factor of $2$. When instead the fixed scale setting is employed,  NLO QCD corrections are reduced to $22\%-32\%$ for $p_{T,\gamma_1\gamma_2}$ and $26\%-42\%$ for $M_{\gamma_1\gamma_2}$ leading to a better agreement with the corresponding LO predictions. On the other hand, the difference between the two NLO results for the central value of $\mu_0$ is only about $2\%$ for the bulk of the distribution. Towards the tails this difference  increases up to $5\%$, showing very good agreement, which is well within the NLO theoretical uncertainties. The latter uncertainties, depending on the phase-space region, are of the order of  $5\%-13\%$. Similar features are observed for $H_T^{phot}$, where the difference between the two NLO results are at most $7\%$ while the NLO scale uncertainties are about $12\%$ for the dynamical and $10\%$ for the fixed scale choice.  Also in this case the fixed scale setting seems to describe this differential cross-section distribution better. Indeed, NLO QCD corrections are reduced from $58\%$ to $30\%$ when $\mu_0=m_t$ is employed. The large differential ${\cal K}$-factor for pure photon, dimensionful observables for $\mu_0=E_T/4$ is driven by the LO predictions.  On the other hand,  we find for the $H_T^{vis}$ observable with  the fixed scale setting perturbative instabilities in the tail of the distribution. In this region the NLO scale uncertainties rapidly increase to about $50\%$ and exceed the LO scale uncertainty bands. In that case, the dynamical scale choice at NLO in QCD is essential and leads to a reduction of the scale uncertainties from $35\%$ at LO to $7\%$ at NLO in QCD as well as rather flat higher-order corrections of about $30\%$. 
\begin{figure}[t!]
    \begin{center}
	\includegraphics[width=0.49\textwidth]{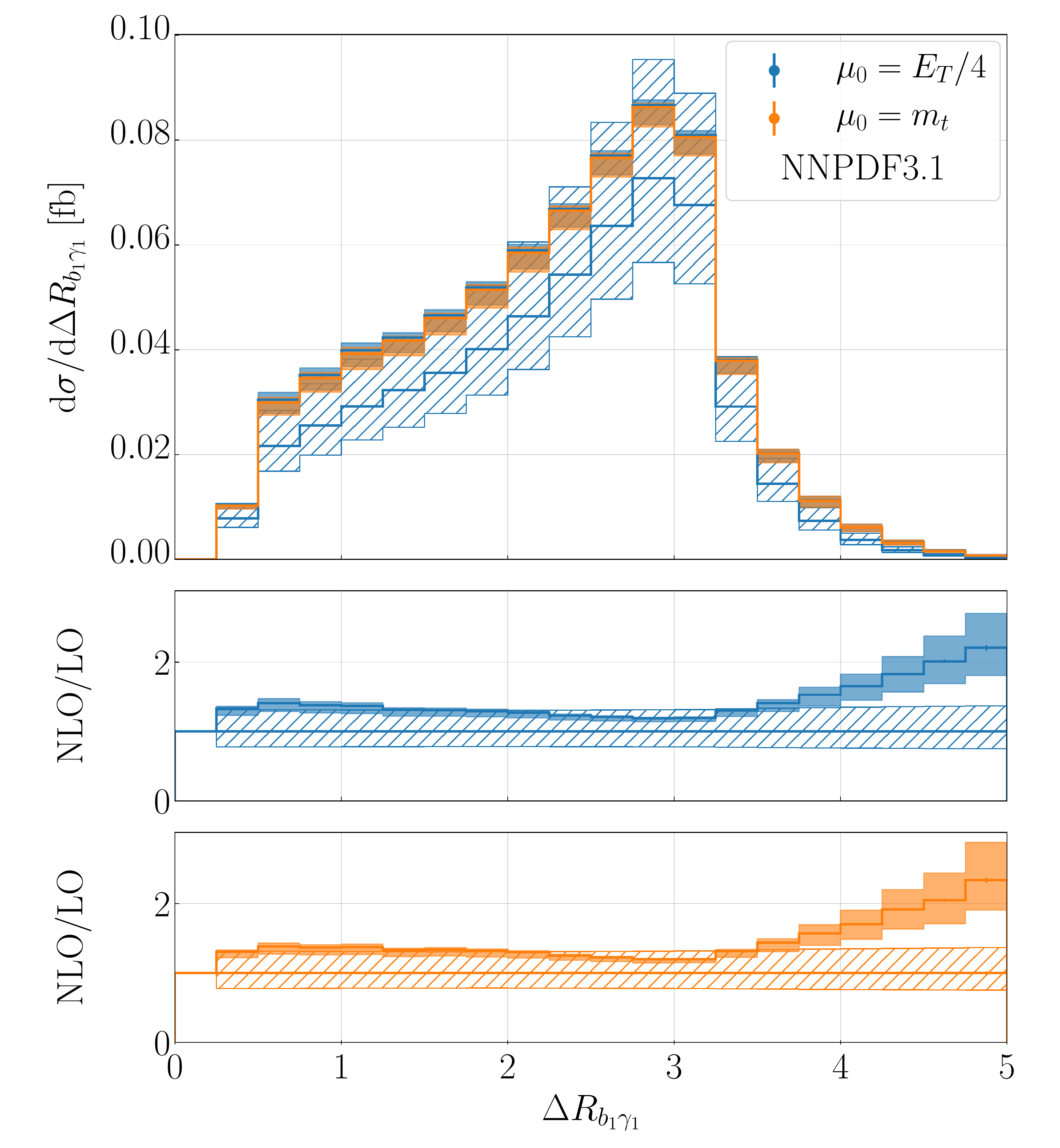}
	\includegraphics[width=0.49\textwidth]{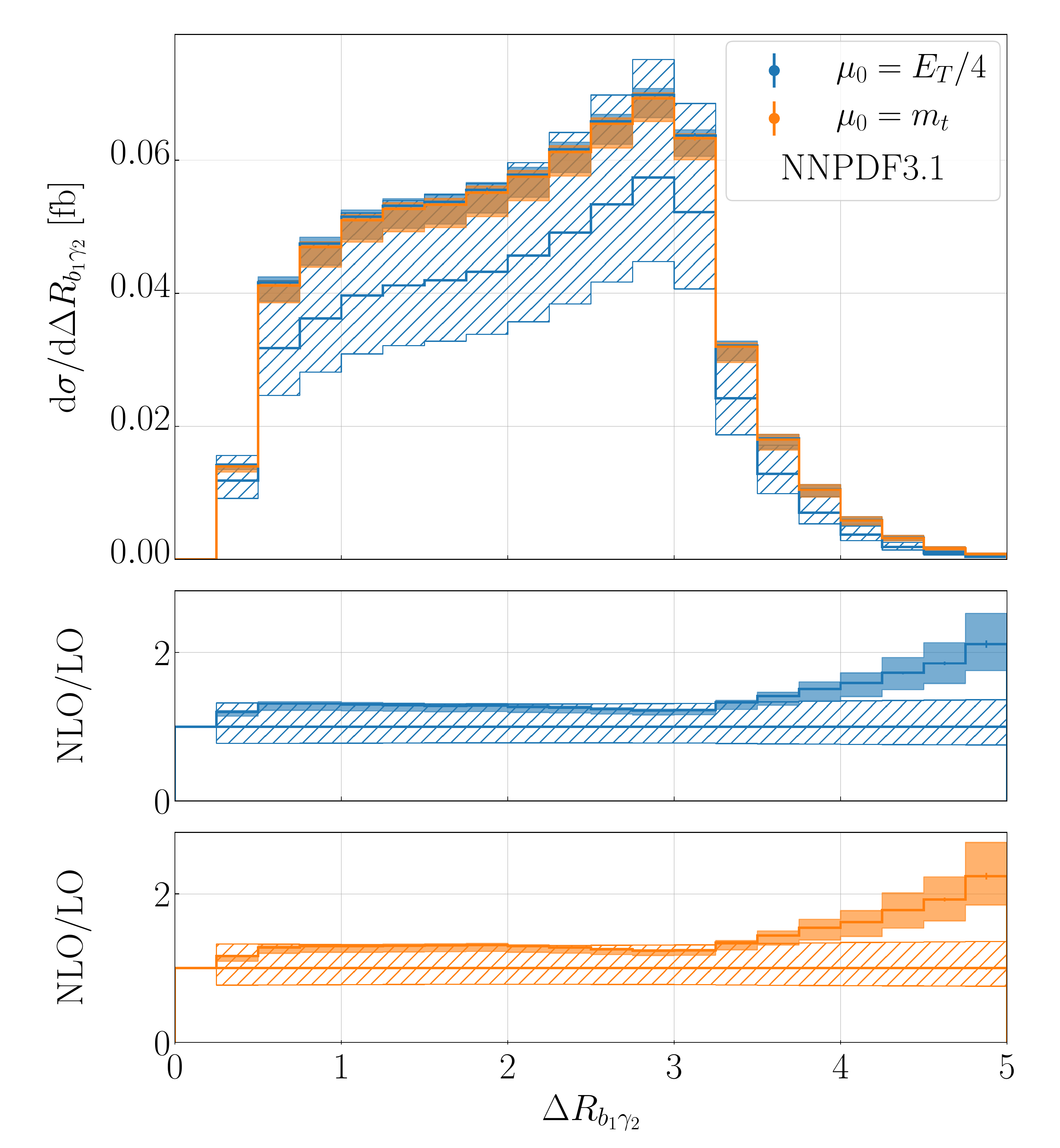}
	\includegraphics[width=0.49\textwidth]{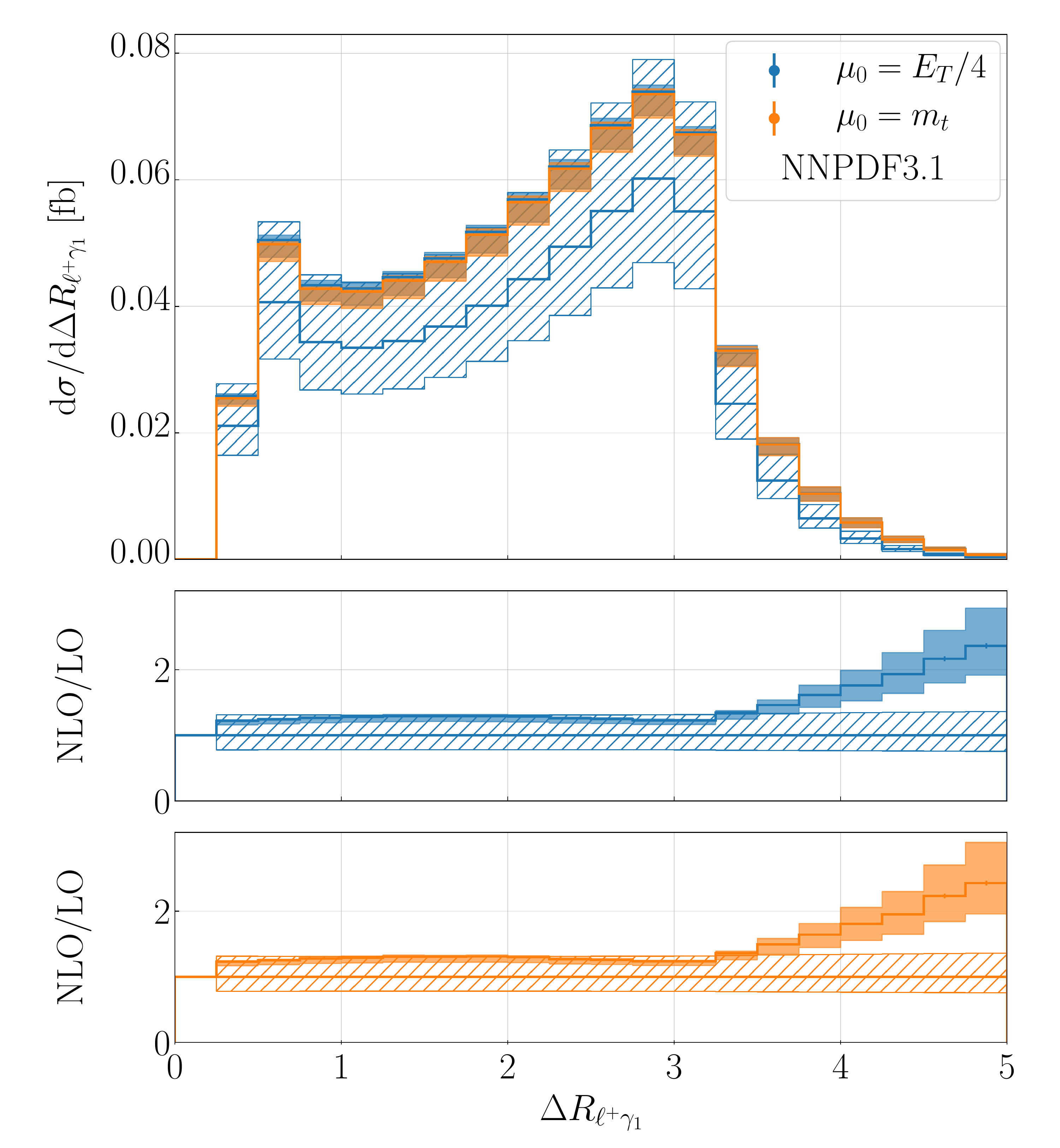}
 	\includegraphics[width=0.49\textwidth]{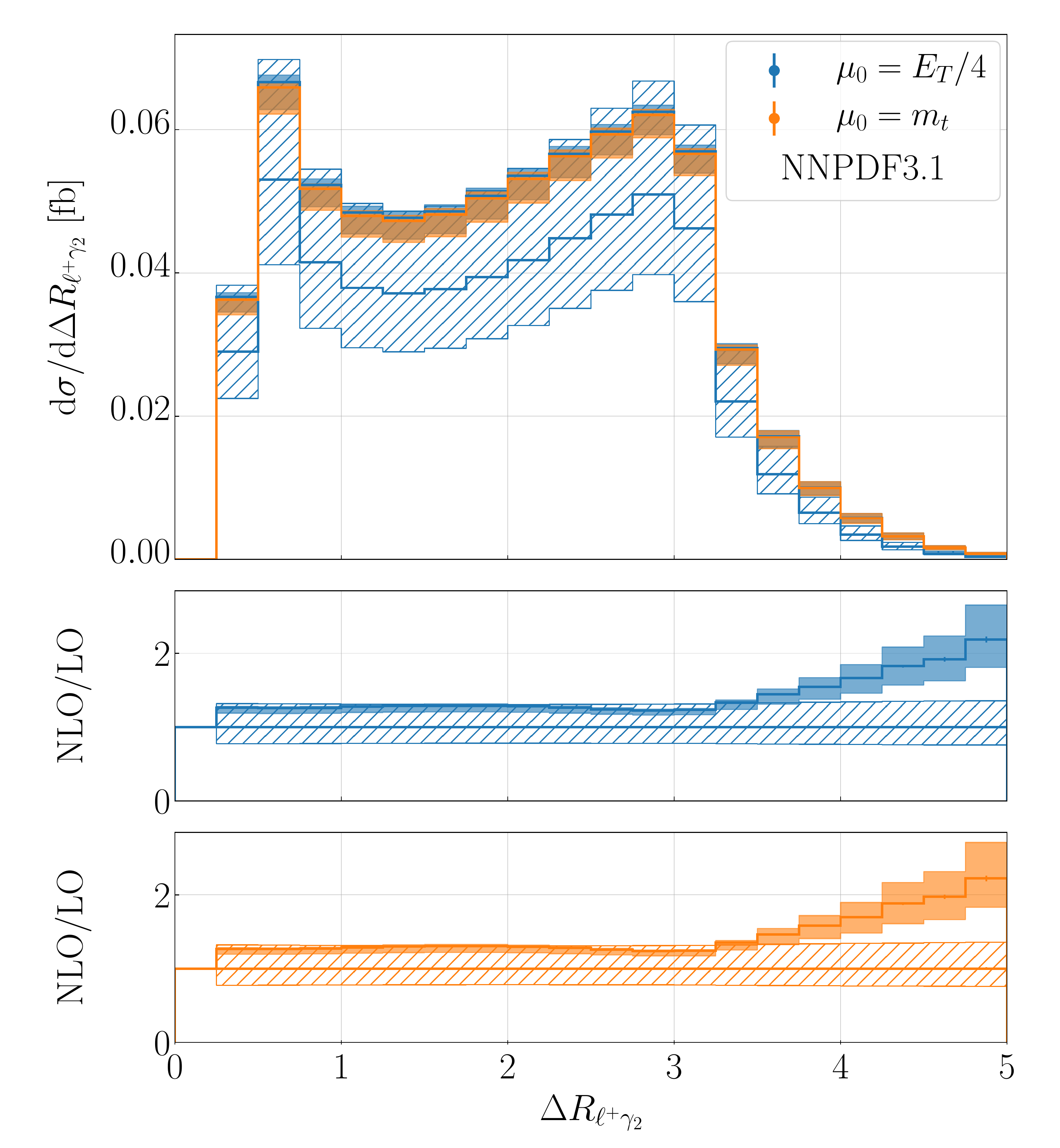}
    \end{center}
    \caption{\label{fig-lep:kfac2} \it Same as Figure \ref{fig-lep:kfac1} but for the observables $\Delta R_{b_1\gamma_1}$, $\Delta R_{b_1\gamma_2}$, $\Delta R_{\ell^+\gamma_1}$ and $\Delta R_{\ell^+\gamma_2}$. }
\end{figure}

We continue with the discussion of NLO QCD corrections at the differential cross-section level but this time we turn our attention to dimensionless observables. Specifically, we analyse  the angular separation between the hardest $b$-jet and the two prompt photons in the rapidity azimuthal angle plane: $\Delta R_{b_1\gamma_1}$, $\Delta R_{b_1\gamma_2}$. In addition, we examine the angular separation between the positively charged lepton and the two prompt photons:  $\Delta R_{\ell^+\gamma_1}$ and $\Delta R_{\ell^+\gamma_2}$. These four observables, that  are shown in Figure \ref{fig-lep:kfac2}, provide important information on the probability of photon emission at different stages of the top-quark decay chain. What all these observables have in common is a peak  around $\Delta R_{ij}\approx 3$, where $i=b_1,\ell^+$ and $j=\gamma_1,\gamma_2$, showing that photons and decay products of top quarks/$W$ bosons are preferably produced in the back-to-back configuration. In addition, a second peak is present for small $\Delta R_{\ell^+\gamma_1}$ and $\Delta R_{\ell^+\gamma_2}$, indicating that photon emission can also originate from the two charged leptons in more collinear configurations.  As we will see later, such configurations are only present for the {\it Mixed} and {\it Decay} contributions. Even though, the second peak at $\Delta R_{ij}\approx 0.4$ is absent  
for $\Delta R_{b_1\gamma_1}$ and $\Delta R_{b_1\gamma_2}$,  the  $\Delta R_{b_1\gamma_2}$ observable is enhanced with respect to $\Delta R_{b_1\gamma_1}$ in this phase-space region, showing the increased probability that the second hardest photon can be emitted from the hardest $b$-jet. The size of NLO QCD corrections is very similar for all four observables and for both scale choices. The largest higher-order effects, up to $40\%$, can be found for $\Delta R_{b_1\gamma_1}$ at the beginning of the spectrum, while  for $\Delta R_{b_1\gamma_1}\approx 3$ NLO QCD corrections of about $20\%-35\%$ are present.  NLO scale uncertainties, on the other hand, are at the level of  $5\%-8\%$ for all four observables. For large angular separation, i.e. for  $\Delta R_{ij} > 4$, the differential ${\cal K}$-factor rapidly rises to more than $2$ and the scale uncertainties increase to $20\%$. These phase-space regions are, however, the least populated. We also note that the differences between theoretical predictions for the dynamical and fixed-scale setting at NLO in QCD do not exceed $2\%-3\%$ for all four differential cross-section distributions.  This can be explained by the dimensionless nature of these observables. Indeed,  they receive contributions from all scales, most notably from those that are sensitive to the threshold for the top-quark pair production. For our scale settings, effects coming from the phase-space regions close to this threshold dominate and the  dynamic $\mu_0=E_T/4$ scale does not substantially alter this behavior.
\begin{figure}[t!]
    \begin{center}
	\includegraphics[width=0.49\textwidth]{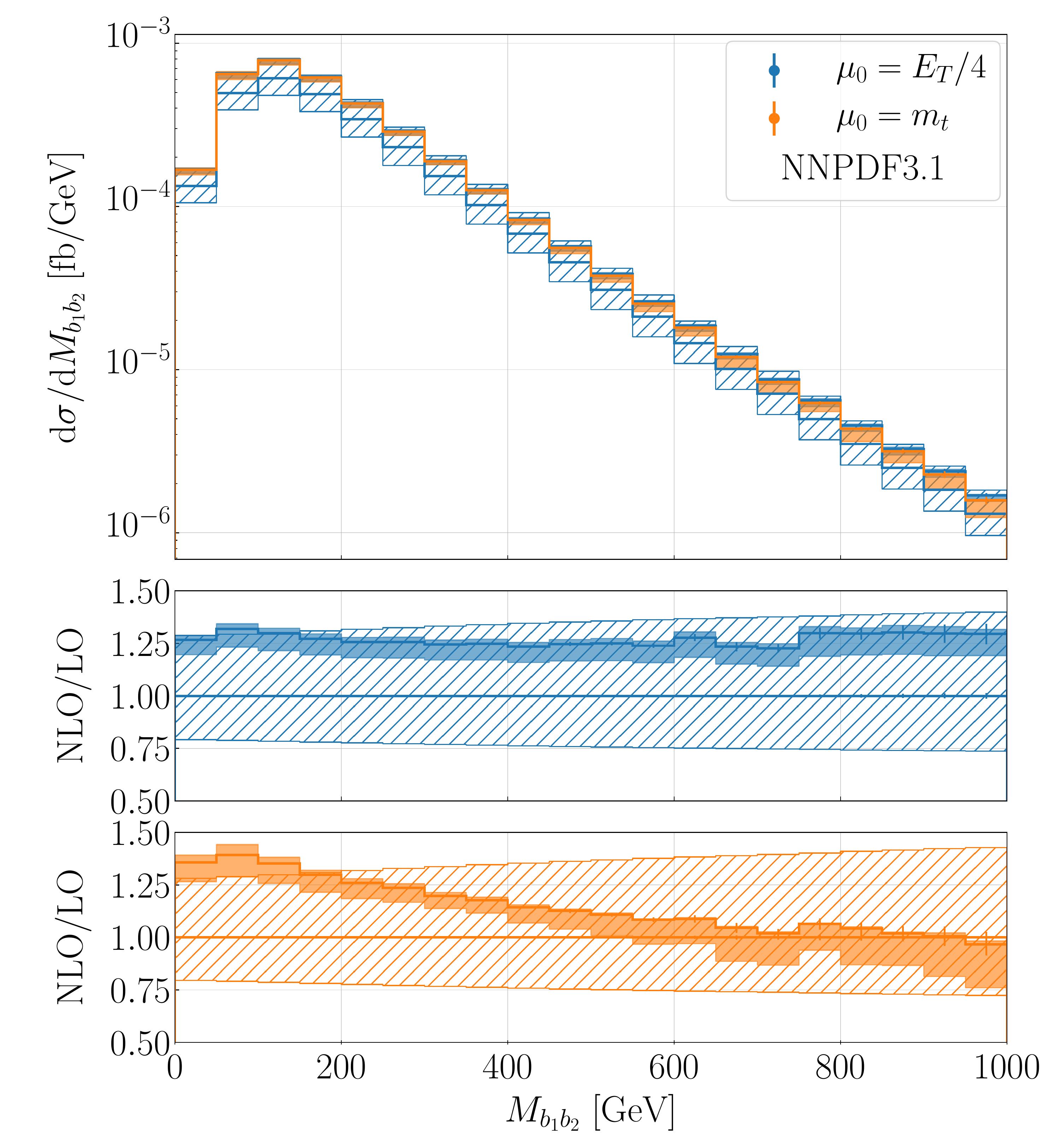}
	\includegraphics[width=0.49\textwidth]{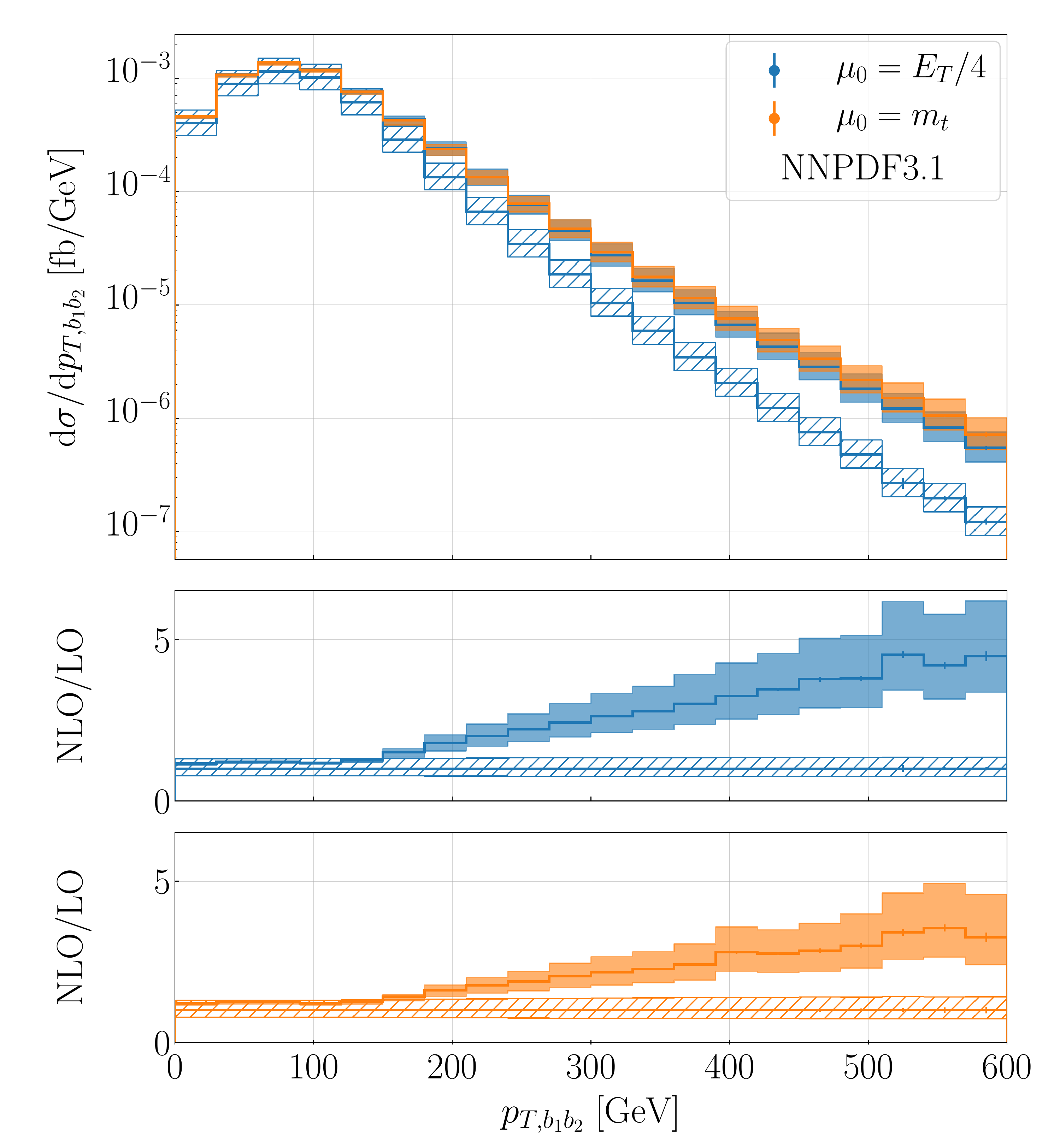}
	\includegraphics[width=0.49\textwidth]{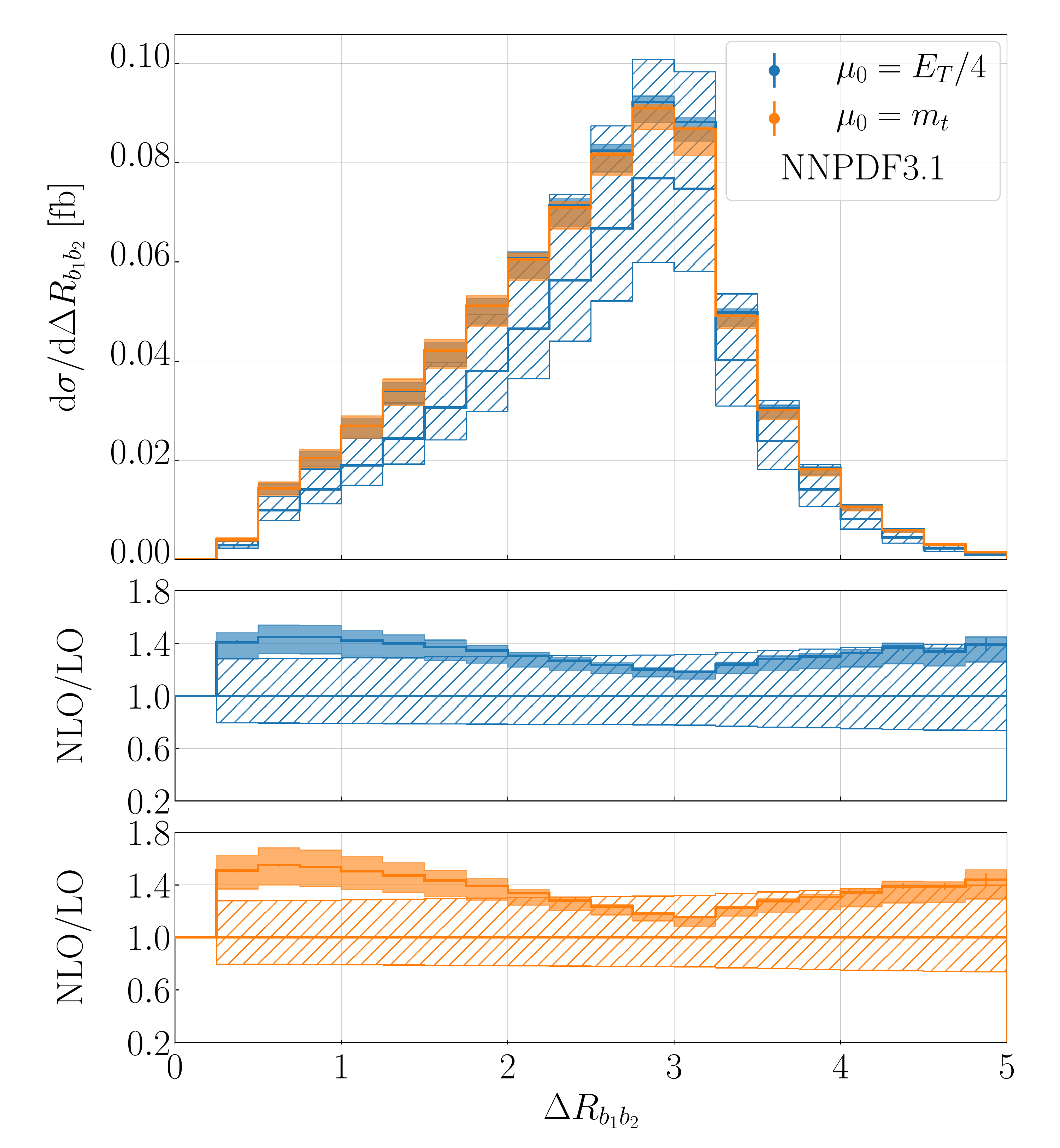}
 	\includegraphics[width=0.49\textwidth]{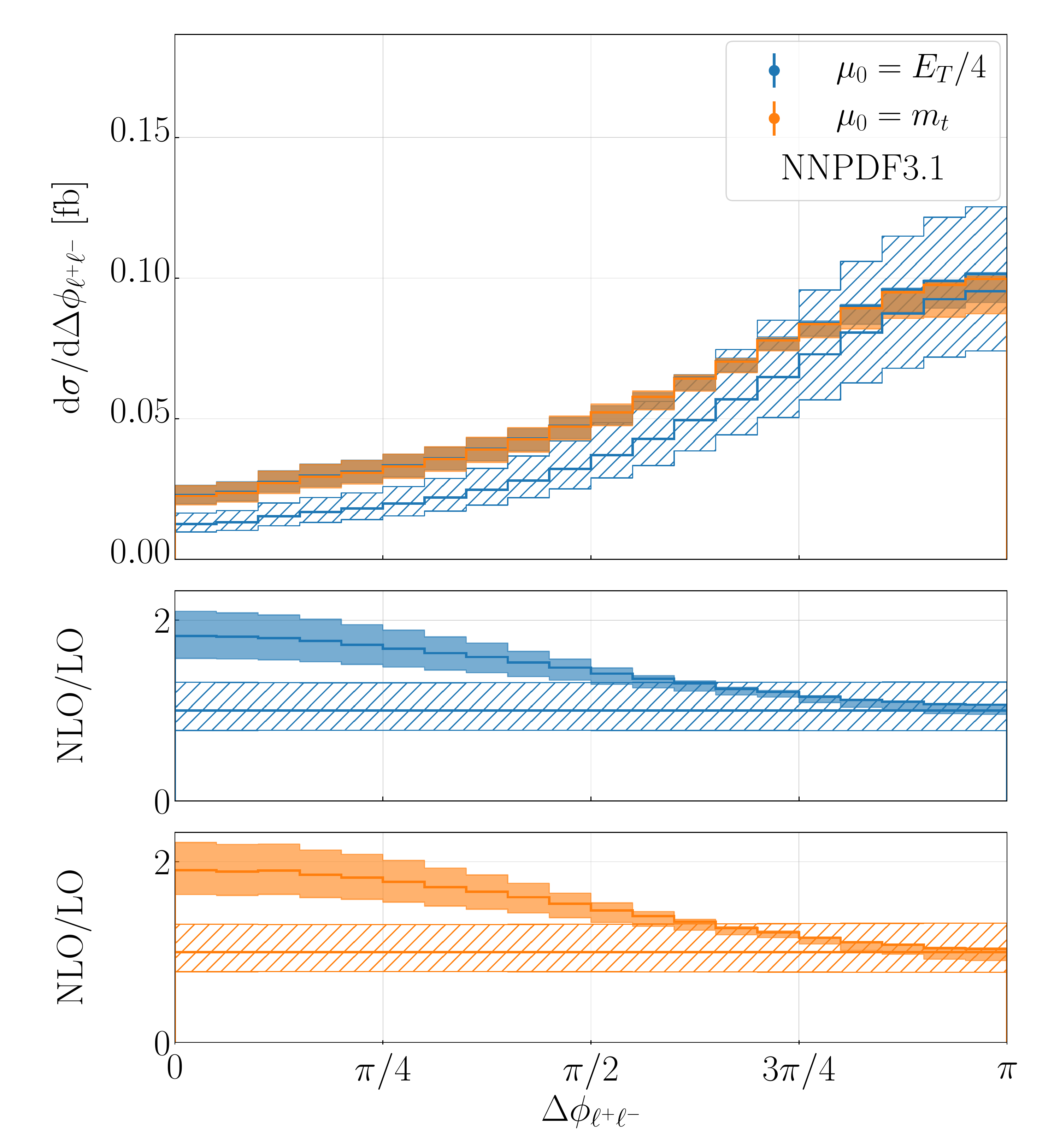}
    \end{center}
    \caption{\label{fig-lep:kfac3} \it Same as Figure \ref{fig-lep:kfac1} but for the observables $M_{b_1b_2}$, $p_{T,b_1b_2}$, $\Delta R_{b_1b_2}$ and $\Delta \Phi_{\ell^+\ell^-}$. }
\end{figure}

In the end, we focus on observables associated with the underlying $t\bar{t}$ production. In particular, we examine the invariant mass and  transverse momentum of the $b_1b_2$ system, denoted as  $M_{b_1b_2}$ and $p_{T,\, b_1b_2}$ respectively. These observables  are displayed in Figure \ref{fig-lep:kfac3}. Also shown are  the angular separation  between the two hardest $b$-jets, $\Delta R_{b_1b_2}$, and the angular difference
between the two charged leptons in the transverse plane, $\Delta \Phi_{\ell^+\ell^-}$. The advantage of the $\Delta \Phi_{\ell^+\ell^-}$ observable lies in the fact that measurements of charged leptons are particularly precise at the LHC due to the excellent lepton energy resolution of the ATLAS and CMS detectors.  Moreover, $\Delta \Phi_{\ell^+\ell^-}$  is sensitive to $t\bar{t}$  spin correlations and can be measured with high precision since the top quarks do not even need to be reconstructed. For the $M_{b_1b_2}$ observable, the NLO QCD corrections for the dynamic scale choice are rather constant at $25\%-30\%$.  The scale uncertainties are reduced from about $40\%$ at LO to $8\%$ at NLO,  showing overall good perturbative convergence. On the other hand, the fixed scale setting leads to larger shape distortions (up to $40\%$) and increased scale uncertainties (more than $20\%$ towards the tails). The behaviour of higher-order QCD corrections for $M_{b_1b_2}$ is very similar to the already analysed $H_T^{vis}$ observable, where  a fixed scale setting also led to deteriorated convergence in the high energy regime. For the transverse momentum of the two $b$-jet system, $p_{T,\,b_1b_2}$, we find huge NLO QCD corrections in the tail of the
distribution that are more than $300\%$ for $\mu_0=E_T/4$ and $200\%$ for $\mu_0=m_t$. Such large higher-order QCD effects have already been found for the $t\bar{t}$ process and other associated $t\bar{t}$ production processes, see e.g. Refs. \cite{Denner:2012yc,Denner:2015yca,Bevilacqua:2019cvp,
Czakon:2020qbd,Stremmer:2021bnk}. They are also present for similar observables like $p_{T}^{miss}=|\vec{p}_{T, \, \nu_\ell}+ \vec{p}_{T,\, \bar{\nu}_\ell}|$ or $p_{T,\,\ell^+\ell^-}$. In general, they occur for observables that are constructed from the decay products of both top quarks.  These huge NLO QCD corrections stem from hard jet radiation recoiling  against the $t\bar{t}$ system, that lifts off the kinematical suppression present at LO. The latter is responsible for the top-quark pair being produced in the back-to-back configuration. While for the process $pp \to t\bar{t}H$ or  $pp \to t\bar{t}Z$  a slight relaxation of the back-to-back configuration in the production of the $t\bar{t}$ system  is already present at LO, leading to less significant NLO QCD corrections, such a relaxation is not apparent in our case.  In addition, we have checked that these NLO QCD corrections are further enhanced for the {\it Mixed} and {\it Decay} contributions. In these two cases a differential $\mathcal{K}$-factor of the order of  $5$ and $10$ could be found, respectively. The theoretical uncertainties due to $\mu_R$ and $\mu_F$ scale dependence at the NLO QCD level are around $40\%$ at high transverse momenta. In addition, the NLO QCD results for the two scale settings lead to differences exceeding $25\%$. NNLO QCD corrections are therefore necessary to accurately predict this observable for the phase-space region  $p_{T, \, b\bar{b}} \gtrsim 150$ GeV, see e.g. Ref. \cite{Czakon:2020qbd}. For the $\Delta R_{b_1b_2}$ observable below $\Delta R_{b_1b_2} < 1$ we find NLO  QCD corrections up to $45\%$ for $\mu_0=E_T/4$ and $55\%$ for $\mu_0=m_t$. In both cases, the higher-order effects  exceed the size of the LO scale uncertainty bands in this region of phase space.  The scale uncertainties are reduced from $28\%-40\%$ at LO to $5\%-10\%$ at NLO for both scale choices, where the smallest theoretical errors are found for the back-to-back configurations at around $\Delta R_{b_1b_2}\approx 3$. Also for $\Delta \Phi_{\ell^+\ell^-}$,  yet another angular distribution that we have analysed,  large NLO QCD corrections up to $80\%-90\%$ are obtained. In that case, higher-order QCD effects are very sensitive to the particular phase-space region. Indeed, they are of the order of  $80\%-90\%$ for $\Delta \Phi_{\ell^+\ell^-} \approx 0$ and are reduced down  to about $5\%$ for $\Delta \Phi_{\ell^+\ell^-} \approx \pi$. The NLO scale uncertainties are of the order of $5\%-16\%$. Finally, at NLO in QCD the two scale settings lead to differences of at 
most $2\%$ for the $\Delta \Phi_{\ell^+\ell^-}$ observable.

%
\subsection{Distribution of prompt photons}
\label{sec:ttaa-lep-diff2}
%

\begin{figure}[t!]
    \begin{center}
	\includegraphics[width=0.49\textwidth]{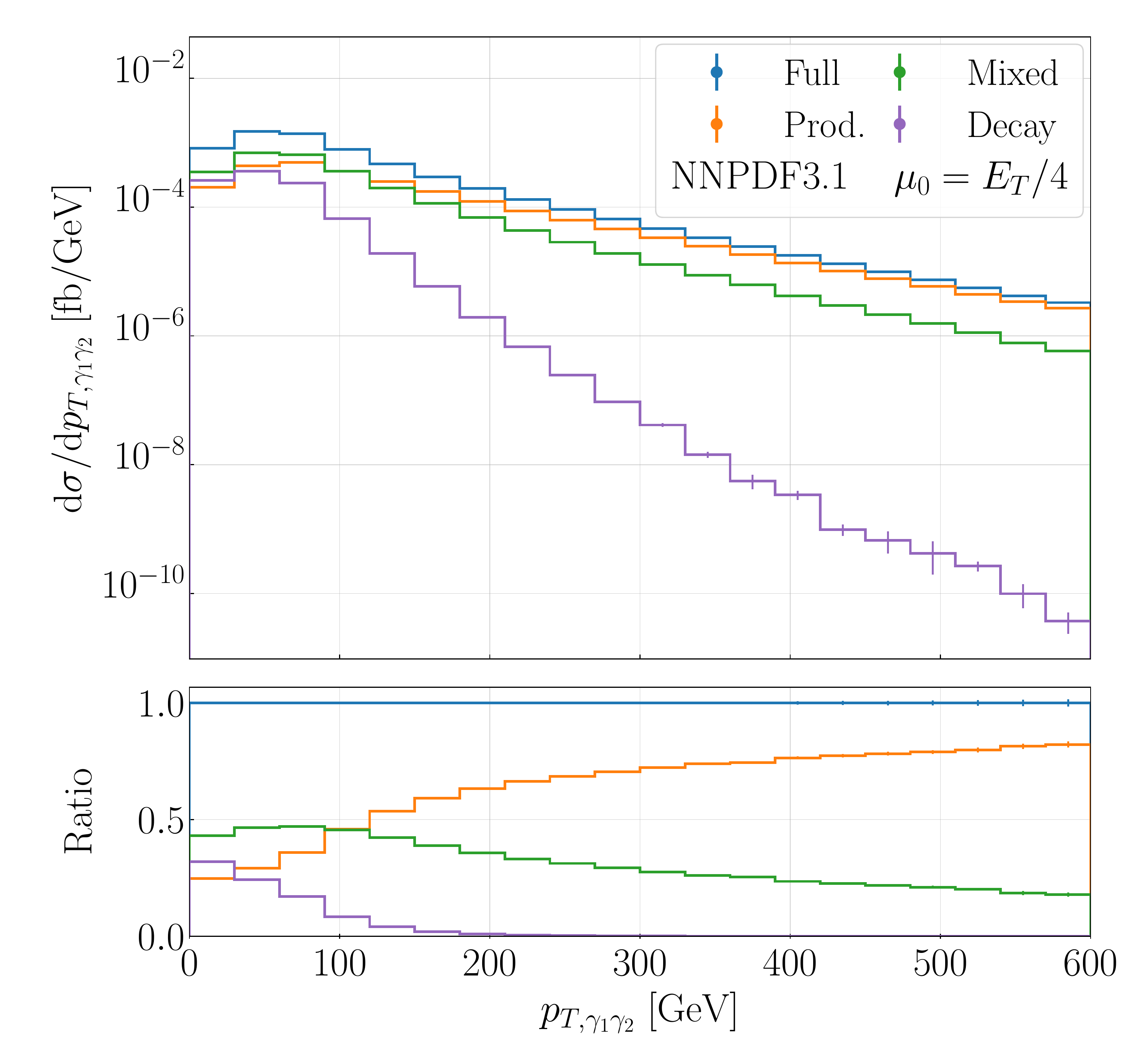}
	\includegraphics[width=0.49\textwidth]{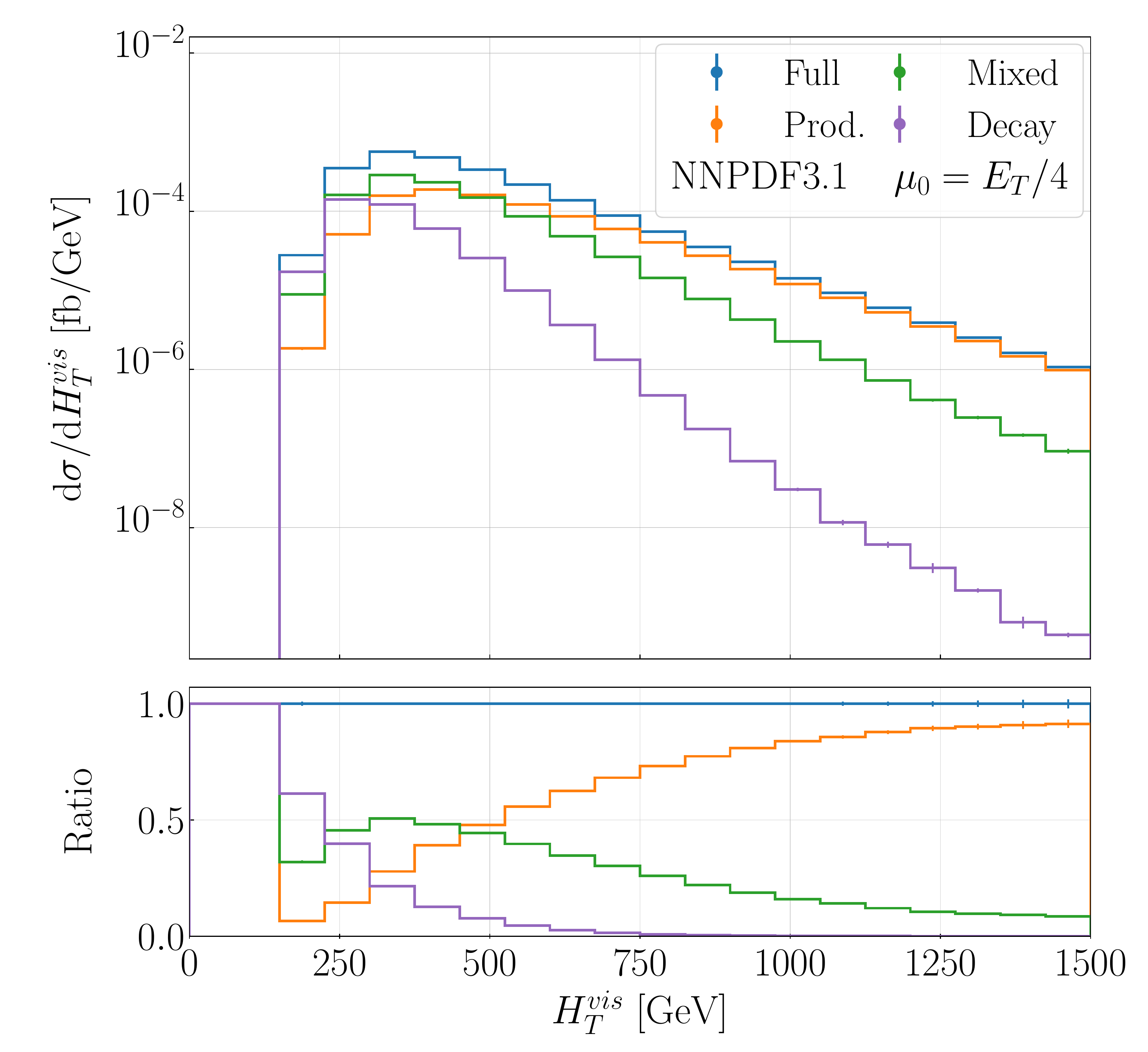}
	\includegraphics[width=0.49\textwidth]{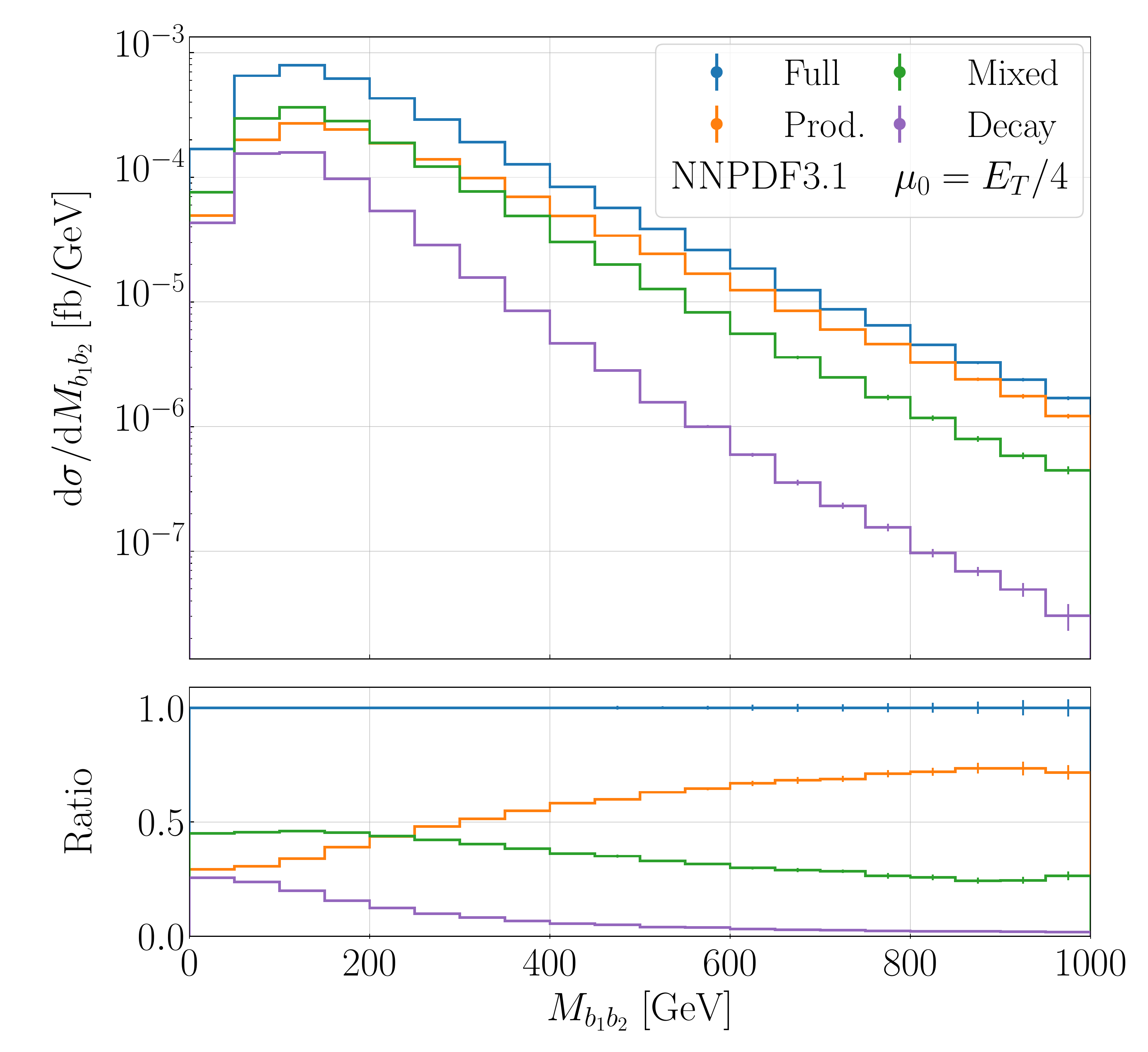}
 	\includegraphics[width=0.49\textwidth]{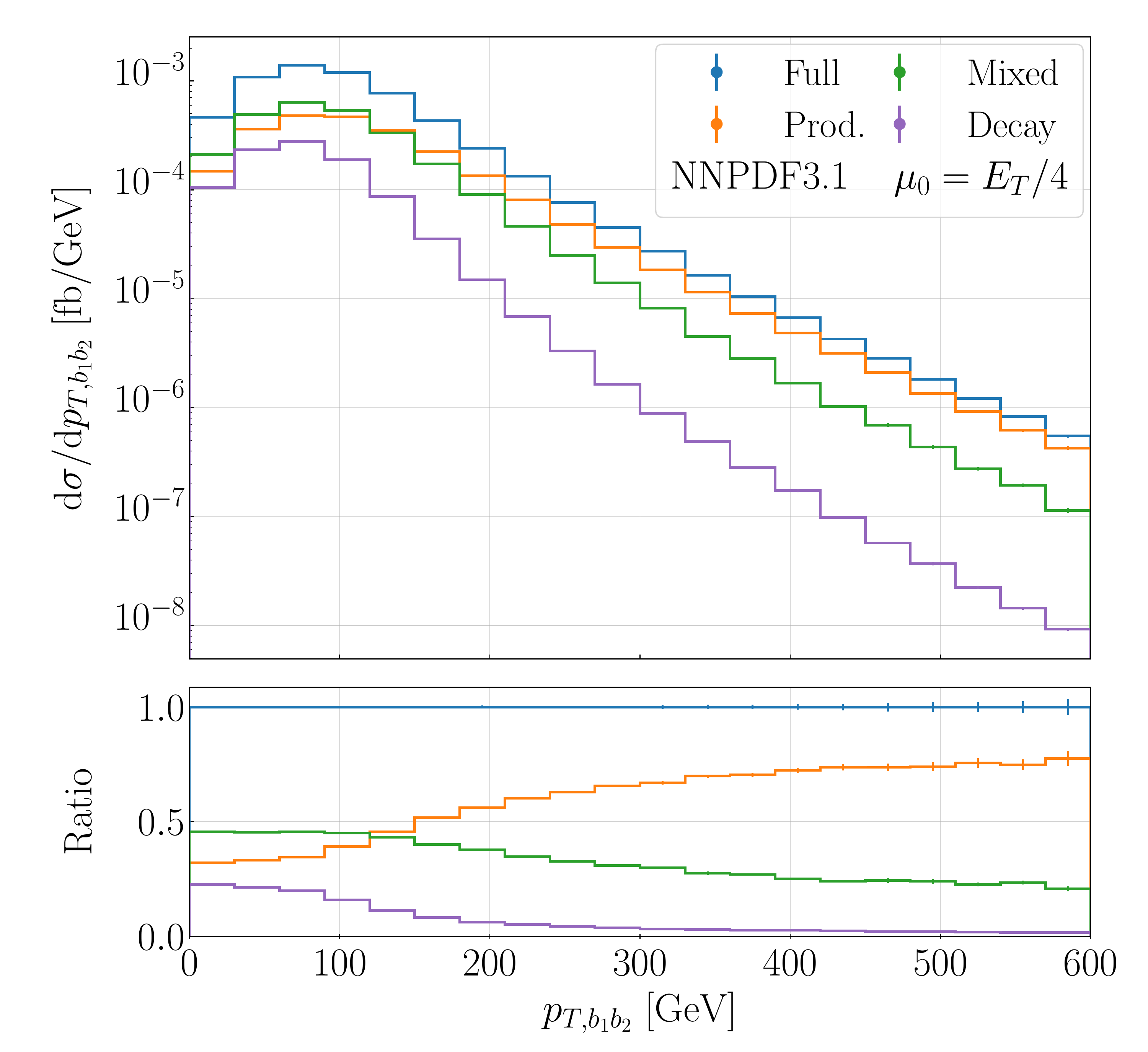}
    \end{center}
    \caption{\label{fig-lep:reg1} \it Differential cross-section distributions at NLO QCD for the observables $p_{T,\gamma_1\gamma_2}$, $H_T^{vis}$, $M_{b_1b_2}$ and $p_{T,b_1b_2}$ for the $pp\to \ell^+\nu_{\ell}\, \ell^-\bar{\nu}_{\ell} \, b\bar{b}\,\gamma\gamma +X$ process at the LHC with $\sqrt{s}=13\textrm{ TeV}$. Theoretical predictions are divided in the three contributions {\it Prod.}, 
    {\it Mixed} and {\it Decay}. They are obtained with $\mu_0=E_T/4$ and the NNPDF3.1 PDF set. The lower panels display the ratio to the full NLO QCD result. MC integration errors are also shown. }
\end{figure}
\begin{figure}[t!]
    \begin{center}
	\includegraphics[width=0.49\textwidth]{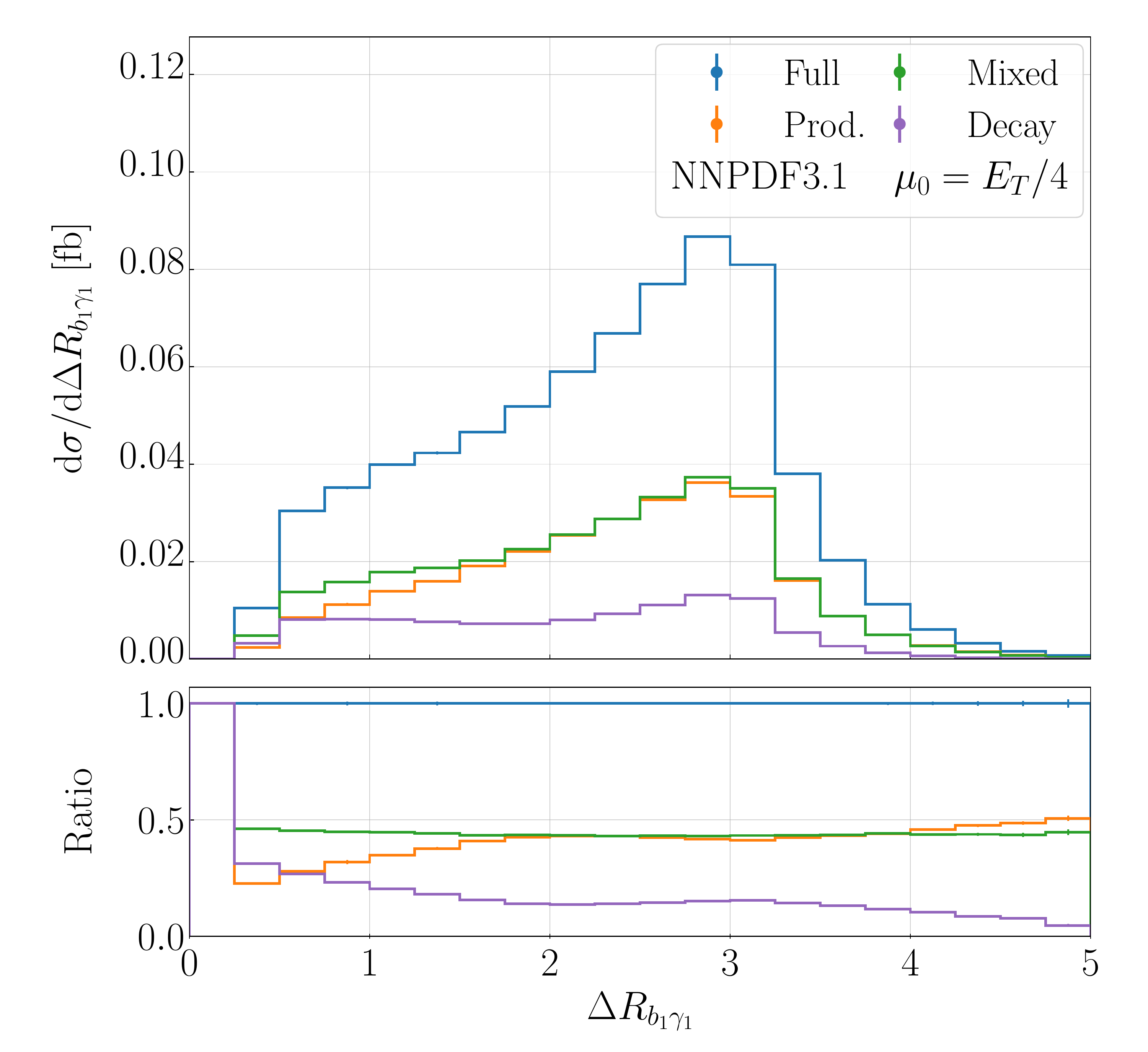}
	\includegraphics[width=0.49\textwidth]{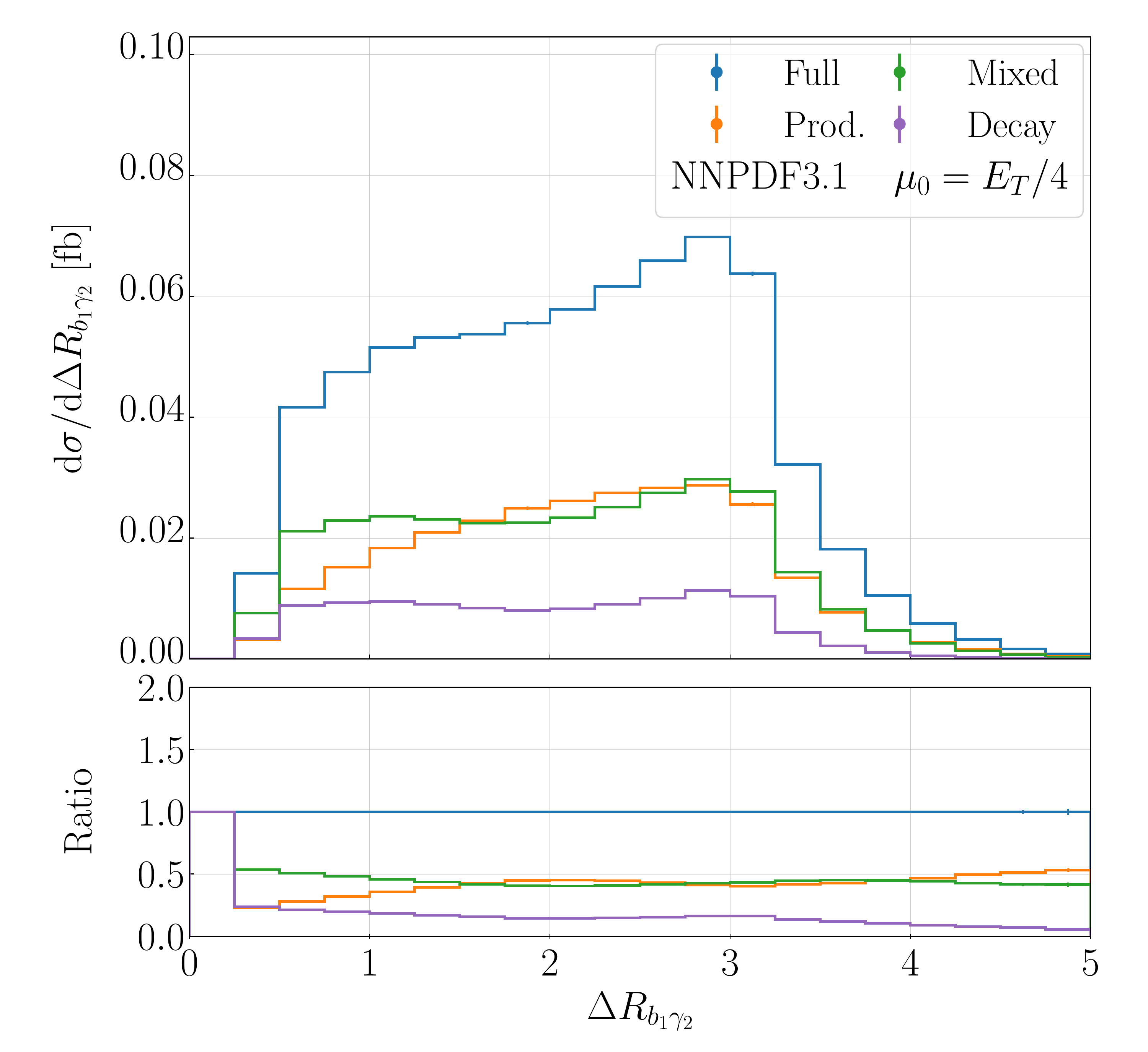}
	\includegraphics[width=0.49\textwidth]{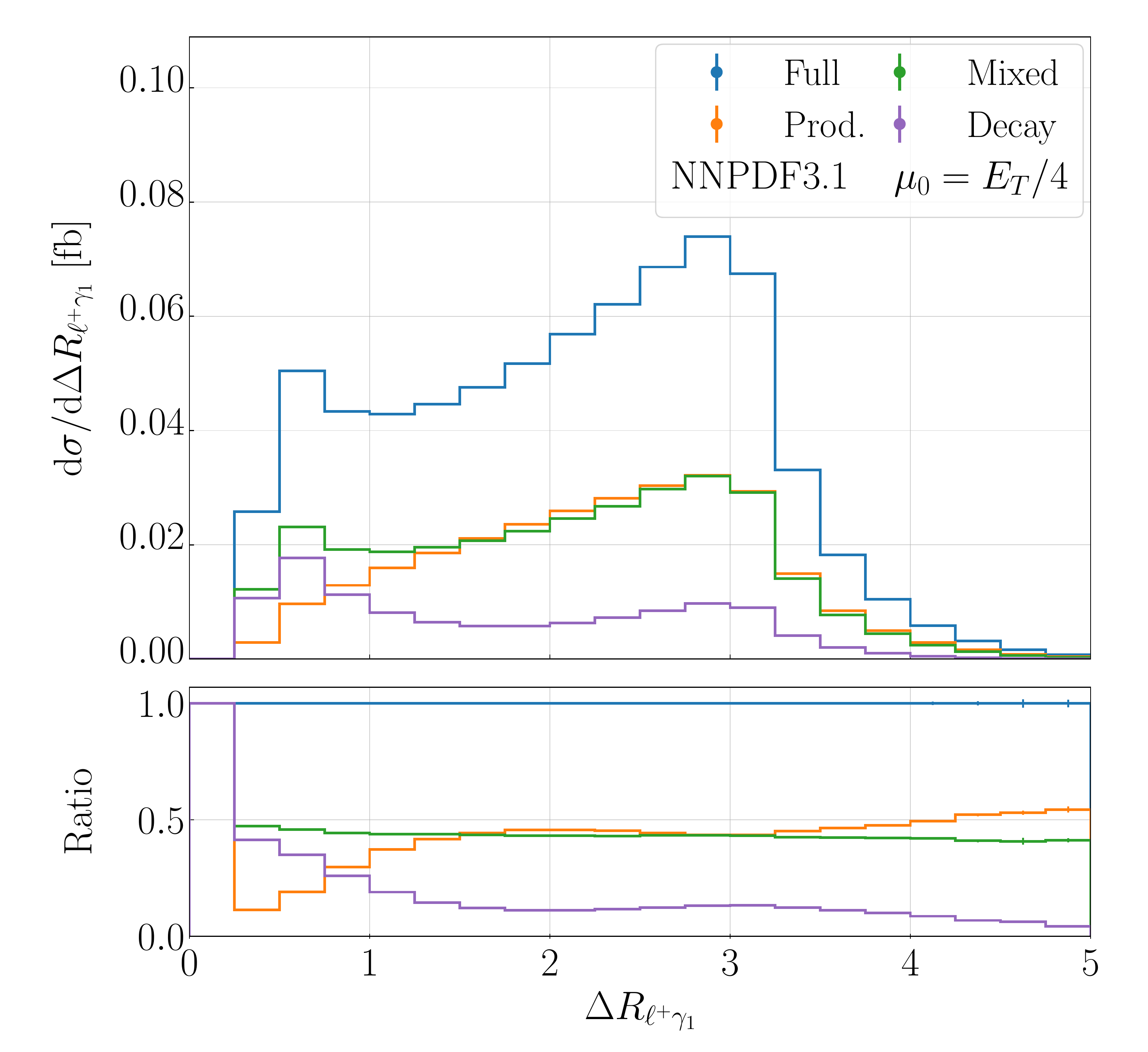}
	\includegraphics[width=0.49\textwidth]{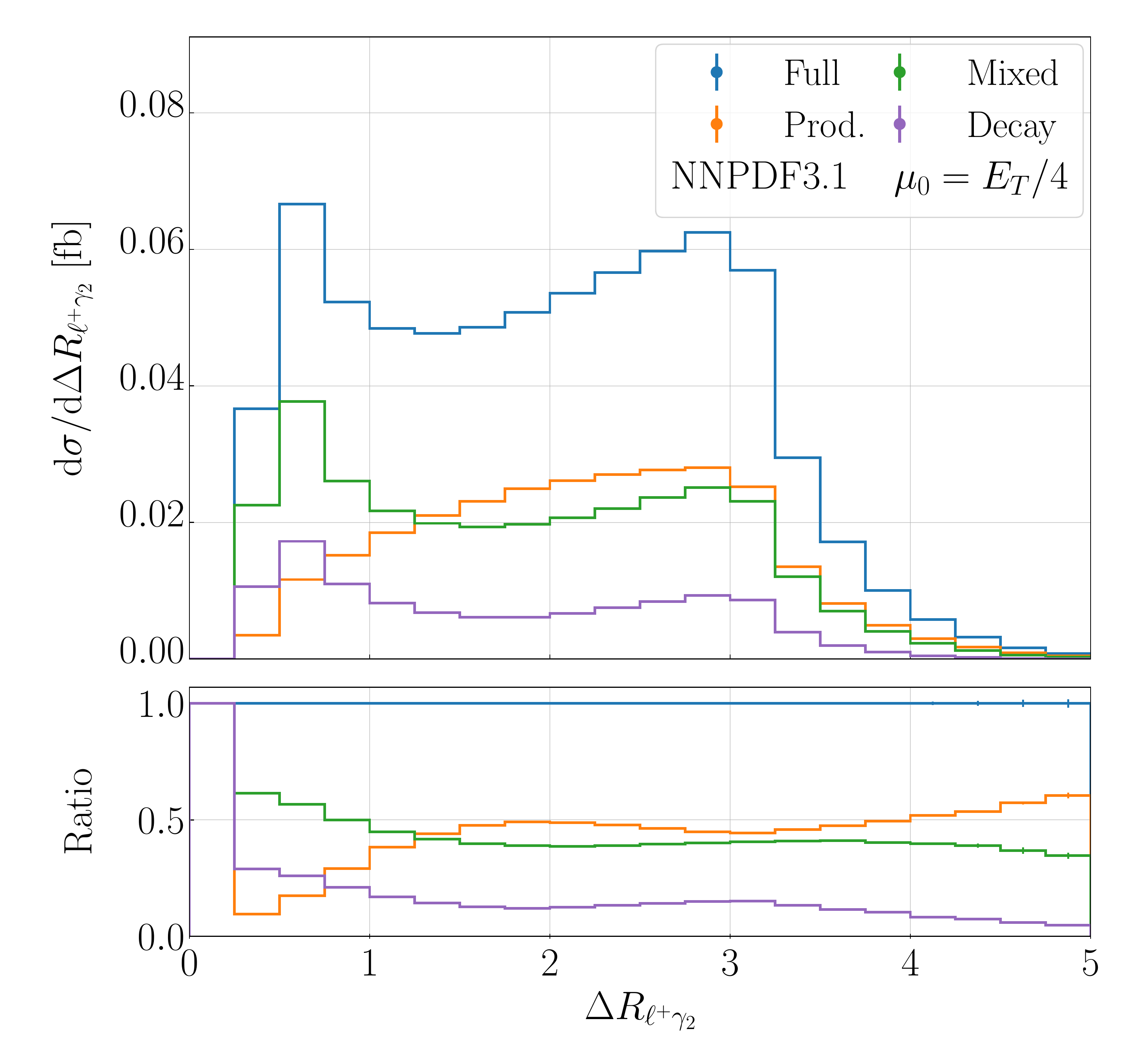}
    \end{center}
    \caption{\label{fig-lep:reg2} \it Same as Figure \ref{fig-lep:reg1} but for the observables $\Delta R_{b_1\gamma_1}$, $\Delta R_{b_1\gamma_2}$, $\Delta R_{\ell^+\gamma_1}$ and $\Delta R_{\ell^+\gamma_2}$.}
\end{figure}

Moving forward, we examine the impact of including photon emissions in top-quark and $W$ gauge boson decays. In Figure \ref{fig-lep:reg1} we present the differential cross-section distribution at NLO in QCD as a function of  $p_{T,\gamma_1\gamma_2}$, $H_T^{vis}$, $M_{b_1b_2}$ and $p_{T, \, b_1b_2}$ 
for the $pp\to \ell^+\nu_{\ell}\, \ell^-\bar{\nu}_{\ell} \, b\bar{b}\, \, \gamma\gamma +X$ process at the LHC with $\sqrt{s}=13$ TeV. We divide our theoretical predictions into the three configurations  {\it Prod.}, {\it Mixed} and {\it Decay}. In each case, the lower panel displays the ratio to the full NLO QCD result.  We employ the dynamical scale setting $\mu_0=E_T/4$ and the NNPDF3.1 PDF set. At the beginning of the $p_{T,\, \gamma_1,\gamma_2}$ spectrum  the {\it Prod.} contribution is the smallest, at about $25\%$ of the full result. It is followed by {\it Decay} with $32\%$ and {\it Mixed.} with $43\%$. The {\it Decay} contribution rapidly decreases for larger values of transverse momenta and already at around $150$ GeV it is smaller than $2\%$, thus, negligible when comparing to the scale uncertainties in this phase-space region. Up to $p_{T, \, \gamma_1,\gamma_2} \approx 150$ GeV, the {\it Mixed} contribution is rather constant and  for 
$p_{T, \, \gamma_1,\gamma_2} > 150$ GeV it slowly decreases to about $18\%$ in the tail. This contribution is not negligible even for higher values of  $p_{T, \, \gamma_1,\gamma_2}$ where the scale uncertainties of about $12\%$ are obtained. The {\it Prod.} contribution increases towards  higher values of $p_{T, \, \gamma_1,\gamma_2}$ and becomes the dominant one from about $p_{T, \, \gamma_1,\gamma_2} \approx 200$ GeV. In short, we can observe a very diverse picture, which depends largely on the available phase space for the three  configurations. The phase space is becoming more restricted as the number of high $p_T$ photons increases in top-quark decays. Overall, for the $H_T^{vis}$ observable we can observe a very similar behaviour for the three resonant contributions as for the $p_{T,\, \gamma_1,\gamma_2}$ spectrum. The {\it Decay} part becomes more enhanced for small values of $H_T^{vis}$, where it dominates this differential cross-section distribution by about $61\%$, however, at around $700$ GeV it is negligible. On the other hand, between $(250-450)$ GeV the {\it Mixed} part dominates with about $45\%-50\%$. Its importance decreases in the tails to about $9\%$ where it becomes comparable in size to the NLO scale uncertainties of about $6\%$.  For  $H_T^{vis} > 450$ GeV the {\it Prod.} contribution starts to increase from $48\%$ to $91\%$ towards the tail. For the standard dimensionful observables, that are  associated with  the underlying  $pp \to t\bar{t}$ production process, like for example $M_{b_1b_2}$ and $p_{T,\, b_1b_2}$, the {\it Decay} contribution is of the similar size as the {\it Prod.} contribution for the phase-space regions below $ \approx 100$ GeV. Even though the {\it Decay} contribution is still negligible in the tails of these distributions $(2\%)$, the decrease is drastically reduced compared to the dimensionful photon observables. In addition, the {\it Mixed} configuration is beginning to play a more important role $(20\%-24\%)$  for larger values of $M_{b_1b_2}$ and $p_{T,\, b_1b_2}$.

At last, we study the composition of photon emissions in the $pp \to t\bar{t}\gamma\gamma$ process for the dimensionless observables $\Delta R_{b_1\gamma_1}$, $\Delta R_{b_1\gamma_2}$, $\Delta R_{\ell^+\gamma_1}$ and $\Delta R_{\ell^+\gamma_2}$, which are all presented in Figure \ref{fig-lep:reg2}. In this way, we are able to study in more detail the probability of photon emission from different stages of the process. For the two observables $\Delta R_{b_1\gamma_1}$ and $\Delta R_{b_1\gamma_2}$ the size of the {\it Mixed} contribution is rather flat independently of the phase-space region and vary between $40\%-54\%$. An opposite trend is found for the other two contributions. The {\it Decay} configuration decreases towards larger values of  $\Delta R_{ij}$ from $31\%$ to $5\%$  and from $24\%$ to $6\%$  for $\Delta R_{b_1\gamma_1}$ and $\Delta R_{b_1\gamma_2}$, respectively. This behavior is compensated by  the increase of the {\it Prod.} contribution from $23\%$ to about $50\%$. We observe that, for the {\it Prod.} contribution  back-to-back configurations are more amplified in both cases, while as the number of photons in top-quark decays increases, photon radiation with a smaller $\Delta R_{ij}$ becomes more and more likely. This enhancement towards smaller values of  $\Delta R_{ij}$ is even more pronounced for the $\Delta R_{ij}$ separations between photons and leptons as demonstrated for $\Delta R_{\ell^+\gamma_1}$ and $\Delta R_{\ell^+\gamma_2}$. In both cases the {\it Mixed} and {\it Decay} contributions have a second peak for small values of $\Delta R_{ij}$ in addition to the one for $\Delta R_{ij} \approx 3$. Consequently, the importance of {\it Mixed} and {\it Decay} is greatly increased for $\Delta R_{ij} < 1.5$ where the full result is dominated by these two configurations with the added contribution of about $90\%$. The {\it Mixed} contribution decreases towards larger values of $\Delta R_{ij}$ from $47\%$ to $41\%$  and from $62\%$ to $35\%$ for $\Delta R_{\ell^+\gamma_1}$ and $\Delta R_{\ell^+\gamma_2}$, respectively. On the other hand,  the {\it Prod.} contribution is only about $10\%$ for smaller values of $\Delta R_{ij}$. However, its importance increases substantially  
towards larger $\Delta R_{ij}$ separation. Specifically, we obtain the contribution of the order of $54\%$ for $\Delta R_{\ell^+\gamma_1}$ and $60\%$ for $\Delta R_{\ell^+\gamma_2}$. Also in this case the {\it Decay} contribution behaves oppositely compared to the {\it Prod.} one and decreases from $41\%$ to $4\%$ towards the end of the $R_{\ell^+\gamma_1}$ spectrum ($29\%$ to $5\%$  for $\Delta R_{\ell^+\gamma_2}$). Concluding, the inclusion of photon bremsstrahlung in top-quark decays is essential for a proper description of angular cross-section distributions. The shape differences between the three resonant contributions are non-trivial as new peaks can arise in the spectra. Thus, the full result cannot be  obtained by simple reweighting of the {\it Prod.} contribution by some fixed factor. Due to the limited phase space for the top-quark decay products in the {\it Decay} configuration, this contribution is heavily suppressed for larger values of $\Delta R_{ij}$, but  it is very important for the more collinear configurations. 

%
\section{Lepton plus jet  channel}
\label{sec:ttaa-semi}
%

%
\subsection{Integrated fiducial cross sections}
\label{sec:ttaa-semi-int}
%

In this section we study the $pp\to t\bar{t}\gamma\gamma$ process in the \jetlep channel at the integrated and differential  cross-section level. Our main goal here is to assess the differences and similarities compared to the \dilep channel discussed in the previous section. We can immediately notice that for the $pp\to \, \ell^-\bar{\nu}_{\ell} \, jj \, b\bar{b}\,\gamma\gamma +X$ process, the two top-quark decay chains: $t\to W^+b \to jj \,b$ and $\bar{t} \to W^- \bar{b} \to \ell^- \bar{\nu}_\ell \, \bar{b}$ are no longer symmetrically treated due to the presence of the different fiducial phase-space cuts. Further differences can be expected as a result of real radiation at NLO in QCD. The real emission part of the calculation can cause large effects, especially in the tails of various differential cross-section distributions. Since the extra radiation  is predominantly produced in the $t\bar{t}$ production stage, it does not suffer from a strongly limited phase space due to the finite mass of the top quark and $W$ gauge boson.
\begin{figure}[t!]
    \begin{center}
	\includegraphics[width=0.49\textwidth]{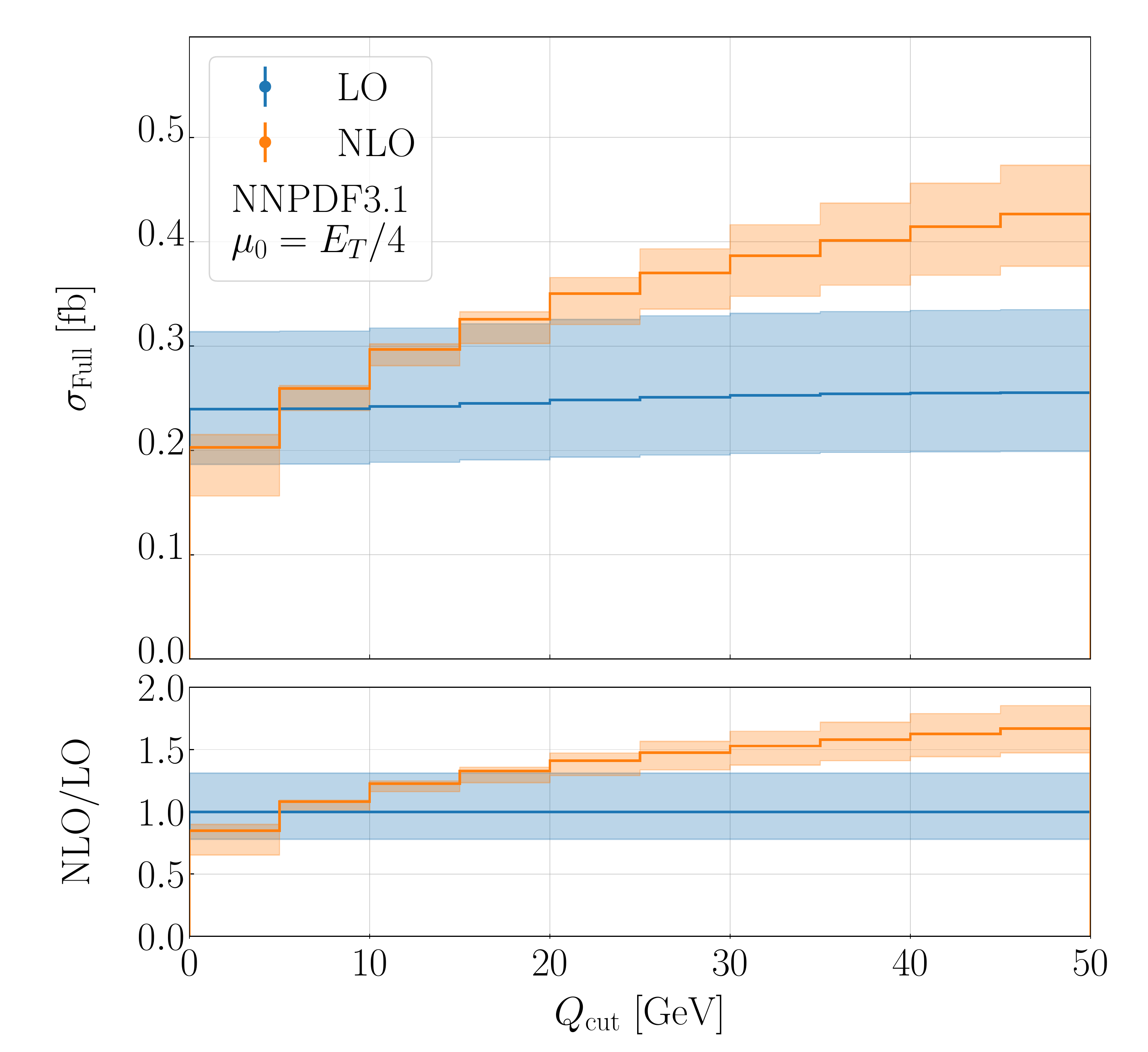}
	\includegraphics[width=0.49\textwidth]{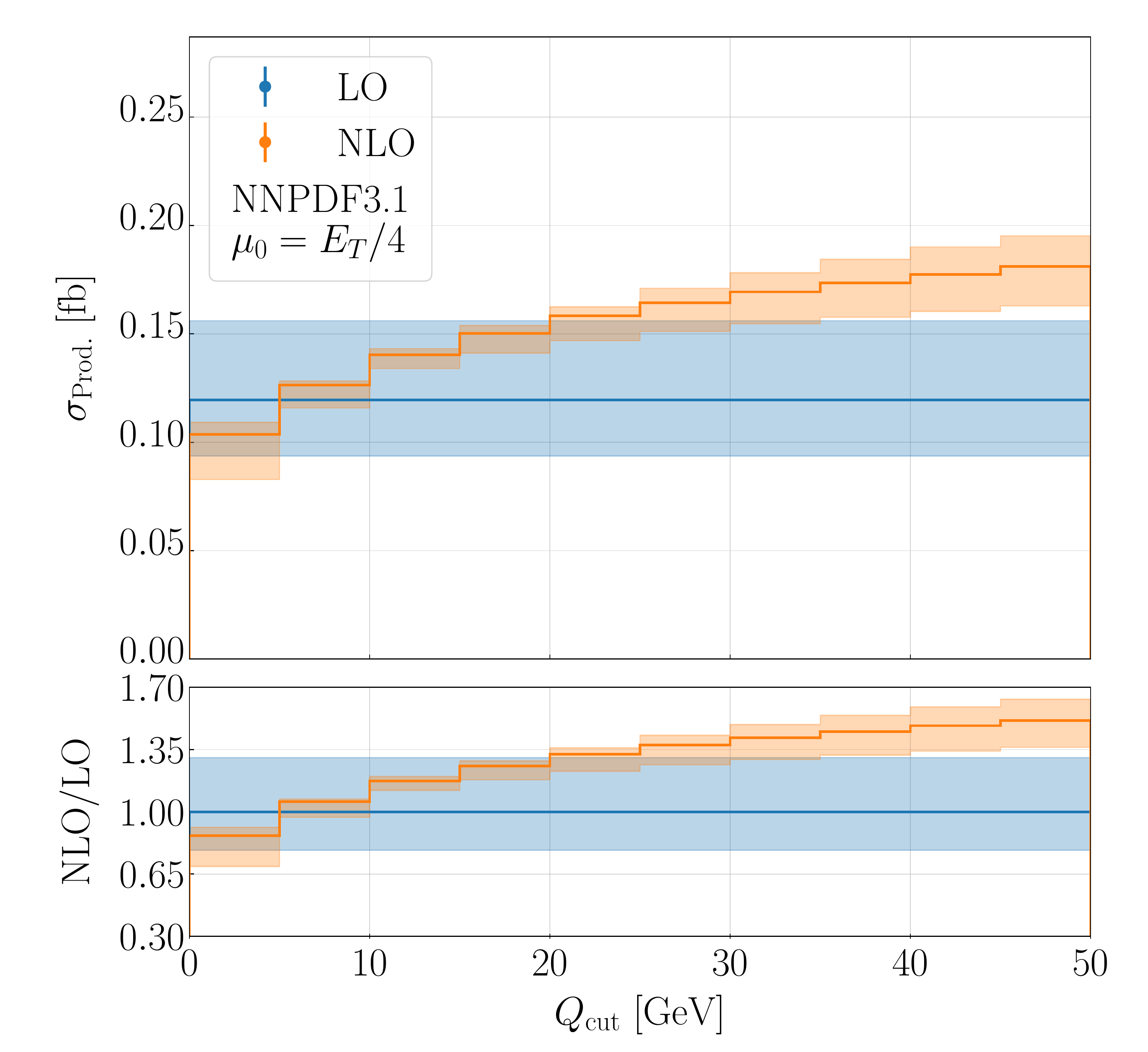}
	\includegraphics[width=0.49\textwidth]{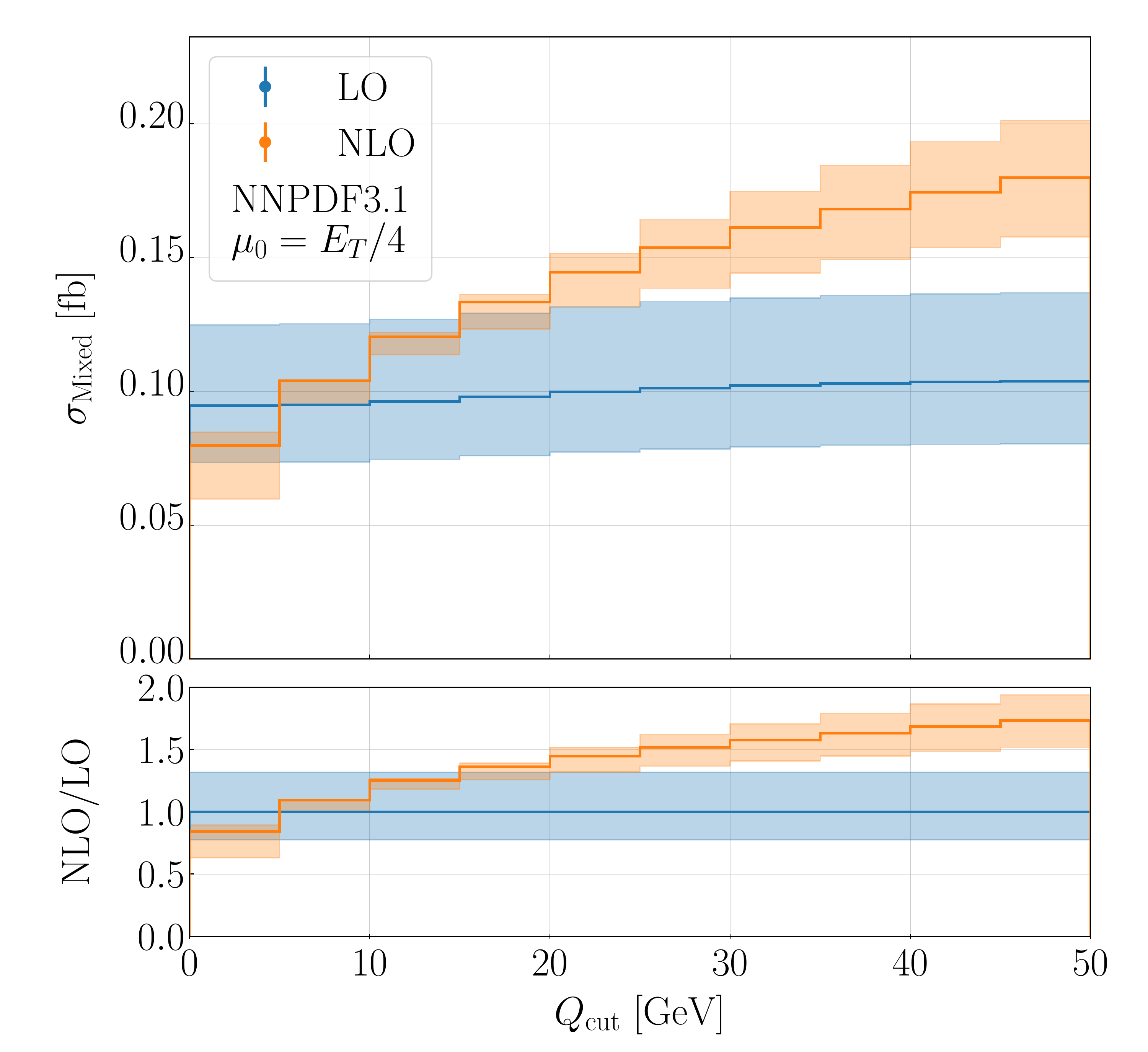}
 	\includegraphics[width=0.49\textwidth]{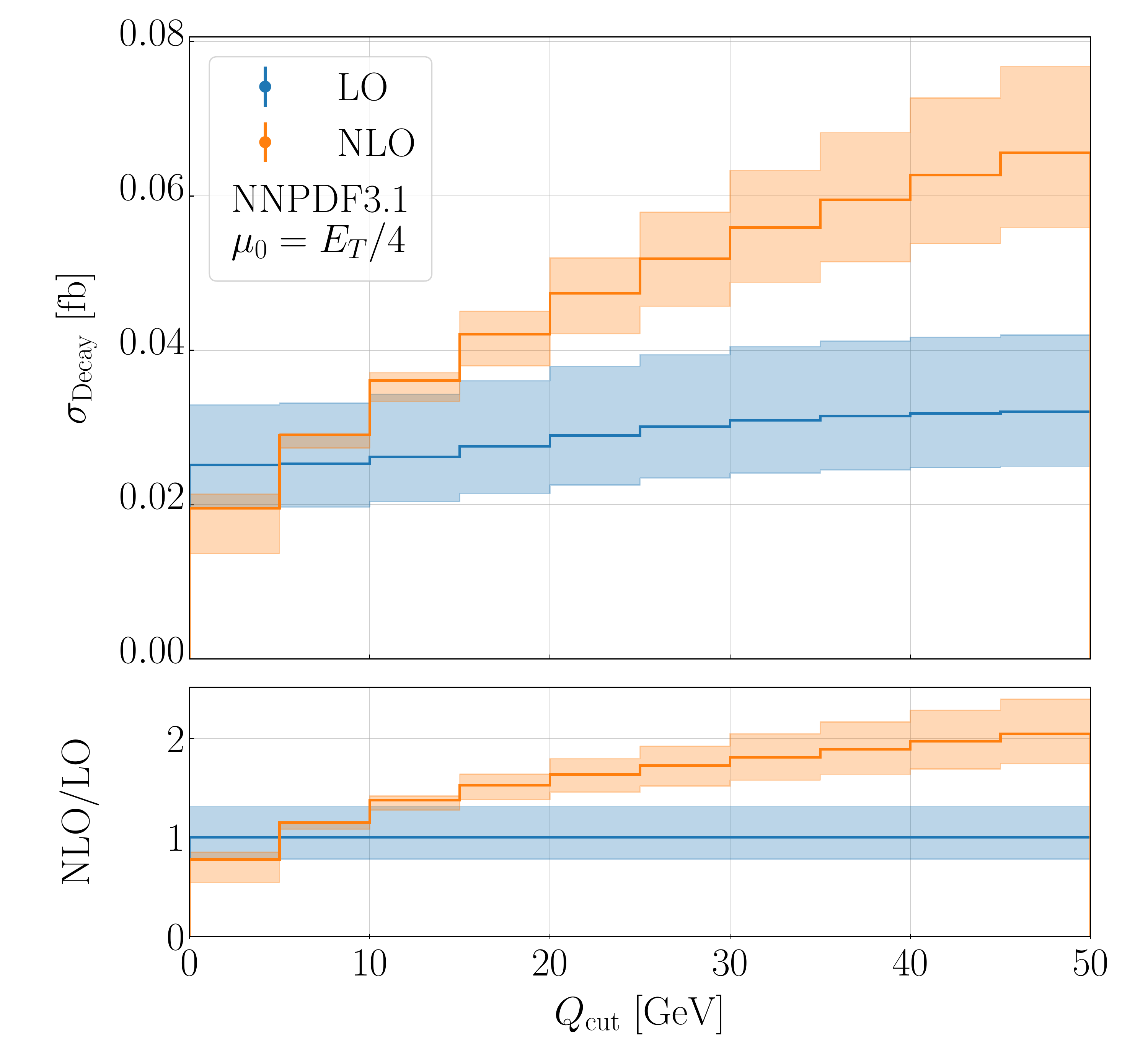}
    \end{center}
    \caption{\label{fig-semi:qcut} \it Integrated fiducial cross sections at LO and NLO QCD for the $pp\to \, \ell^-\bar{\nu}_{\ell} \, jj \, b\bar{b}\,\gamma\gamma +X$ process at the LHC with $\sqrt{s}=13~{\rm TeV}$ as a function of the $Q_{\rm cut}$ parameter defined as $|m_{W}-M_{jj}|<Q_{\rm cut}$. Results are shown for the full process and the three resonant contributions {\it Prod.}, {\it Mixed} and {\it Decay}.   The theoretical uncertainties from the 7-point scale variation  are also provided. The NNPDF3.1 PDF set and the dynamical scale $\mu_0=E_T/4$ are employed. The lower panel displays  the ${\cal K}$-factor with the uncertainty band and the relative scale uncertainties of the LO cross section.}
\end{figure}
As a first step, we examine the impact of the invariant mass cut defined in Eq. \eqref{eq_qcut} on the LO and NLO cross section for the $pp\to \, \ell^-\bar{\nu}_{\ell} \, jj \, b\bar{b}\,\gamma\gamma +X$ process at the LHC with $\sqrt{s}=13$ TeV. Theoretical predictions, with the corresponding scale uncertainties, are presented in Figure \ref{fig-semi:qcut} as a function of $Q_{\rm cut}$, defined according to 
\begin{equation}
\label{eq_qcut2}
|m_{W}-M_{jj}|<Q_{\rm cut}.
\end{equation}
We vary the $Q_{\rm cut}$ parameter in the range of $Q_{\rm cut} \in (5 -50)$ GeV in steps of $5~\rm{GeV}$. We note that our final choice of $Q_{\rm cut}=15$ GeV  corresponds to the result in the third bin. We use the dynamical scale setting, $\mu_0=E_T/4$, and the NNPDF3.1 PDF set, but similar conclusions can be drawn from the result with the fixed scale choice, $\mu_0=m_t$. The integrated fiducial cross section is shown for the full result as well as for the three resonant contribution {\it Prod.}, {\it Mixed} and {\it Decay}. The differences between the two extreme cases $Q_{\rm cut}=5$ GeV  and no cut ($Q_{\rm cut}\to \infty$) is about $7\%$ for the full integrated cross section at LO showing a rather minor dependence on this cut. At NLO in QCD, however, the situation drastically changes and huge higher-order QCD corrections are found for large values of the $Q_{cut}$ parameter. In particular, we find NLO QCD corrections of about $67\%$ for $Q_{\rm cut}=50$ GeV,  which further increase up to $140\%$ if no cut is applied. For $Q_{\rm cut} \le 25$ GeV  the uncertainty bands of the LO and NLO QCD predictions start to overlap. Only for $Q_{\rm cut} \le 15$ GeV the NLO QCD prediction is within the LO scale uncertainty. These large higher-order QCD corrections are associated with kinematical configurations in which the two light jets from the hadronically decaying $W$ gauge boson are recombined into a single jet. Such kinematical configurations are not present at LO since we are interested in the resolved topology where the two light jets are always present. On the other hand, at NLO in QCD such configurations are indeed possible due to the additional light jet from the real corrections. The latter light jet, when resolved and passes all the cuts, can act as the second decay product of the hadronically decaying $W$ gauge boson. As demonstrated in Figure \ref{fig-semi:qcut}, the size of the real emission contribution can be drastically reduced by imposing the 
$|m_{W}-M_{jj}|< 15$ GeV cut. When examining the {\it Prod.} contribution  separately, we note that at LO this contribution is insensitive to the $M_{jj}$ cut as no photons are emitted in top-quark decays, so we always have $M_{jj}=m_W$. At the NLO level in QCD we again observe large higher-order QCD corrections which are, however, less pronounced than for the full result. On the other hand, major corrections are visible for the {\it Mixed} and {\it Decay} contributions due to photon radiation inside the top-quark decay. Already at LO the differences between the two extreme case $Q_{\rm cut}=5$ GeV and $Q_{\rm cut} \to \infty$ amount to $9\%$ for the {\it Mixed} configuration and to $22\%$ for the {\it Decay} one. At NLO QCD the relaxation due to additional radiation becomes even stronger compared to the {\it Prod.} case due to the more limited LO phase space caused by photon radiation in the decays. This leads to a huge increase in the integrated fiducial cross section and the $\mathcal{K}$-factor. Indeed, we obtain ${\cal K}= 2.55$ for {\it Mixed} and ${\cal K}= 3.16$ for {\it Decay} if no $M_{jj}$ cut is applied. Finally, the {\it Decay} part is affected the most as the NLO QCD prediction lies within the LO scale uncertainty only for $Q_{\rm cut}<10$ GeV. Even for small values of the $Q_{\rm cut}$ cut like $Q_{\rm cut}<15$ GeV large NLO QCD corrections are clearly noticeable.
\begin{table*}[t!]
    \centering
    \renewcommand{\arraystretch}{1.5}
    \begin{tabular}{l@{\hskip 10mm}ll@{\hskip 10mm}ll@{\hskip 10mm}l}
        \hline\noalign{\smallskip}
        $\mu_0$ &  &  & LO & NLO &${\cal K} = \sigma_{\text{NLO}} / \sigma_{\text{LO}}$ \\
        \noalign{\smallskip}\midrule[0.5mm]\noalign{\smallskip}
        \multirow{4}{*}{$E_T/4$}   
        & $\sigma_{\rm Full}$ & [fb] & $ 0.24214(4)^{+31.1\%}_{-22.0\%} $ & $ 0.2973(3)^{+1.9\%}_{-5.4\%} $ & $ 1.23 $ \\
        & $\sigma_{\rm Prod.}$ & [fb] & $ 0.11960(3)^{+30.5\%}_{-21.6\%} $ & $ 0.1405(2)^{+2.1\%}_{-4.6\%} $ & $ 1.17 $ \\
        & $\sigma_{\rm Mixed}$ & [fb] & $ 0.09632(3)^{+31.9\%}_{-22.5\%} $ & $ 0.1205(2)^{+1.5\%}_{-5.7\%} $ & $ 1.25 $ \\
        & $\sigma_{\rm Decay}$ & [fb] & $ 0.026230(9)^{+30.9\%}_{-22.1\%} $ & $ 0.03629(7)^{+3.3\%}_{-7.7\%} $ & $ 1.38 $ \\

        \noalign{\smallskip}\midrule[0.5mm]\noalign{\smallskip}
        \multirow{4}{*}{$m_t$}   
        & $\sigma_{\rm Full}$ & [fb] & $ 0.23898(4)^{+31.2\%}_{-22.1\%} $ & $ 0.2948(3)^{+1.6\%}_{-5.4\%} $ & $ 1.23 $ \\
        & $\sigma_{\rm Prod.}$ & [fb] & $ 0.12107(3)^{+31.0\%}_{-21.8\%} $ & $ 0.1402(2)^{+1.8\%}_{-4.2\%} $ & $ 1.16 $ \\
        & $\sigma_{\rm Mixed}$ & [fb] & $ 0.09340(3)^{+31.8\%}_{-22.4\%} $ & $ 0.1193(3)^{+1.4\%}_{-6.0\%} $ & $ 1.28 $ \\
        & $\sigma_{\rm Decay}$ & [fb] & $ 0.024500(9)^{+30.4\%}_{-21.8\%} $ & $ 0.03534(7)^{+4.3\%}_{-8.2\%} $ & $ 1.44 $ \\

        \noalign{\smallskip}\hline\noalign{\smallskip}
    \end{tabular}
    \caption{\label{tab:int-semi-scale} \it Integrated  fiducial cross section at LO and NLO QCD for the $pp\to \, \ell^-\bar{\nu}_{\ell} \, jj \, b\bar{b}\,\gamma\gamma +X$ process at the LHC with $\sqrt{s}=13~{\rm TeV}$. Results are given with the $|m_{W}-M_{jj}|<15$ GeV cut 
     for two scale settings $\mu_0=E_T/4$ and $\mu_0=m_t$ as well as for the three contributions: {\it Prod.}, {\it Mixed} and {\it Decay}. The NNPDF3.1 PDF set is employed. The theoretical uncertainties from the 7-point scale variation and MC integration errors (in parenthesis) are also 
     displayed. }
\end{table*}

In Table \ref{tab:int-semi-scale} the integrated fiducial cross section at LO and NLO QCD is shown for the $pp \to t\bar{t}\gamma\gamma$ process in the \jetlep channel with the additional $|m_{W}-M_{jj}|<15$ GeV cut. Similarly to the  \dilep channel, also in this case the two different scale settings $\mu_0=E_T/4$ and $\mu_0=m_t$ are examined and the NLO NNPDF3.1 PDF set is employed.  The full integrated fiducial $pp$ cross section is dominated by the {\it Prod.} contribution with $50\%$ at LO and $48\%$ at NLO QCD. This is in contrast to the \dilep channel which was dominated by the {\it Mixed} configuration. In the \jetlep case, the {\it Mixed} contribution amounts to about $40\%$ both at LO and NLO QCD, while the {\it Decay} part is about $10\%$ at LO and $12\%$ at NLO QCD. We note, however, that after omitting the cut on the invariant mass of the two light jets, the {\it Mixed}  contribution becomes dominant at NLO  in QCD and amounts to $43\%$ compared to the {\it Prod.} contribution with $40\%$. Thus, without the $M_{jj}$ cut the size of the {\it Mixed} and {\it Prod.} contributions is the same as in the \dilep channel. Returning to the results shown in Table \ref{tab:int-semi-scale}, we find higher-order QCD corrections of $23\%$ for the full result for both scale settings. For the different resonant contributions the $\mathcal{K}$-factors  vary widely from $1.16$ to $1.44$. For the {\it Prod.} and {\it Mixed} contributions the NLO QCD corrections are within the LO scale uncertainty bands. For the {\it Decay} contribution, on the other hand, where we have  ${\cal K} =1.38$ for $\mu_0=E_T/4$ and ${\cal K}=1.44$ for $\mu_0=m_t$, higher order QCD corrections exceed the LO uncertainty bands that are of the order of $30\%$. The scale uncertainties of the full result are reduced by a factor of $6$ from $31\%$ at LO to  $5\%$ at NLO QCD for both scale choices. The full integrated fiducial cross section differs between the dynamical and fixed scale setting by about $1\%$ at LO and less than $1\%$ at NLO QCD. Similarly to the \dilep decay channel, the largest differences between the two scale choices are found for the {\it Decay} contribution with about $7\%$ at LO and $3\%$ at NLO QCD. Thus, at the integrated fiducial cross-section level both scales are equivalent.
\begin{table*}[t!]    
    \centering
    \renewcommand{\arraystretch}{1.5}
    \begin{tabular}{l@{\hskip 10mm}l}
        \hline\noalign{\smallskip}
        & $\sigma_{\rm Full}^{\rm NLO}$ [fb] \\
        \noalign{\smallskip}\midrule[0.5mm]\noalign{\smallskip}
        $\epsilon_\gamma=1.0$ & $ 0.2973(3)^{+1.9\%}_{-5.4\%} $ \\
        $\epsilon_\gamma=0.5$ & $ 0.2832(7)^{+1.5\%}_{-4.2\%} $ \\
        $E_{T\,\gamma}\,\epsilon_\gamma=10~{\rm GeV}$ &  $ 0.2666(8)^{+1.0\%}_{-7.2\%} $ \\
        \noalign{\smallskip}\hline\noalign{\smallskip}
    \end{tabular}
    \caption{\label{tab:int-semi-iso} \it Integrated cross section at NLO QCD for the $pp\to \, \ell^-\bar{\nu}_{\ell} \, jj \, b\bar{b}\,\gamma\gamma +X$ process at the LHC with $\sqrt{s}=13~{\rm TeV}$. Results are given for different parameter choices of the smooth photon isolation prescription defined in Eq. \eqref{eq_iso} but with  $n=1$. They are presented for the dynamical scale $\mu_0=E_T/4$ employing the NNPDF3.1 PDF set. Theoretical uncertainties from the 7-point scale variation and MC integration error are also displayed. }
\end{table*}

At last, in Table \ref{tab:int-semi-iso} we show the integrated fiducial cross section at NLO in QCD using  different parameter choices for the smooth photon isolation prescription defined in Eq. \eqref{eq_iso}. Results are given for the $\mu_0=E_T/4$ scale setting and the NNPDF3.1 PDF set. In particular, the first prediction corresponds to our default choice with $n=1$ and $\epsilon_\gamma=1.0$. For the other two results we do not change the parameter $n$, but rather modify the coefficient  $E_{T\,\gamma}\,\epsilon_\gamma$ in front of the right hand side of Eq. \eqref{eq_iso}. Thus, in the second case we set $\epsilon_\gamma=0.5$ and for third parameter choice we use $E_{T\,\gamma}\,\epsilon_\gamma=10~{\rm GeV}$. It is important to address the dependence on these parameters as various values are employed in literature for processes with prompt photons,  see e.g. \cite{Bern:2011pa,Melnikov:2011ta,Campbell:2017dqk,Bevilacqua:2018woc,Chen:2019zmr,Chawdhry:2019bji,Gehrmann:2020oec,Chawdhry:2021hkp,Badger:2021ohm,Chen:2022gpk,Badger:2023mgf}. Especially,  when many photons and jets are present in the final state, the dependence on these parameters might be non-negligible and could affect comparisons between theoretical predictions and experimental results.  In the \jetlep decay channel the integrated fiducial cross section is reduced by about $5\%$ if we set $\epsilon_\gamma=0.5$. A larger reduction of about $10\%$ is observed for the last parameter setting $E_{T\,\gamma}\,\epsilon_\gamma=10~{\rm GeV}$. Indeed, these substantial differences are due to the high number of jets (up to $5$) and/or photons $(2)$ in the final state. Moreover, these effects are similar in size or even larger than the corresponding NLO scale uncertainties for this process and therefore of high relevance. In the \dilep decay channel the dependence on these parameters is  smaller,  but still not negligible as the differences up to $3\%$ and $6\%$, respectively, can be observed. Thus, these effects are at most as large as the corresponding  NLO scale uncertainties, which are of the order of $6\%$.

%
\subsection{Differential fiducial cross sections}
\label{sec:ttaa-semi-diff}
%

\begin{figure}[t!]
    \begin{center}
	\includegraphics[width=0.49\textwidth]{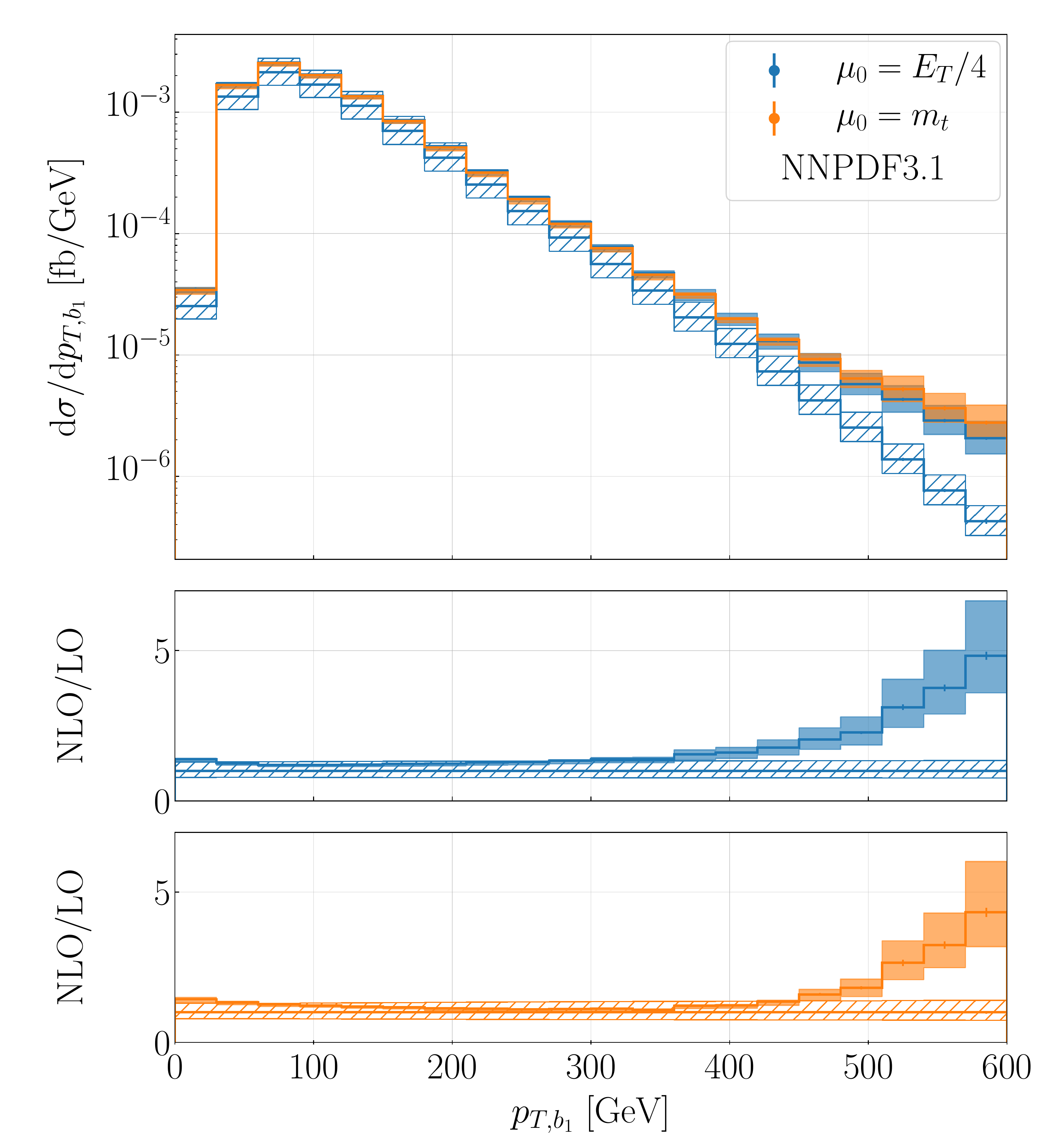}
	\includegraphics[width=0.49\textwidth]{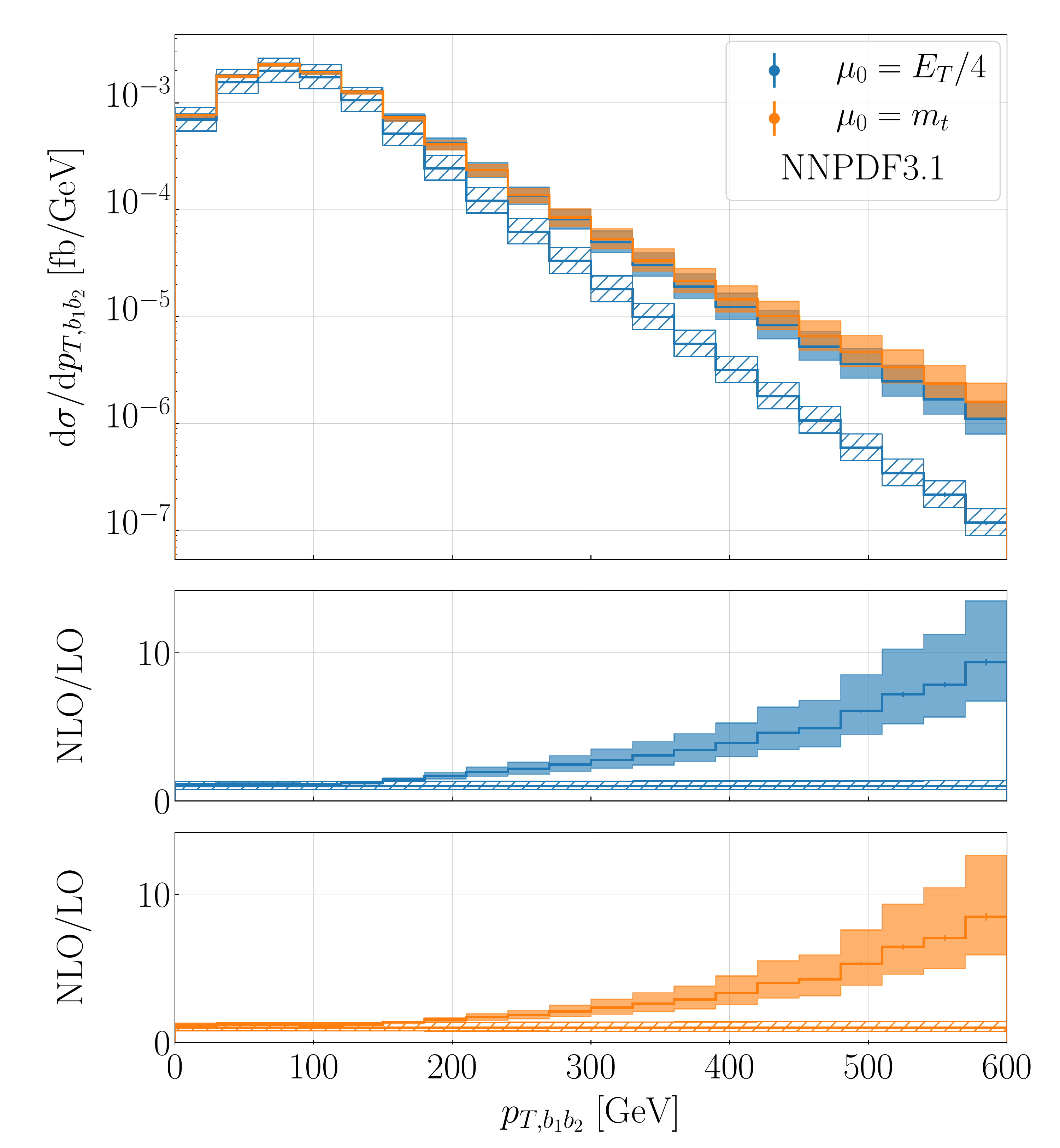}
	\includegraphics[width=0.49\textwidth]{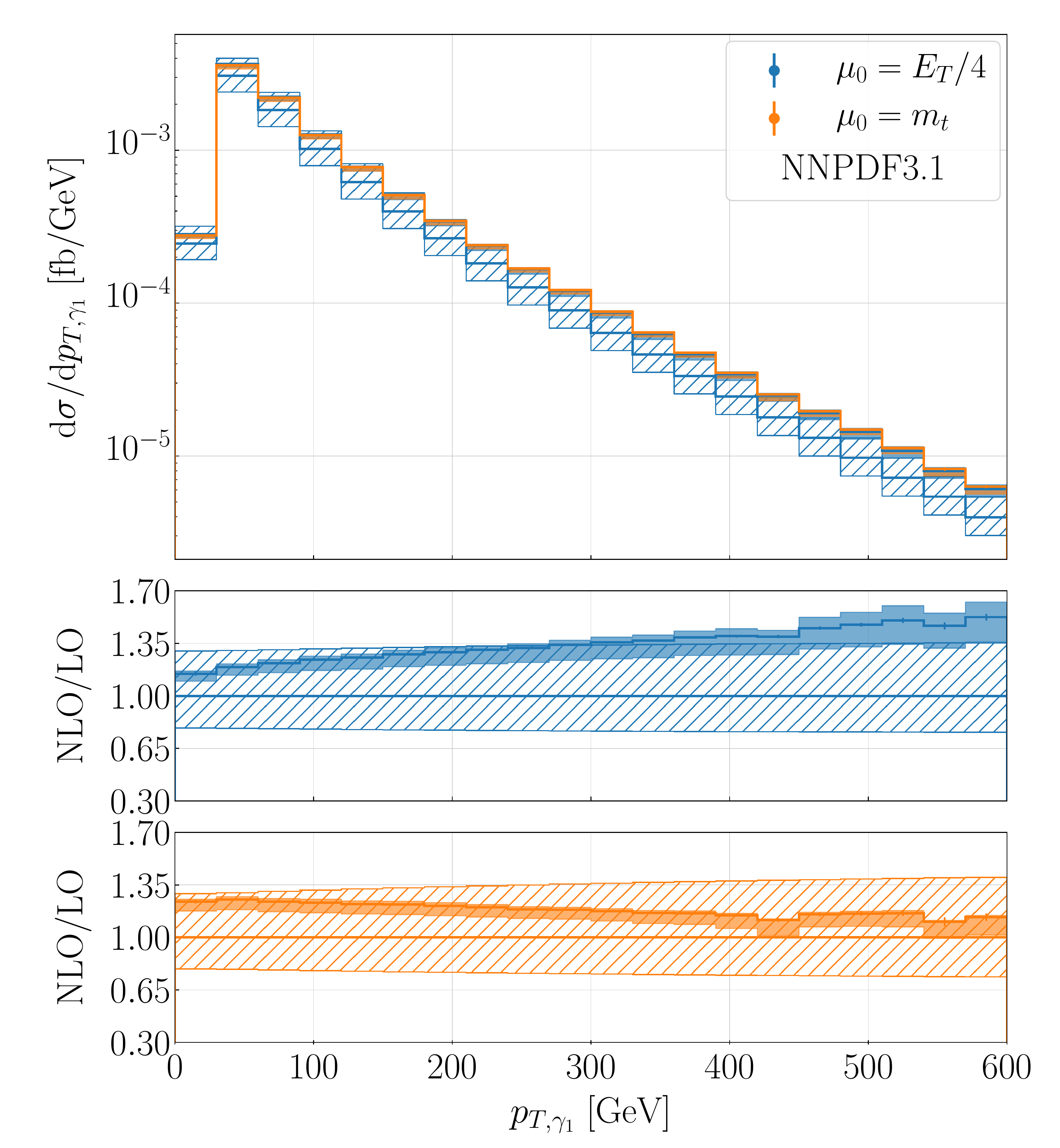}
	\includegraphics[width=0.49\textwidth]{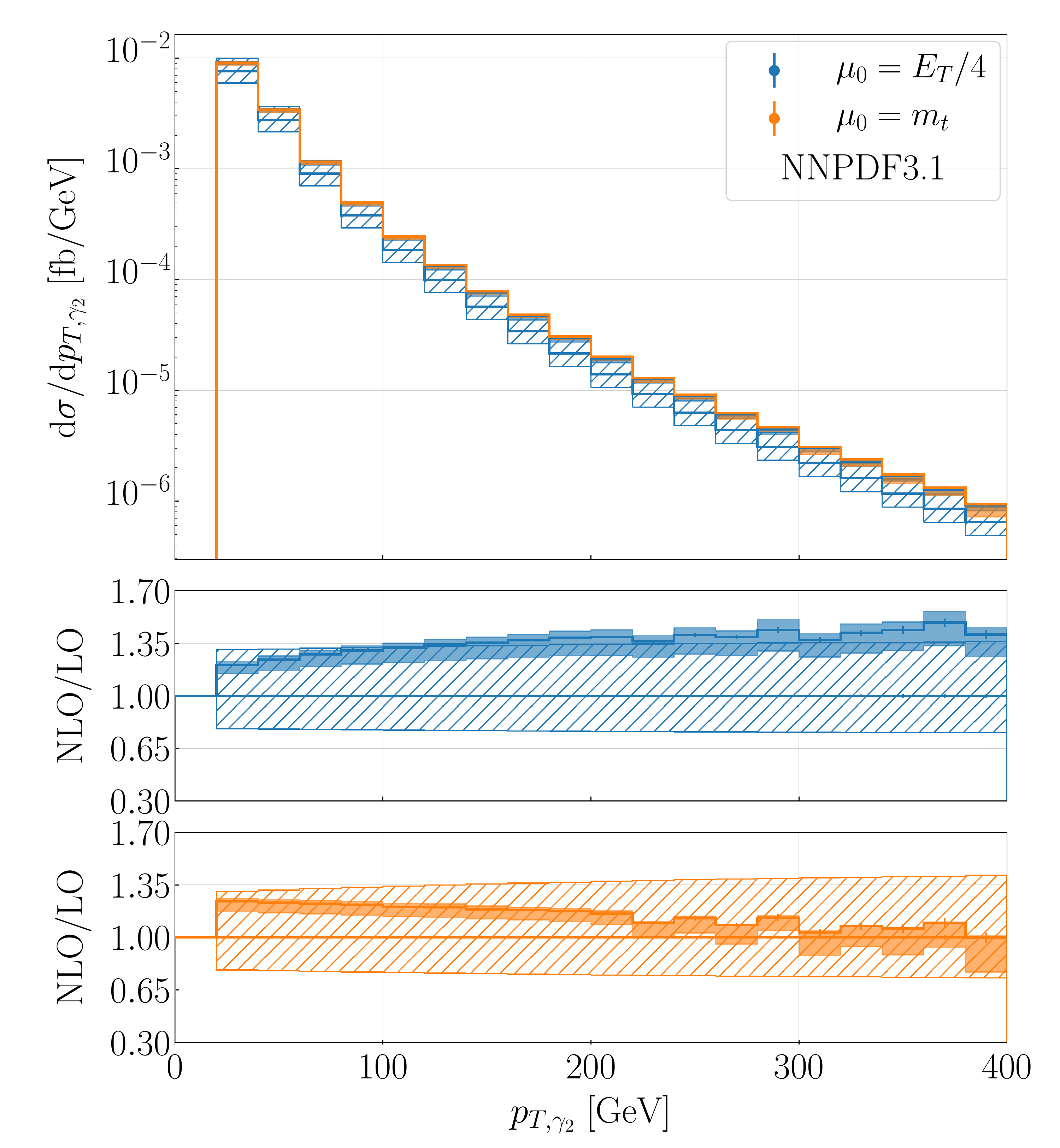}
    \end{center}
    \caption{\label{fig-semi:kfac1} \it Differential cross-section distributions for the observables $p_{T,b_1}$, $p_{T,b_1b_2}$, $p_{T,\gamma_1}$ and $p_{T,\gamma_2}$ for the $pp\to jj\, \ell^-\bar{\nu}_{\ell} \, b\bar{b}\,\gamma\gamma +X$ process at the LHC with $\sqrt{s}=13$ TeV. Results are presented for $\mu_0=E_T/4$ (blue) 
    and $\mu_0=m_t$ (orange) at NLO (solid) and LO (dashed) using the NNPDF3.1 PDF set. The two lower panels display the differential ${\cal K}$-factor for both scale choices with the uncertainty band  and the relative scale uncertainties of the LO cross section. }
\end{figure}
\begin{figure}[t!]
    \begin{center}
	\includegraphics[width=0.49\textwidth]{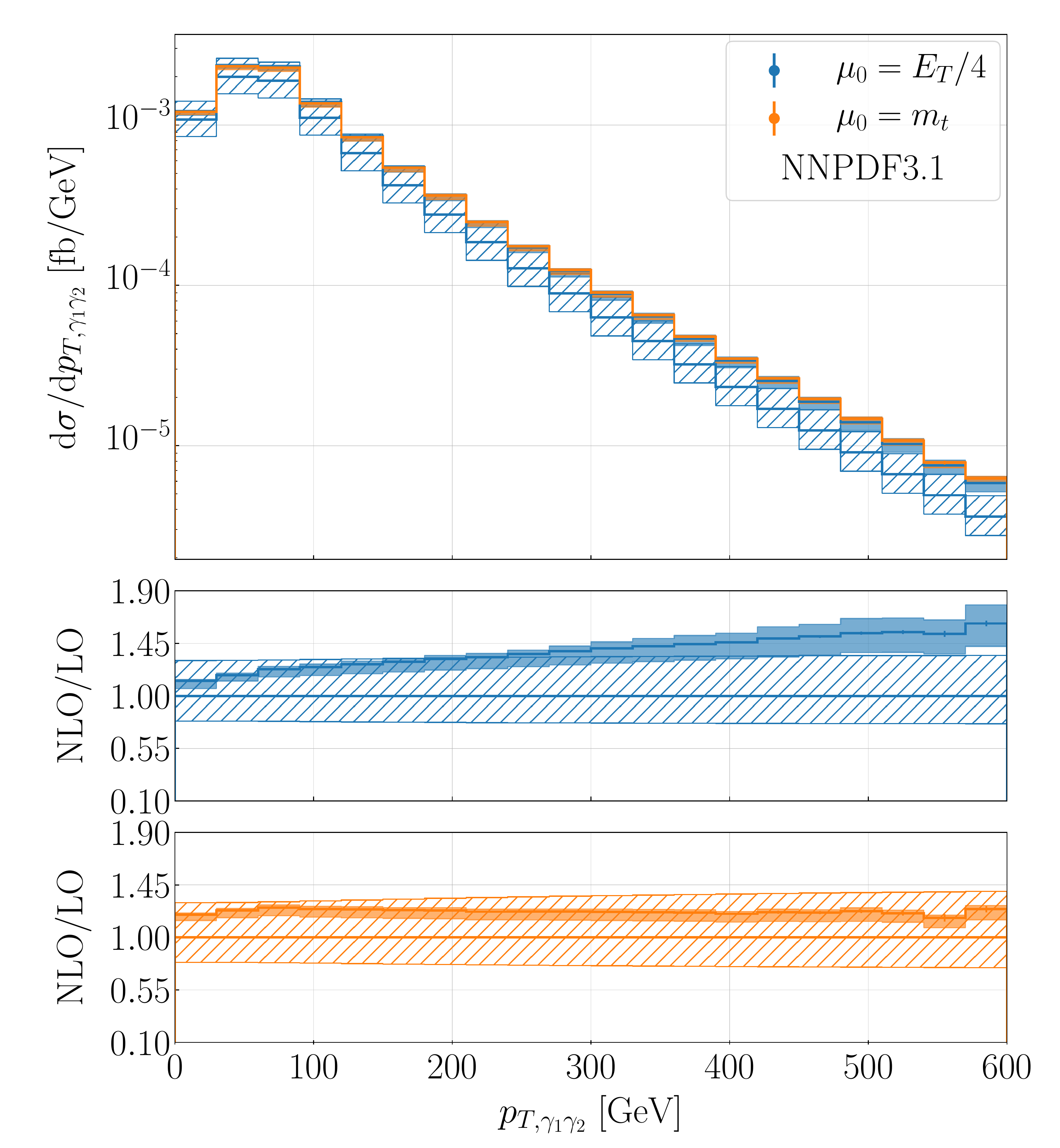}
	\includegraphics[width=0.49\textwidth]{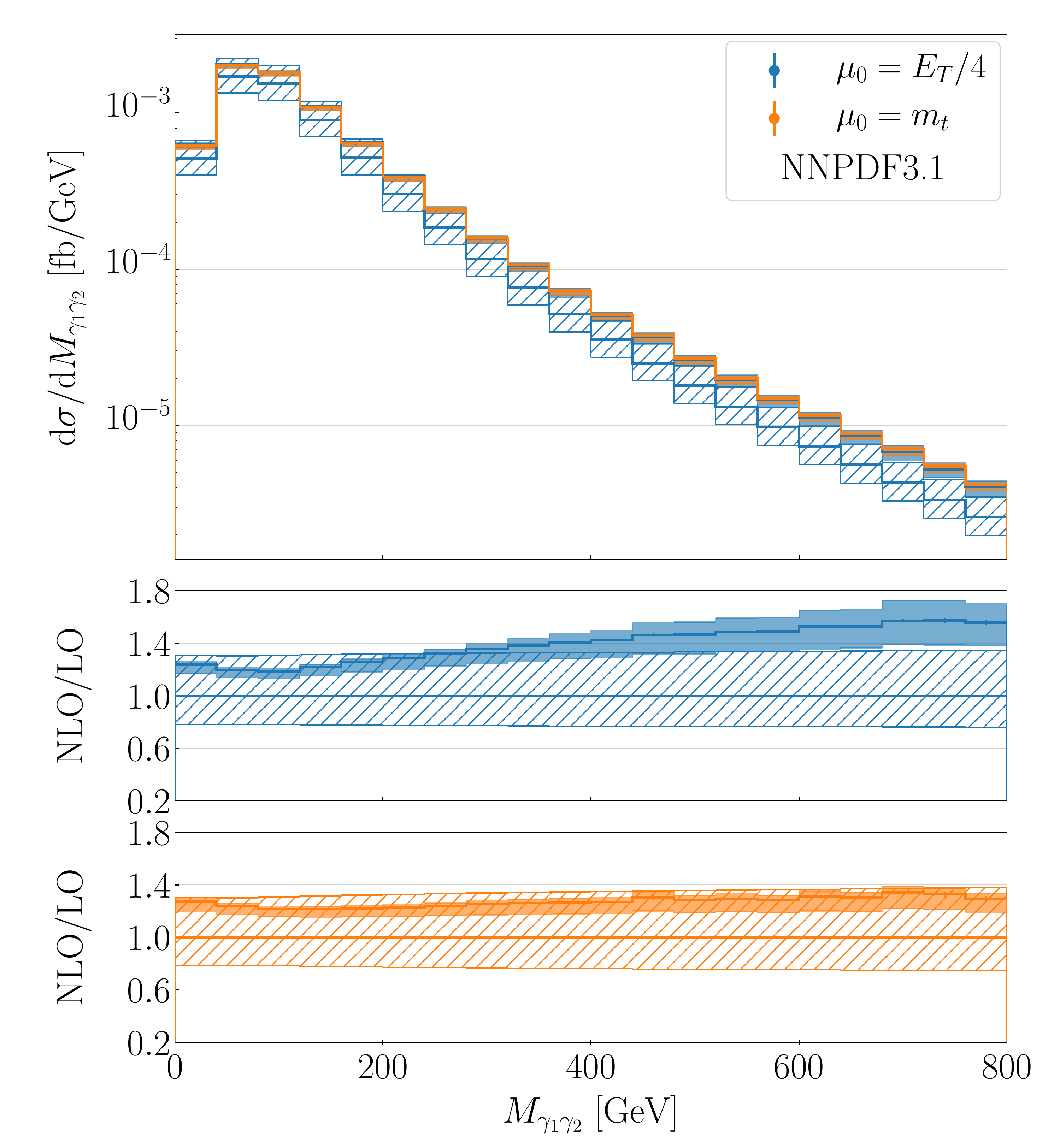}
	\includegraphics[width=0.49\textwidth]{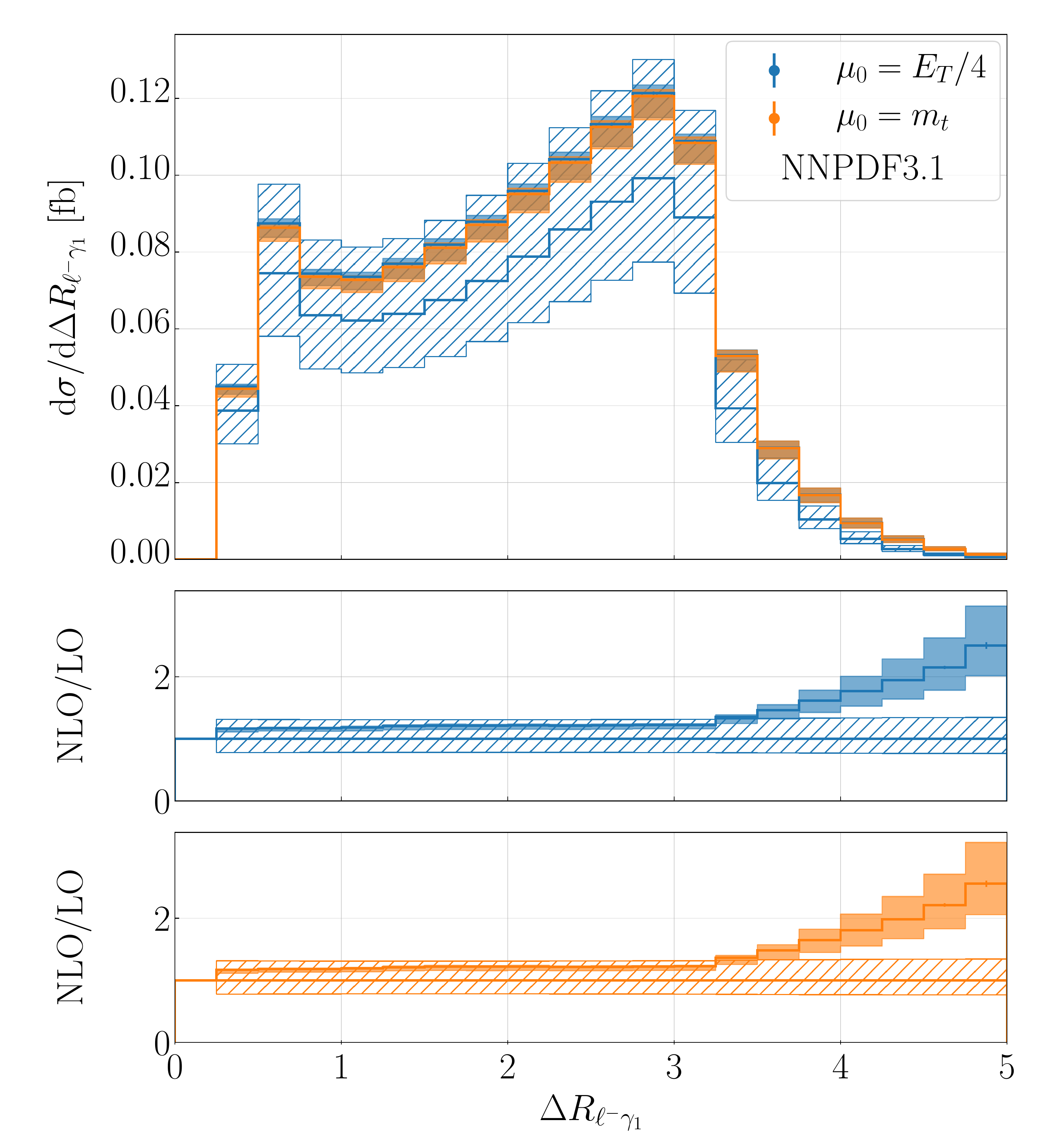}
	\includegraphics[width=0.49\textwidth]{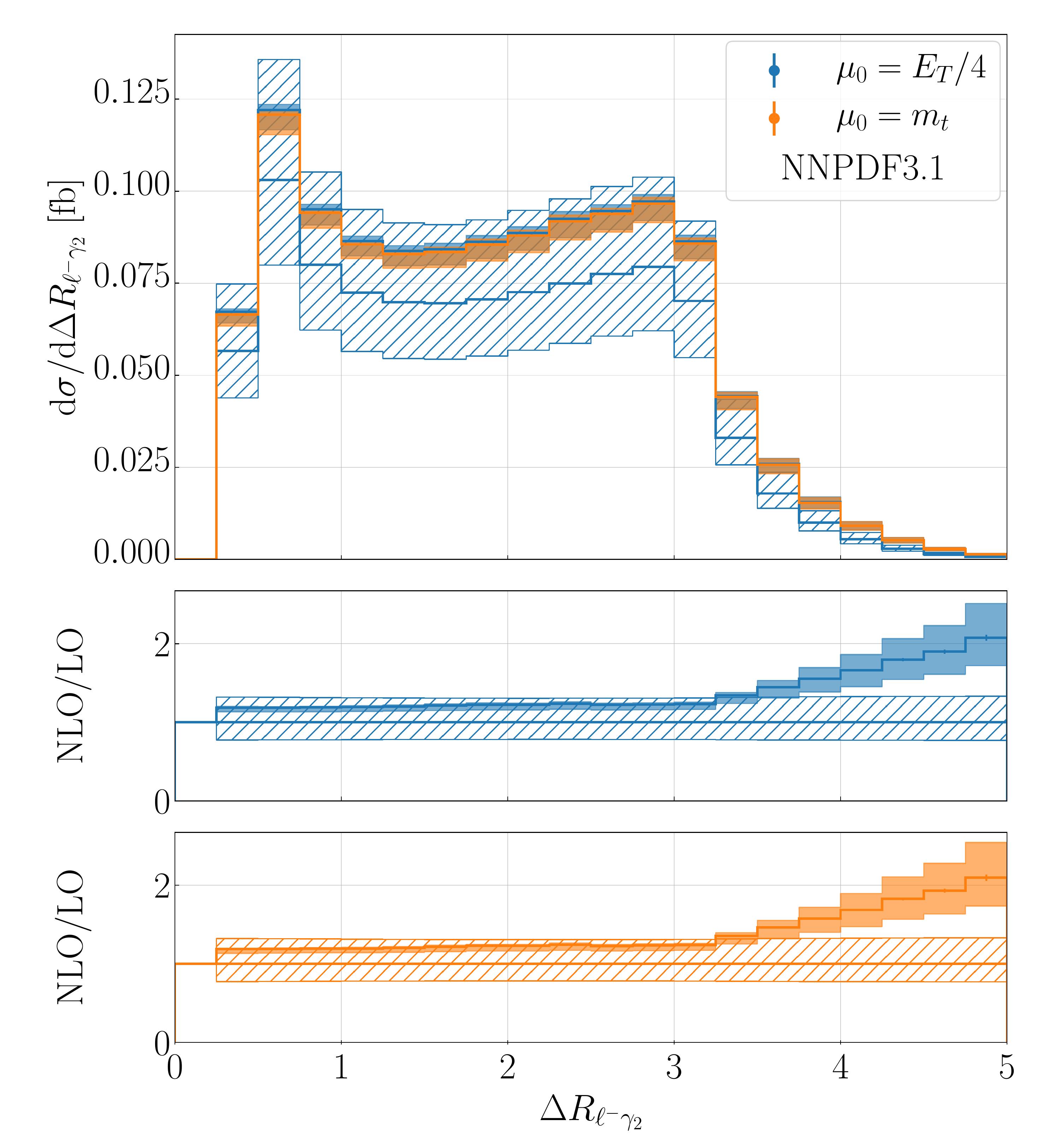}
    \end{center}
    \caption{\label{fig-semi:kfac2} \it Same as Figure \ref{fig-semi:kfac1} but for the observables $p_{T,\gamma_1\gamma_2}$, $M_{\gamma_1\gamma_2}$, $\Delta R_{\ell^-\gamma_1}$ and $\Delta R_{\ell^-\gamma_2}$. }
\end{figure}
\begin{figure}[t!]
    \begin{center}
	\includegraphics[width=0.49\textwidth]{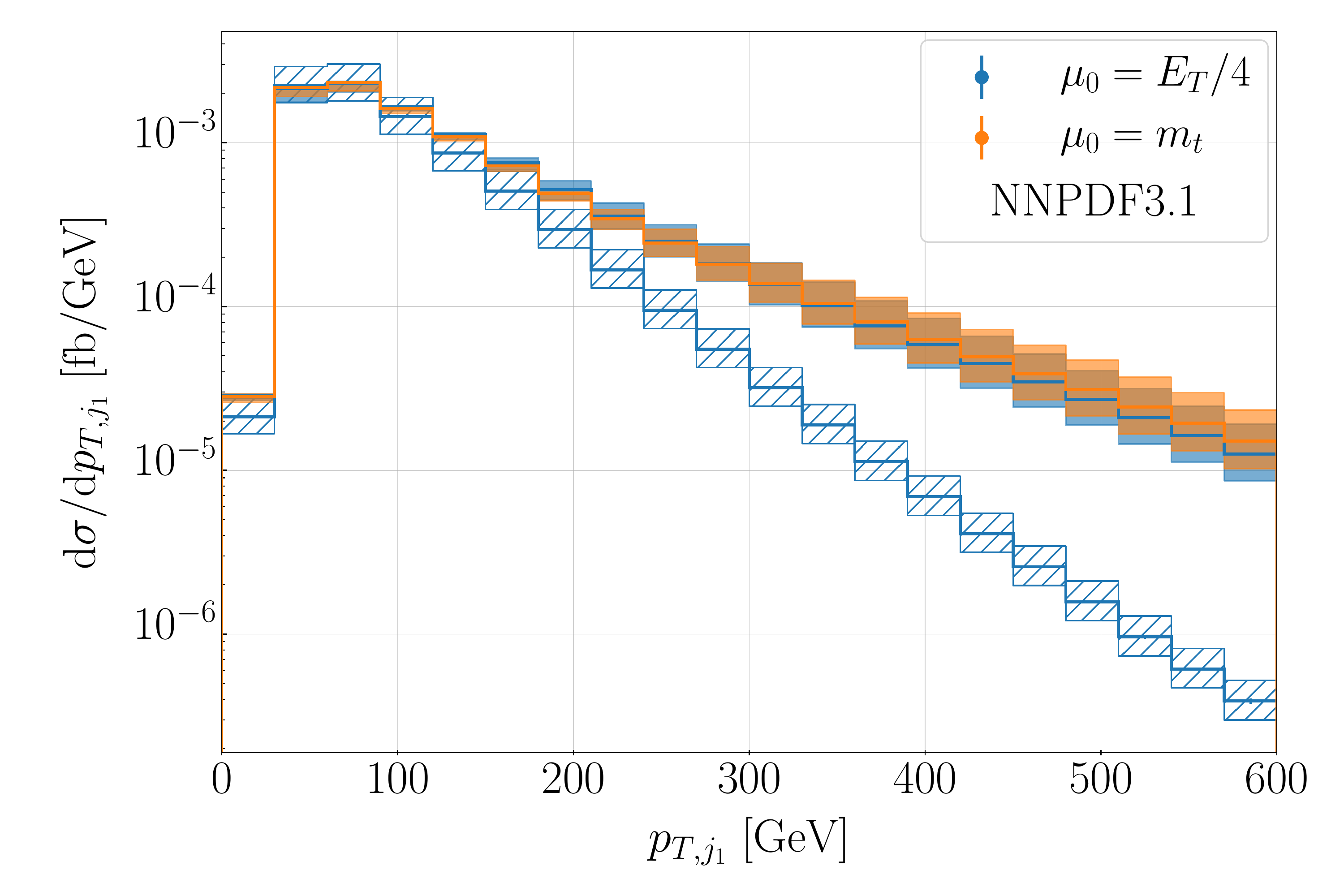}
	\includegraphics[width=0.49\textwidth]{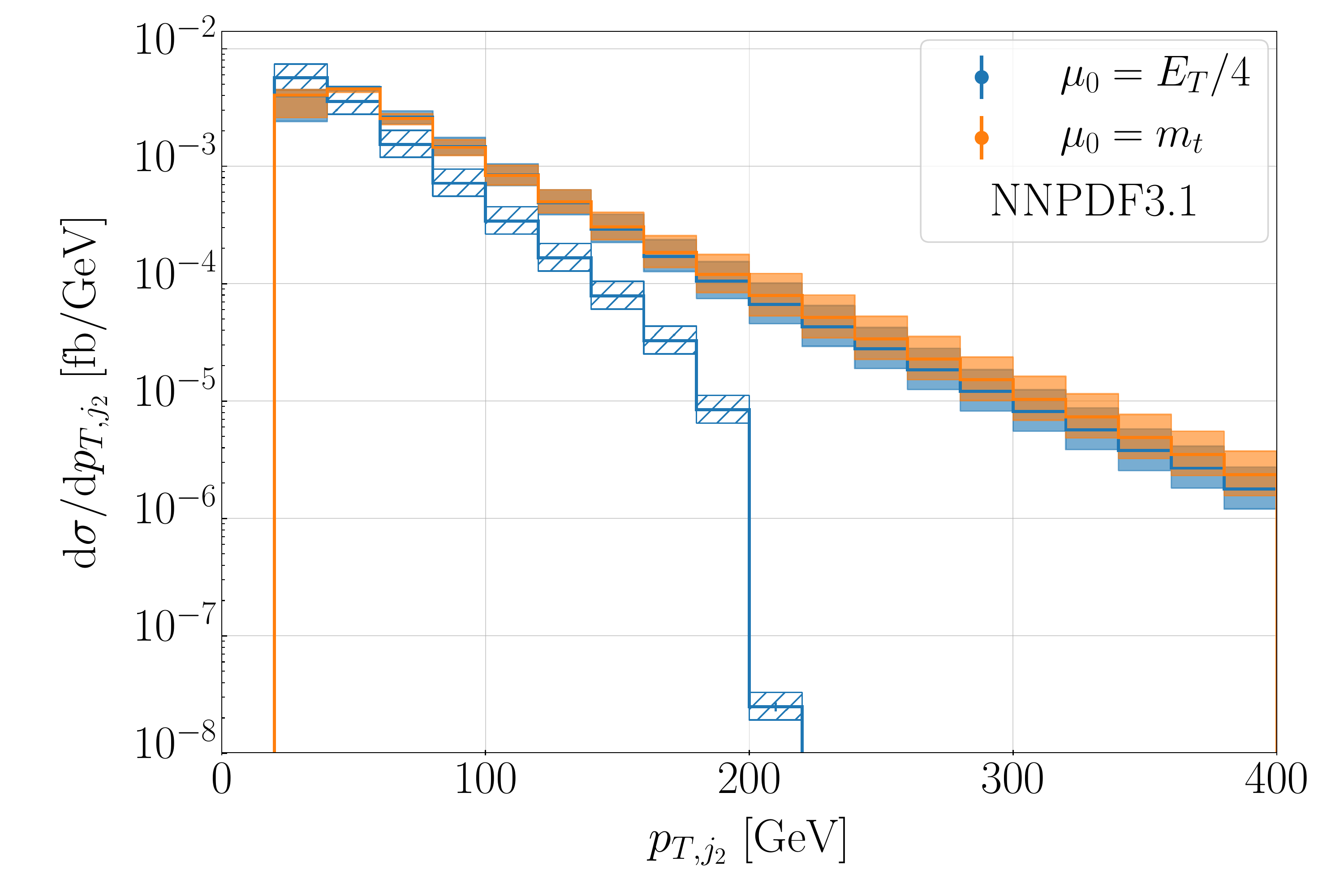}
	\includegraphics[width=0.49\textwidth]{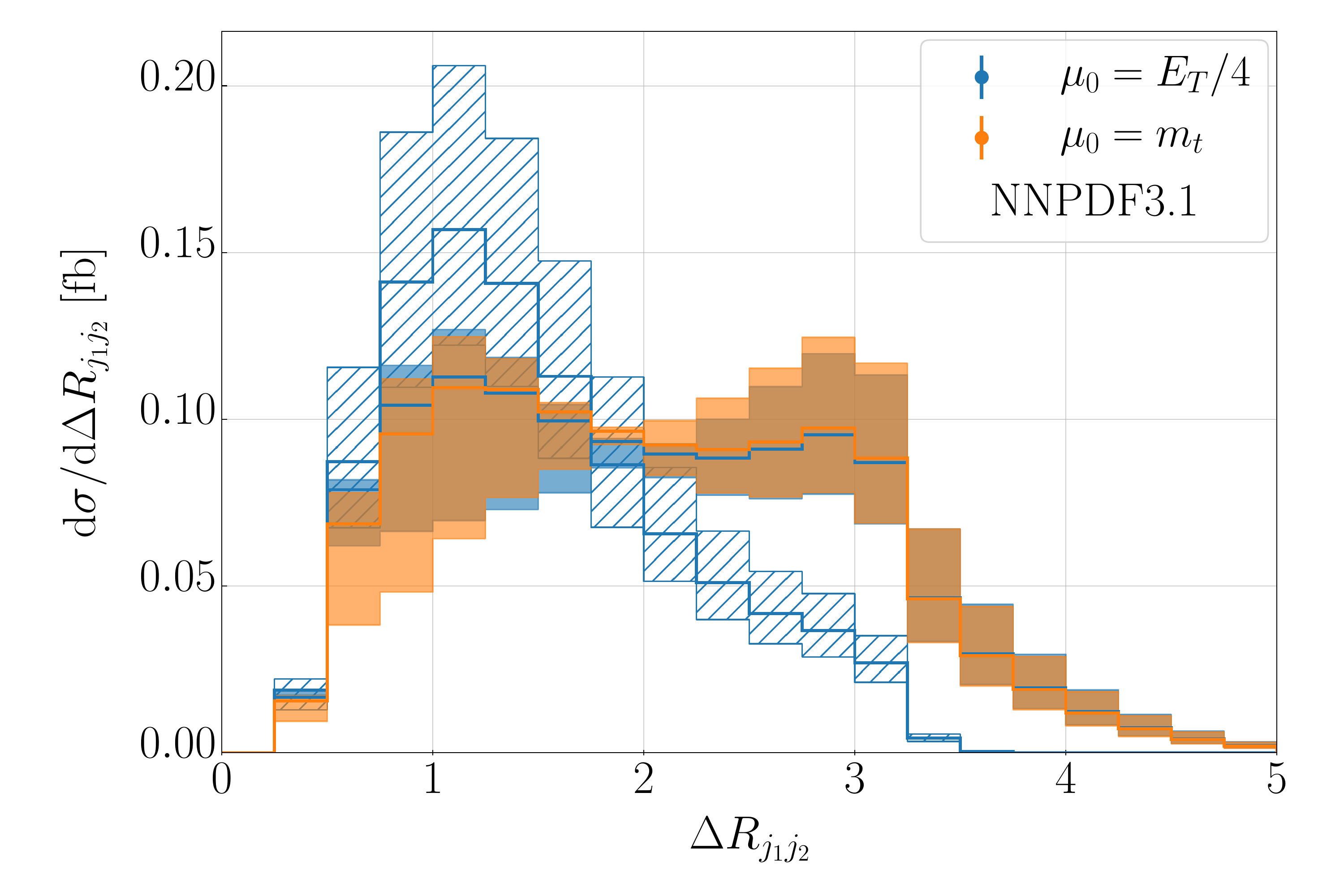}
	\includegraphics[width=0.49\textwidth]{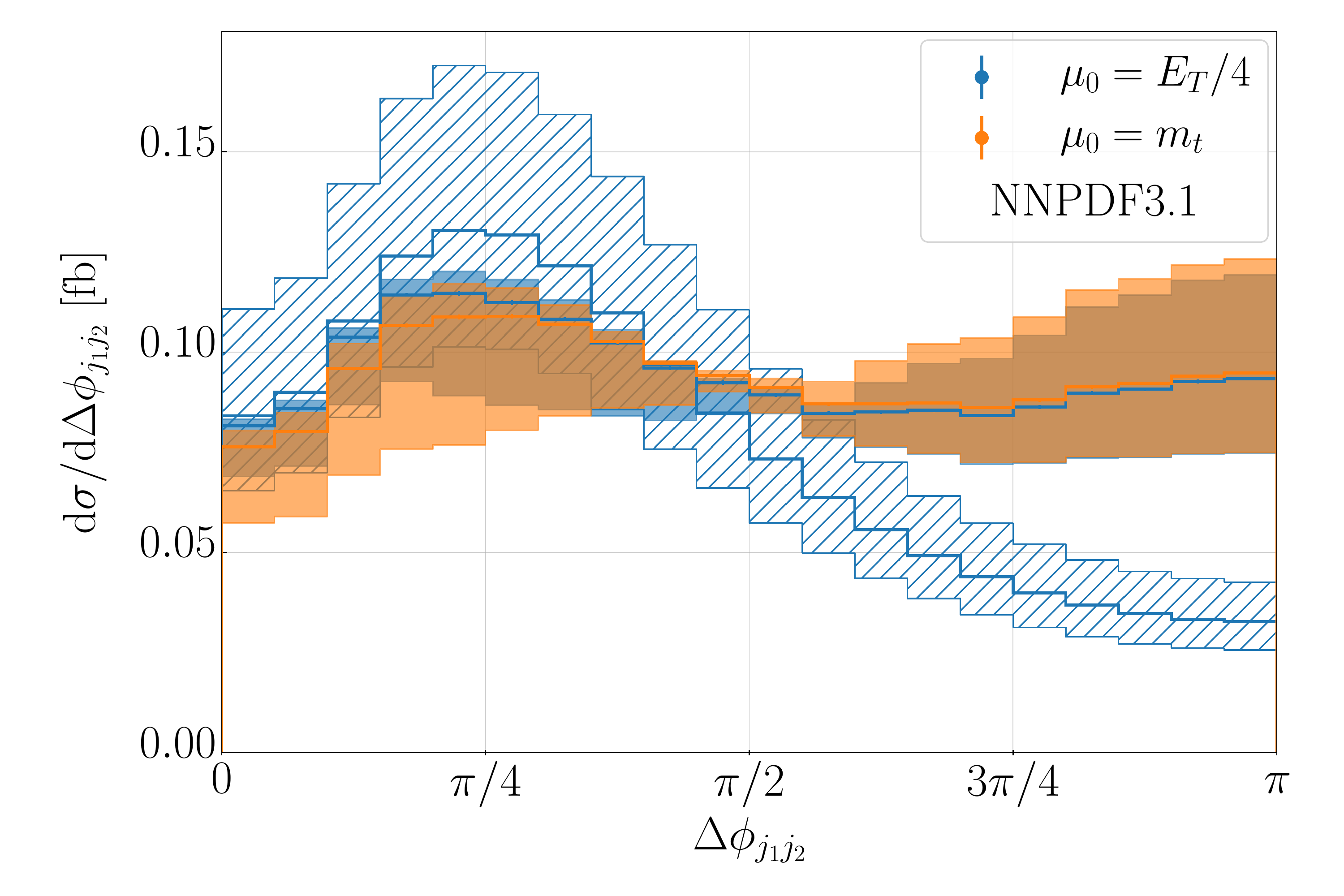}
    \end{center}
    \caption{\label{fig-semi:kfac3} \it Same as Figure \ref{fig-semi:kfac1} but without ratio plots for the observables $p_{T,j_1}$, $p_{T,j_2}$, $\Delta R_{j_1j_2}$ and $\Delta \Phi_{j_1j_2}$. }
\end{figure}

We continue our discussion of the \jetlep channel with the presentation of the results at the  differential cross-section level. First, we examine the size of NLO QCD corrections for similar observables that have been studied in the \dilep channel to directly assess the differences and similarities between the two decay channels. In Figure \ref{fig-semi:kfac1} we show the differential cross-section distributions as well as differential $\mathcal{K}$-factors for the following observables: $p_{T,b_1}$, $p_{T,b_1b_2}$, $p_{T,\gamma_1}$ and $p_{T,\gamma_2}$ at LO (dashed) and NLO QCD (solid) for the scales $\mu_0=E_T/4$ (blue) and $\mu_0=m_t$ (orange) employing the NNPDF3.1 PDF set. The lower plots show again the differential $\mathcal{K}$-factor for both scale choices together
with the corresponding uncertainty bands of the LO and NLO QCD predictions. We find huge NLO QCD corrections for the first two observables. In particular, higher-order QCD corrections of more than $350\%$  and $800\%$ are found in the tails of the  distributions for $p_{T,b_1}$ and $p_{T,b_1b_2}$, respectively, when the dynamical scale setting $\mu_0=E_T/4$ is employed. The fixed scale choice $\mu_0=m_t$ does not alter this behavior. As already explained for the $p_{T,b_1b_2}$ distribution in the case of  the \dilep channel, these large NLO QCD corrections are due to real radiation recoiling against the $t\bar{t}$ system. However, in the \jetlep channel we can have up to three hard unflavoured jets which enhance the size of NLO QCD corrections for these observables even further. Furthermore, we find that NLO QCD scale uncertainties in these high $p_T$ phase-space regions are up to $40\%-50\%$ while the differences between the two scale choices are within the range of $35\%-45\%$. On the other hand, for  $p_T < (150-200)$ GeV the NLO QCD corrections are reduced to $20\%-25\%$ for $p_{T,b_1}$ and $15\%-20\%$ for $p_{T,b_1b_2}$. In addition, both NLO results are within the LO scale uncertainty bands and the scale uncertainties are reduced from $30\%$ at LO to $7\%$ at NLO.  This phase-space region represents the bulk of the distribution and it is very important for current measurements at the LHC, while the high $p_T$ tails are not accessible yet.  Turning to  photon observables, we find that the differential $\mathcal{K}$-factor increases towards the tails for both $p_{T,\gamma_1}$ and $p_{T,\gamma_2}$ when the dynamical scale is employed. Similarly to the \dilep decay channel NLO QCD corrections up to $50\%$ are reduced to $25\%$ when the fixed scale setting is used instead. These differences are nevertheless driven by the substantial changes in the LO predictions. Indeed, the two NLO QCD results differ at most by $5\%$ between the two scale choices for both observables.  For the $p_{T,\gamma_1}$ observable both scale settings lead to rather similar scale uncertainties that are of the order of $5\%-11\%$.  On the other hand, for $p_{T,\gamma_2}$  scale uncertainties up to $15\%-20\%$  are  obtained for $\mu_0=m_t$, while for the dynamical scale setting these uncertainties are at most $10\%$. Thus,  the dynamical scale is necessary in the tails of dimensionful observables.

In the next step, we study  $p_{T,\gamma_1\gamma_2}$, $M_{\gamma_1\gamma_2}$, $\Delta R_{\ell^-\gamma_1}$ and $\Delta R_{\ell^-\gamma_2}$, which are  presented in Figure \ref{fig-semi:kfac2}. For the transverse momentum and the invariant mass of the two photon system, denoted as $p_{T,\gamma_1\gamma_2}$ and $M_{\gamma_1\gamma_2}$ respectively, NLO QCD corrections up to about $60\%$ have been found for the dynamical scale setting. The scale uncertainties are reduced from $30\%-35\%$ at LO to about $10\%-12\%$ at NLO in QCD. For the fixed scale choice higher-order QCD corrections are reduced to $20\%-25\%$ for $p_{T,\gamma_1\gamma_2}$ and are in the range of $20\%-35\%$ for $M_{\gamma_1\gamma_2}$. Consequently, $\mu_0=m_t$ provides better agreement with the LO predictions and overall flatter ${\cal K}$-factors for both distributions. In addition, scale uncertainties have been slightly reduced to just under $10\%$. Thus, in the case of dimensional photon observables also for the \jetlep decay channel we find a similar behavior of higher-order QCD effects as in the \dilep one. Indeed, also here larger differential $\mathcal{K}$-factors are found when employing the dynamical scale setting. However, both $\mu_0=E_T/4$ and  $\mu_0=m_t$ lead to 
equivalent results at NLO in QCD. For the angular separation between the photons and the negatively charged lepton: $\Delta R_{\ell^-\gamma_1}$ and $\Delta R_{\ell^-\gamma_2}$, we again find two distinct configurations, namely collinear and back-to-back configurations.  Compared to the \dilep channel the peak at small $\Delta R_{\ell^- \gamma_1}$ and $\Delta R_{\ell^- \gamma_2}$ is enhanced here due to the different event selection  for the top-quark decay products in the two decay channels.  Indeed, the definition of the fiducial phase space in the \jetlep decay channel highly suppresses photon bremsstrahlung in the hadronically decaying $W$ gauge boson.  As in the \dilep decay channel also here moderate NLO QCD corrections of about $15\%-25\%$ are found for both scale choices when $\Delta R_{ij} \in (0.4,3)$. In this range of $\Delta R_{ij}$ the scale uncertainties are below $10\%$.  On the other hand,  for $\Delta R_{ij}>  3$ higher-order corrections rapidly increase up to $100\%-150\%$ and the scale uncertainties are of the order of $20\%-25\%$.

Finally, we examine  the kinematics of the light jets. Thus, in the following we are focusing  on the truly new effects that are only visible in the \jetlep decay channel. In Figure \ref{fig-semi:kfac3} we display the transverse momentum of the hardest and the second hardest light jet, denoted as $p_{T,j_1}$ and $p_{T,j_2}$ respectively. Also shown are the angular separation between the first and the second hardest light jet as well as the angular difference between them in the transverse plane, denoted as $\Delta R_{j_1j_2}$ and $\Delta \Phi_{j_1j_2}$ respectively. No ratio plots are provided for these observables due to extreme values appearing in the corresponding differential $\mathcal{K}$-factors. In the case of the transverse momentum of the hardest light jet,  for $p_{T,\,j_1}<150$ GeV, we find NLO QCD corrections ranging from $-5\%$ to $35\%$ when  the dynamical scale setting $\mu_0=E_T/4$ is employed. In this phase-space region the NLO scale uncertainties vary between $6\%$ and $17\%$. For $p_{T,\,j_1} \ge 150$ GeV the NLO QCD prediction becomes significantly harder compared to the LO distribution. In particular, the NLO QCD prediction becomes larger by up to a factor of about $30$ than the LO one. The scale uncertainties in the high $p_T$ region are of the order of $50\%$. In addition, at the NLO QCD level the difference  between the two scale settings can be even up to $20\%$, that is still within the large NLO uncertainty bands. As we have already discussed, at LO the phase space of the two light jets, especially in the high $p_T$ region, is restricted due to the production mechanism as both light jets are originating from the $W$ gauge boson decay. However, at NLO QCD a light jet can also be produced in the $t\bar{t}$ production stage. This light jet, if resolved and passes all the cuts, is not affected by the kinematical restriction and can lead to huge enhancements in the tail of the $p_{T,\,j_1}$ distribution. As this phase-space region is only LO accurate, not only large scale uncertainties but also substantial differences between the two scale choices can be observed. For the transverse momentum of the second hardest jet, $p_{T,\,j_2}$, this effect is even more pronounced as has already been discussed e.g. in Ref. \cite{Denner:2017kzu} for the $pp \to t\bar{t}+X$ process. Indeed, at $p_{T,\,j_2} \approx 200$ GeV, the LO predictions decrease sharply to become even zero for  $p_{T,\,j_2} \ge 220$ GeV. Due to additional radiation at NLO QCD the restriction on the second hardest jet is lifted and thus the distribution is no longer zero for $p_{T,\, j_2}>220$ GeV.  For the same reason also the angular distributions $\Delta R_{j_1j_2}$ and $\Delta \Phi_{j_1j_2}$  show significant shape distortions at NLO QCD.  At LO the distribution of $\Delta R_{j_1j_2}$ peaks strongly at about $\Delta R_{j_1j_2}\approx 1$. Afterwards it rapidly decreases and is heavily suppressed for large values of  $\Delta R_{j_1j_2}$. 
The situation drastically changes at NLO in QCD. The peak at about $\Delta R_{j_1j_2}\approx 1$ is significantly reduced and a second peak can be found at  $\Delta R_{j_1j_2}\approx 3$. In addition, the huge suppression for large $\Delta R_{j_1j_2}$ does not occur anymore. For this observable large scale uncertainties up to $50\%$ are found for most parts of the distribution. Finally, for the $\Delta \Phi_{j_1j_2}$ distribution we find  again large differences between the LO and NLO spectra. At LO  a clear peak at $\pi/4$ is found that is  caused by the production mechanism of the two light jets at this order.  At NLO QCD the spectrum is rather flat over the entire range and the peak at $\pi/4$ is substantially reduced.  NLO QCD scale uncertainties in most parts of the distribution are up to $28\%$. Concluding, all distributions based on the kinematics of the hardest (light) jets receive large contributions from real radiation at NLO in QCD and NNLO QCD corrections would be necessary for more precise predictions of these specific observables in the phase-space regions that are kinematically restricted at LO.    In the absence of NNLO calculations for the $pp \to t\bar{t}\gamma\gamma$ process, however, other options should instead be explored. For example, a redefinition of the fiducial phase space or the inclusion of a jet veto might mitigate the large observed higher-order effects. In both cases, special and detailed studies are needed to clarify the issue.  We leave such studies for future investigation. 

%
\subsection{Distribution of prompt photons}
\label{sec:ttaa-semi-diff2}
%

%
\begin{figure}[t!]
    \begin{center}
	\includegraphics[width=0.49\textwidth]{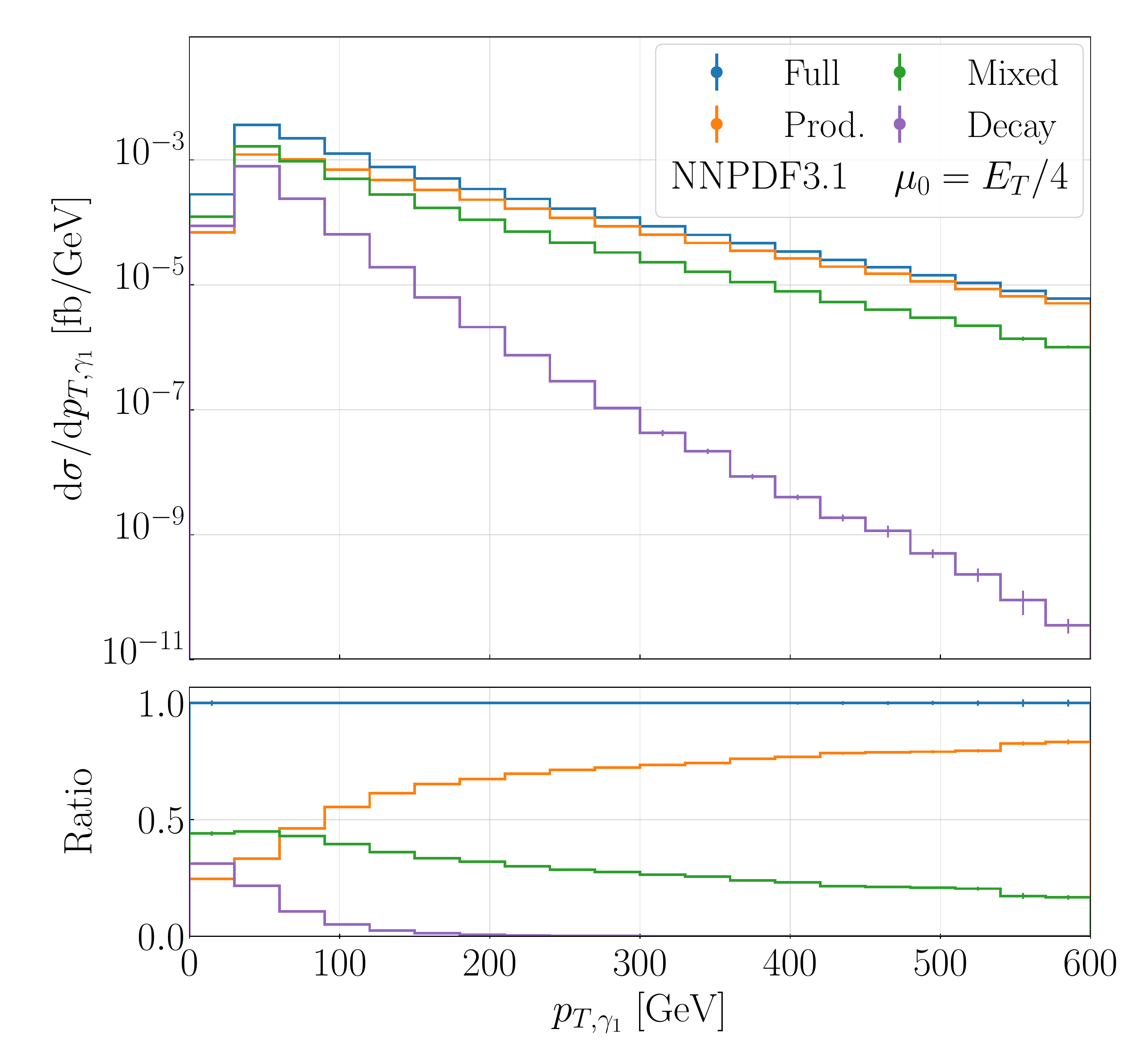}
	\includegraphics[width=0.49\textwidth]{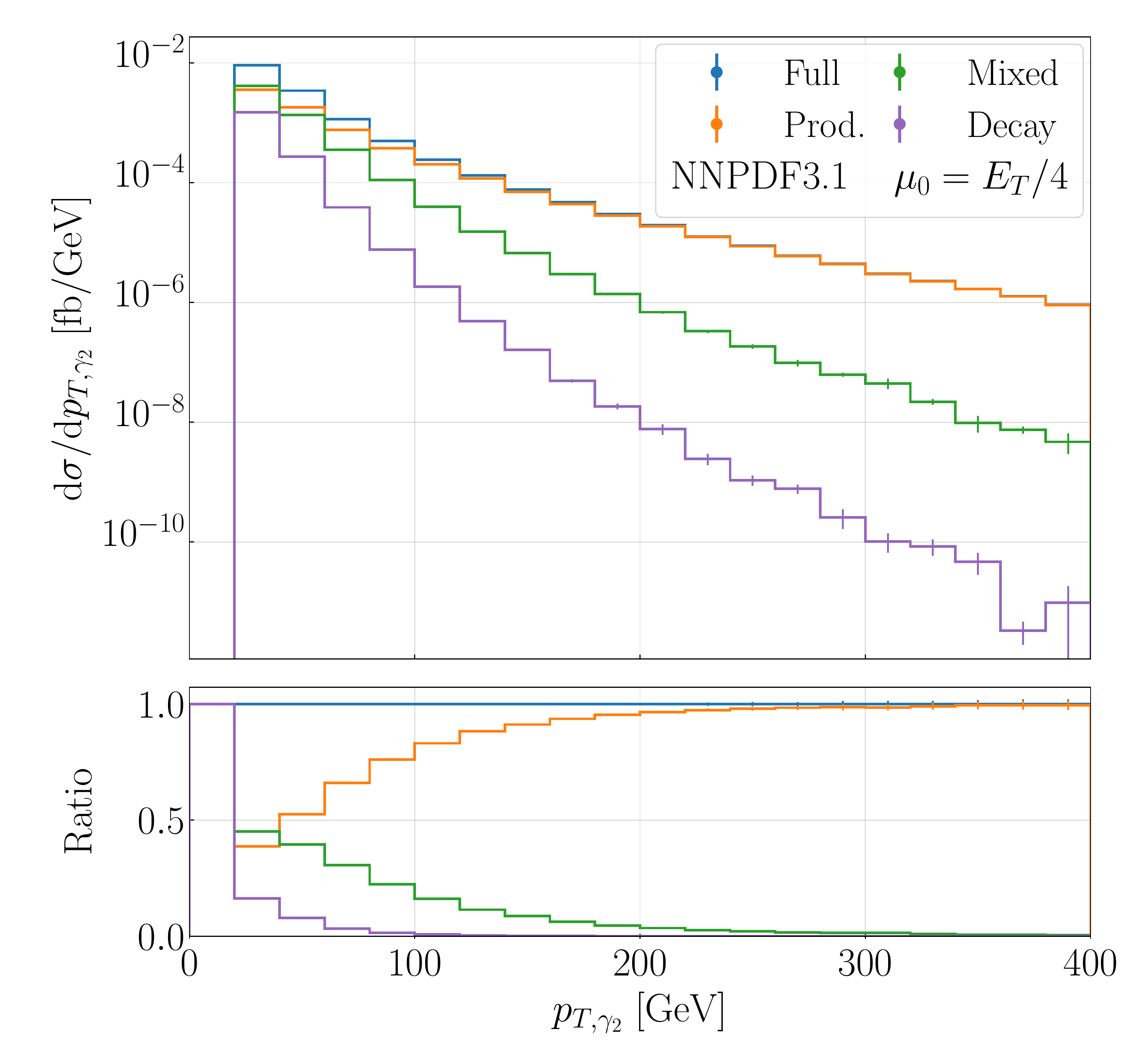}
	\includegraphics[width=0.49\textwidth]{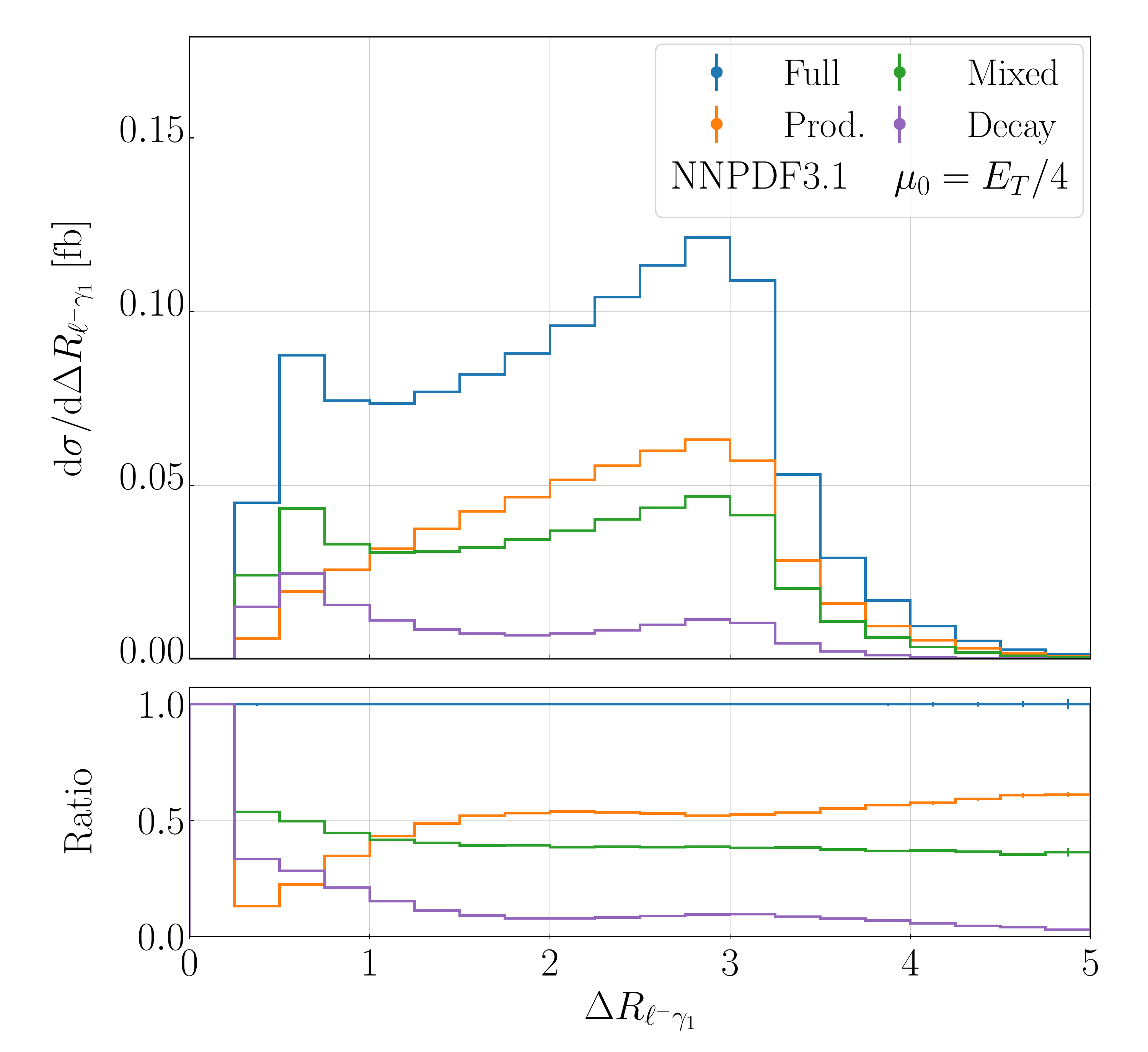}
 	\includegraphics[width=0.49\textwidth]{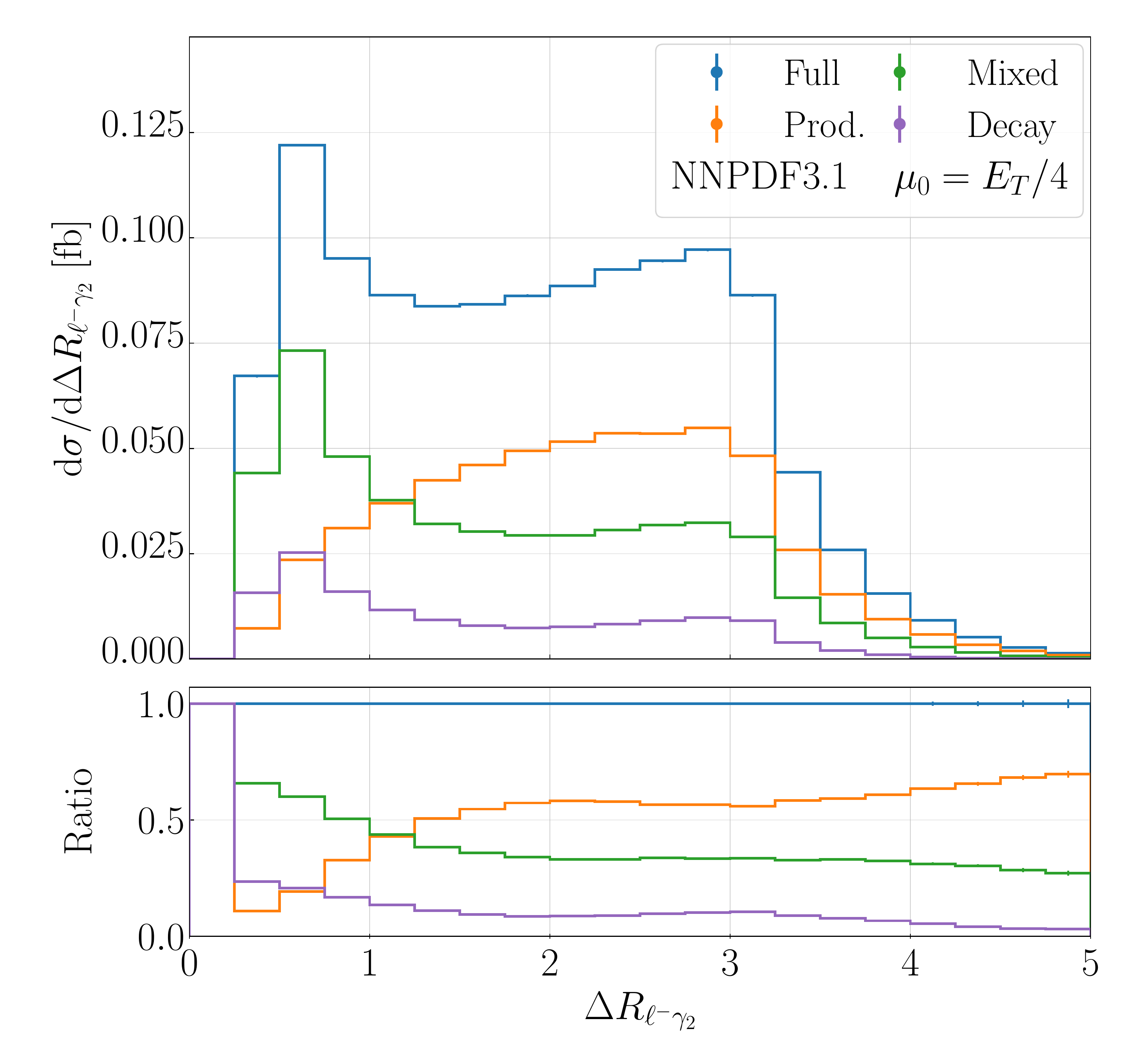}
    \end{center}
    \caption{\label{fig-semi:reg1} \it Differential cross-section distributions at NLO QCD for the observables $p_{T,\gamma_1}$, $p_{T,\gamma_2}$, $\Delta R_{\ell^-\gamma_1}$ and $\Delta R_{\ell^-\gamma_2}$ for the $pp\to jj\, \ell^-\bar{\nu}_{\ell} \, b\bar{b}\,\gamma\gamma +X$ process at the LHC with $\sqrt{s}=13$ TeV. Theoretical predictions are divided in the three contributions {\it Prod.}, {\it Mixed} and {\it Decay}. They are obtained with $\mu_0=E_T/4$ and the NNPDF3.1 PDF set. The lower panels display the ratio to the full NLO QCD result. MC integration errors are also shown. }
\end{figure}
As in the case of the \dilep channel, for the \jetlep decay channel we also study the distribution of photons in the $pp \to t\bar{t}\gamma\gamma$ process. In Figure \ref{fig-semi:reg1} we show the differential cross-section distribution  at NLO QCD as a function of $p_{T,\,\gamma_1}$, $p_{T,\,\gamma_2}$, $\Delta R_{\ell^-\gamma_1}$ and $\Delta R_{\ell^-\gamma_2}$. We employ  $\mu_0=E_T/4$ and the NNPDF3.1 PDF set. Theoretical predictions are once more divided into  the following three contributions:  {\it Prod.}, {\it Mixed} and {\it Decay}. As expected from the \dilep decay channel we find again a very similar picture of the size of the different resonant contributions. In particular, we observe that the {\it Mixed} contribution is the largest one in the small $p_T$ region with about $45\%$ for $p_{T,\,\gamma_1}$ and $p_{T,\,\gamma_2}$. Even the {\it Decay} contribution becomes comparable in size to the {\it Prod.} contribution  for $p_{T,\,\gamma_1}$ in this phase-space region and amounts to $31\%$. However, this contribution reduces rapidly towards larger values of $p_{T, \, \gamma}$ and amounts to less than $1\%$ for $p_{T,\,\gamma_1}>180$  GeV and $p_{T, \,\gamma_2}>100$ GeV.  Nevertheless, the {\it Decay} contribution is non-negligible in the phase space region that is  relevant for current measurements at the LHC. The {\it Mixed} contribution also decreases towards the tails of both distributions but remains at the level of  $17\%$ for $p_{T,\,\gamma_1}$ even when $p_{T,\,\gamma_1} \approx 600$ GeV. Contrary, for $p_{T,\gamma_2}$ we find that the {\it Mixed} contribution is below $1\%$ for $p_{T,\gamma_2} > 320$ GeV and, therefore, phenomenologically negligible. Indeed, in the high $p_{T, \, \gamma}$ tails the full distribution is dominated by the {\it Prod.} contribution. Similarly to the \dilep channel, also in this case a peak for small values of the $\Delta R_{\ell^-\gamma_1}$ and $\Delta R_{\ell^-\gamma_2}$ separation is entirely  driven by the {\it Mixed} and {\it Decay} contributions, which are  at the $54\%-66\%$ and $23\%-33\%$ level,  respectively. The {\it Prod.} contribution is actually the smallest one in this phase-space region, at the level of $11\%-13\%$ only. However, around $\Delta R_{\ell^-\gamma} \approx 3$ it increases up to $52\%$ and $57\%$ for $\Delta R_{\ell^-\gamma_1}$ and  $\Delta R_{\ell^-\gamma_2}$, respectively. In this phase-space region the counterparts {\it Mixed} and {\it Decay} are at the level of $40\%$ and $10\%$, respectively.

%
\section{Summary}
\label{sec:sum}
%

In this paper we presented the calculation of NLO QCD corrections to the $pp\to t\bar{t}\gamma\gamma$ process with realistic final states at the LHC with $\sqrt{s}=13$ TeV. In particular, we considered the \dilep as well as  \jetlep decay channel of the top-quark pair. The decays of the top quarks and $W$ bosons are handled in the NWA preserving spin correlations. In contrast to previous calculations of $pp\to t\bar{t}\gamma\gamma$ available in the literature, in this paper for the first time photon radiation and NLO QCD corrections have been consistently included in the production as well as the decays of  the $t\bar{t}$ pair.  An important finding of this paper is the magnitude of NLO QCD corrections  for the \jetlep decay channel due to kinematical configurations in which the two light jets from the hadronic decaying $W$ boson are recombined into one jet. Such contributions are not present at LO in our calculation since our event selection focus on resolved topologies. They are, however, possible at NLO QCD due to extra radiation. These large higher-order corrections can be substantially reduced by  requiring that  the invariant mass of at least one light-jet pair meets the following criterion  $|m_W-M_{jj}|< 15$ GeV. Similar large NLO QCD effects have already been observed for the $pp \to t\bar{t}$ process in Ref. \cite{Denner:2017kzu}. Two different scale settings, a dynamical and fixed one, have been used in this calculation. The former choice is based on the kinematics of the top-quark pair and the two photons $(\mu_0=E_T/4)$,  whereas the latter one is motivated by a natural scale for the $pp\to t\bar{t}\gamma\gamma$ process that is the mass of the top quark $(\mu_0=m_t)$. At the integrated fiducial cross-section level NLO QCD corrections for the $pp \to t\bar{t}\gamma\gamma$ process are moderate, at the level of $30\%$ and $23\%$, respectively for the \dilep and  \jetlep decay channel. On the other hand, the theoretical uncertainties from the scale variation are reduced from about $30\%$ at LO to $5\%$ at NLO QCD. 

The second important finding of this paper comprises the   decomposition of photon radiation for the $pp\to t\bar{t}\gamma\gamma$ process. At the integrated fiducial cross-section level the {\it Prod.} contribution amounts $40\%$ in the \dilep and $48\%$ in the \jetlep decay channel only.  Thus, the incorporation of photons in top-quark and $W$ boson decays leads to an increase  of the cross section by more than a factor of two. We have found that the large size of the {\it Mixed} and {\it Decay} contributions are due to a suppression of photon emission in  the $gg$ initiated process of the {\it Prod.} part, which is partially relaxed for the other two contributions.  

 The third important finding of this paper is the analysis of the dependence on various parameter choices in the smooth photon isolation prescription as introduced by Frixione in Ref. \cite{Frixione:1998jh}.  When many photons and jets
are present in the final state, the dependence on these parameters  is non-negligible and might affect comparisons between theoretical predictions and experimental results.  We have found that when varying   $n$, $\epsilon_\gamma$ and the coefficient $E_{T\, \gamma}\, \epsilon_\gamma$ for the $pp\to t\bar{t}\gamma\gamma$ process in the \jetlep  decay
channel, the integrated fiducial cross section has been changed by up to $10\%$. This change is larger than the corresponding NLO scale uncertainties
for this process and therefore of high relevance. In the \dilep  decay channel the dependence on these
parameters has been smaller, up to $6\%$ only, but still  as large as the corresponding NLO scale uncertainties. We plan to carry out dedicated studies on this topic 
not only at the integrated level, but also at the differential cross-section level in the near future. 

As a second source of theoretical uncertainties, we have studied the size of internal PDF uncertainties and the relative differences among the cross-section results obtained with various PDF sets.  In particular, we obtained PDF uncertainties in the range of $1\%-2\%$ for the three PDF sets NNPDF3.1, MSHT20 and CT18. The largest differences between PDF sets have been found for NNPDF3.1 and CT18 with about $2.4\%$. Concluding, the size of scale uncertainties remain the dominant source of theoretical systematics.

At the differential cross-section level NLO QCD corrections of up to $70\%$ have been observed for dimensionful observables related to photon kinematics for $\mu_0=E_T/4$. Furthermore,  these higher-order effects exceed the LO uncertainty bands. The use of the fixed scale, $\mu_0=m_t$, leads to a reduction of higher order corrections to a moderate range and places the NLO results within the LO uncertainty bands. However, only minor differences exist between the two predictions for the central value of the scale at NLO in QCD. In addition, we have seen that in general the use of a dynamical scale setting for dimensionalful observables is essential, as the fixed scale can lead to perturbative instabilities in high $p_T$ tails of certain distributions. In the \jetlep channel large shape differences have been found between LO and NLO predictions for several distributions describing the kinematics of the first and second hardest (light) jet. These differences are due to the additional radiation at NLO in QCD, which is not kinematically constrained due to the production mechanism, unlike the light jets that appear at LO. This becomes especially visible for $p_{T,\, j_2}$ which is kinematically limited at LO by the finite mass of the $W$ boson and the cut on the invariant mass of a jet pair.

In the next step,  we have assessed the size of the {\it Prod.},  {\it Decay} and  {\it Mixed} contribution  at the differential cross-section level. We have found that for dimensionful observables the {\it Mixed} contribution becomes the most important one in the small $p_T$ region. Even the {\it Decay} contribution can be comparable in size with the {\it Prod.} one in that phase-space region. However, in high $p_T$ tails the {\it Decay} contribution becomes phenomenologically negligible and the {\it Prod.} part is starting to dominate the full result. Nevertheless,  the {\it Mixed} contribution is generally significant even in high $p_T$ tails, where it can amount to more than $20\%$.  For angular distributions it is crucial to incorporate photon radiation in the decays of the top quark and $W$ boson.  Otherwise, entire new peaks may be missed, such as those clearly visible in the case of the angular separation of prompt  photons and charged leptons. Finally, we conclude by saying that the inclusion of photon radiation and NLO QCD corrections in top-quark decays is essential for LHC physics. Both effects would play a crucial role in the direct measurement of  the $pp\to t\bar{t}\gamma\gamma$ process and are equally important for reliable background modelling in the light of precise measurements of the $pp\to t\bar{t}H$ signal process in the $H\to \gamma\gamma$ decay channel.

\acknowledgments{
The work  was supported by the Deutsche Forschungsgemeinschaft (DFG) under grant 396021762 $-$ TRR 257: {\it P3H - Particle Physics Phenomenology after the Higgs Discovery}. Support by a grant of the Bundesministerium f\"ur Bildung und Forschung (BMBF) is additionally acknowledged.

Simulations were performed with computing resources granted by RWTH
Aachen University under project {\tt p0020216}. }


\begin{thebibliography}{10}

\bibitem{CMS:2018uxb}
{\scshape CMS} collaboration, \emph{{Observation of $t\bar{t}H$ production}},
  \href{https://doi.org/10.1103/PhysRevLett.120.231801}{\emph{Phys. Rev. Lett.}
  {\bfseries 120} (2018) 231801}
  [\href{https://arxiv.org/abs/1804.02610}{{\ttfamily 1804.02610}}].

\bibitem{ATLAS:2018mme}
{\scshape ATLAS} collaboration, \emph{{Observation of Higgs boson production in
  association with a top quark pair at the LHC with the ATLAS detector}},
  \href{https://doi.org/10.1016/j.physletb.2018.07.035}{\emph{Phys. Lett. B}
  {\bfseries 784} (2018) 173}
  [\href{https://arxiv.org/abs/1806.00425}{{\ttfamily 1806.00425}}].

\bibitem{ATLAS:2016ifi}
{\scshape ATLAS} collaboration, \emph{{Test of CP Invariance in vector-boson
  fusion production of the Higgs boson using the Optimal Observable method in
  the ditau decay channel with the ATLAS detector}},
  \href{https://doi.org/10.1140/epjc/s10052-016-4499-5}{\emph{Eur. Phys. J. C}
  {\bfseries 76} (2016) 658}
  [\href{https://arxiv.org/abs/1602.04516}{{\ttfamily 1602.04516}}].

\bibitem{CMS:2016tad}
{\scshape CMS} collaboration, \emph{{Combined search for anomalous pseudoscalar
  $HVV$ couplings in $VH(H \to b \bar b)$ production and $H \to VV$ decay}},
  \href{https://doi.org/10.1016/j.physletb.2016.06.004}{\emph{Phys. Lett. B}
  {\bfseries 759} (2016) 672}
  [\href{https://arxiv.org/abs/1602.04305}{{\ttfamily 1602.04305}}].

\bibitem{ATLAS:2017azn}
{\scshape ATLAS} collaboration, \emph{{Measurement of the Higgs boson coupling
  properties in the $H\rightarrow ZZ^{*} \rightarrow 4\ell$ decay channel at
  $\sqrt{s}$ = 13 TeV with the ATLAS detector}},
  \href{https://doi.org/10.1007/JHEP03(2018)095}{\emph{JHEP} {\bfseries 03}
  (2018) 095} [\href{https://arxiv.org/abs/1712.02304}{{\ttfamily
  1712.02304}}].

\bibitem{ATLAS:2018hxb}
{\scshape ATLAS} collaboration, \emph{{Measurements of Higgs boson properties
  in the diphoton decay channel with 36 fb$^{-1}$ of $pp$ collision data at
  $\sqrt{s} = 13$ TeV with the ATLAS detector}},
  \href{https://doi.org/10.1103/PhysRevD.98.052005}{\emph{Phys. Rev. D}
  {\bfseries 98} (2018) 052005}
  [\href{https://arxiv.org/abs/1802.04146}{{\ttfamily 1802.04146}}].

\bibitem{CMS:2019ekd}
{\scshape CMS} collaboration, \emph{{Measurements of the Higgs boson width and
  anomalous $HVV$ couplings from on-shell and off-shell production in the
  four-lepton final state}},
  \href{https://doi.org/10.1103/PhysRevD.99.112003}{\emph{Phys. Rev. D}
  {\bfseries 99} (2019) 112003}
  [\href{https://arxiv.org/abs/1901.00174}{{\ttfamily 1901.00174}}].

\bibitem{CMS:2019jdw}
{\scshape CMS} collaboration, \emph{{Constraints on anomalous $HVV$ couplings
  from the production of Higgs bosons decaying to $\tau$ lepton pairs}},
  \href{https://doi.org/10.1103/PhysRevD.100.112002}{\emph{Phys. Rev. D}
  {\bfseries 100} (2019) 112002}
  [\href{https://arxiv.org/abs/1903.06973}{{\ttfamily 1903.06973}}].

\bibitem{Gunion:1996xu}
J.~F. Gunion and X.-G. He, \emph{{Determining the CP nature of a neutral Higgs
  boson at the LHC}},
  \href{https://doi.org/10.1103/PhysRevLett.76.4468}{\emph{Phys. Rev. Lett.}
  {\bfseries 76} (1996) 4468}
  [\href{https://arxiv.org/abs/hep-ph/9602226}{{\ttfamily hep-ph/9602226}}].

\bibitem{Demartin:2014fia}
F.~Demartin, F.~Maltoni, K.~Mawatari, B.~Page and M.~Zaro, \emph{{Higgs
  characterisation at NLO in QCD: CP properties of the top-quark Yukawa
  interaction}},
  \href{https://doi.org/10.1140/epjc/s10052-014-3065-2}{\emph{Eur. Phys. J. C}
  {\bfseries 74} (2014) 3065}
  [\href{https://arxiv.org/abs/1407.5089}{{\ttfamily 1407.5089}}].

\bibitem{Mileo:2016mxg}
N.~Mileo, K.~Kiers, A.~Szynkman, D.~Crane and E.~Gegner, \emph{{Pseudoscalar
  top-Higgs coupling: exploration of CP-odd observables to resolve the sign
  ambiguity}}, \href{https://doi.org/10.1007/JHEP07(2016)056}{\emph{JHEP}
  {\bfseries 07} (2016) 056}
  [\href{https://arxiv.org/abs/1603.03632}{{\ttfamily 1603.03632}}].

\bibitem{Gritsan:2016hjl}
A.~V. Gritsan, R.~R\"ontsch, M.~Schulze and M.~Xiao, \emph{{Constraining
  anomalous Higgs boson couplings to the heavy flavor fermions using matrix
  element techniques}},
  \href{https://doi.org/10.1103/PhysRevD.94.055023}{\emph{Phys. Rev. D}
  {\bfseries 94} (2016) 055023}
  [\href{https://arxiv.org/abs/1606.03107}{{\ttfamily 1606.03107}}].

\bibitem{Demartin:2016axk}
F.~Demartin, B.~Maier, F.~Maltoni, K.~Mawatari and M.~Zaro, \emph{{tWH
  associated production at the LHC}},
  \href{https://doi.org/10.1140/epjc/s10052-017-4601-7}{\emph{Eur. Phys. J. C}
  {\bfseries 77} (2017) 34} [\href{https://arxiv.org/abs/1607.05862}{{\ttfamily
  1607.05862}}].

\bibitem{AmorDosSantos:2017ayi}
S.~Amor Dos~Santos et~al., \emph{{Probing the CP nature of the Higgs coupling
  in $t{\bar t}h$ events at the LHC}},
  \href{https://doi.org/10.1103/PhysRevD.96.013004}{\emph{Phys. Rev. D}
  {\bfseries 96} (2017) 013004}
  [\href{https://arxiv.org/abs/1704.03565}{{\ttfamily 1704.03565}}].

\bibitem{Bernreuther:2018ynm}
W.~Bernreuther, L.~Chen and Z.-G. Si, \emph{{Differential decay rates of
  CP-even and CP-odd Higgs bosons to top and bottom quarks at NNLO QCD}},
  \href{https://doi.org/10.1007/JHEP07(2018)159}{\emph{JHEP} {\bfseries 07}
  (2018) 159} [\href{https://arxiv.org/abs/1805.06658}{{\ttfamily
  1805.06658}}].

\bibitem{Goncalves:2018agy}
D.~Gon\c{c}alves, K.~Kong and J.~H. Kim, \emph{{Probing the top-Higgs Yukawa CP
  structure in dileptonic $ t\overline{t}h $ with M$_{2}$-assisted
  reconstruction}}, \href{https://doi.org/10.1007/JHEP06(2018)079}{\emph{JHEP}
  {\bfseries 06} (2018) 079}
  [\href{https://arxiv.org/abs/1804.05874}{{\ttfamily 1804.05874}}].

\bibitem{Martini:2021uey}
T.~Martini, R.-Q. Pan, M.~Schulze and M.~Xiao, \emph{{Probing the CP structure
  of the top quark Yukawa coupling: Loop sensitivity versus on-shell
  sensitivity}}, \href{https://doi.org/10.1103/PhysRevD.104.055045}{\emph{Phys.
  Rev. D} {\bfseries 104} (2021) 055045}
  [\href{https://arxiv.org/abs/2104.04277}{{\ttfamily 2104.04277}}].

\bibitem{Hermann:2022vit}
J.~Hermann, D.~Stremmer and M.~Worek, \emph{{$ \mathcal{CP} $ structure of the
  top-quark Yukawa interaction: NLO QCD corrections and off-shell effects}},
  \href{https://doi.org/10.1007/JHEP09(2022)138}{\emph{JHEP} {\bfseries 09}
  (2022) 138} [\href{https://arxiv.org/abs/2205.09983}{{\ttfamily
  2205.09983}}].

\bibitem{Barman:2021yfh}
R.~K. Barman, D.~Gon\c{c}alves and F.~Kling, \emph{{Machine learning the Higgs
  boson-top quark CP phase}},
  \href{https://doi.org/10.1103/PhysRevD.105.035023}{\emph{Phys. Rev. D}
  {\bfseries 105} (2022) 035023}
  [\href{https://arxiv.org/abs/2110.07635}{{\ttfamily 2110.07635}}].

\bibitem{Bahl:2021dnc}
H.~Bahl and S.~Brass, \emph{{Constraining $ \mathcal{CP} $-violation in the
  Higgs-top-quark interaction using machine-learning-based inference}},
  \href{https://doi.org/10.1007/JHEP03(2022)017}{\emph{JHEP} {\bfseries 03}
  (2022) 017} [\href{https://arxiv.org/abs/2110.10177}{{\ttfamily
  2110.10177}}].

\bibitem{Azevedo:2022jnd}
D.~Azevedo, R.~Capucha, A.~Onofre and R.~Santos, \emph{{CP-violation,
  asymmetries and interferences in $ t\overline{t}\phi $}},
  \href{https://doi.org/10.1007/JHEP09(2022)246}{\emph{JHEP} {\bfseries 09}
  (2022) 246} [\href{https://arxiv.org/abs/2208.04271}{{\ttfamily
  2208.04271}}].

\bibitem{CMS:2020cga}
{\scshape CMS} collaboration, \emph{{Measurements of $t\bar{t}H$ Production and
  the CP Structure of the Yukawa Interaction between the Higgs Boson and Top
  Quark in the Diphoton Decay Channel}},
  \href{https://doi.org/10.1103/PhysRevLett.125.061801}{\emph{Phys. Rev. Lett.}
  {\bfseries 125} (2020) 061801}
  [\href{https://arxiv.org/abs/2003.10866}{{\ttfamily 2003.10866}}].

\bibitem{ATLAS:2020ior}
{\scshape ATLAS} collaboration, \emph{{$CP$ Properties of Higgs Boson
  Interactions with Top Quarks in the $t\bar{t}H$ and $tH$ Processes Using $H
  \rightarrow \gamma\gamma$ with the ATLAS Detector}},
  \href{https://doi.org/10.1103/PhysRevLett.125.061802}{\emph{Phys. Rev. Lett.}
  {\bfseries 125} (2020) 061802}
  [\href{https://arxiv.org/abs/2004.04545}{{\ttfamily 2004.04545}}].

\bibitem{Alwall:2014hca}
J.~Alwall, R.~Frederix, S.~Frixione, V.~Hirschi, F.~Maltoni, O.~Mattelaer
  et~al., \emph{{The automated computation of tree-level and next-to-leading
  order differential cross sections, and their matching to parton shower
  simulations}}, \href{https://doi.org/10.1007/JHEP07(2014)079}{\emph{JHEP}
  {\bfseries 07} (2014) 079} [\href{https://arxiv.org/abs/1405.0301}{{\ttfamily
  1405.0301}}].

\bibitem{Kardos:2014pba}
A.~Kardos and Z.~Tr\'ocs\'anyi, \emph{{Hadroproduction of $t\bar{t}$ pair with
  two isolated photons with PowHel}},
  \href{https://doi.org/10.1016/j.nuclphysb.2015.05.032}{\emph{Nucl. Phys. B}
  {\bfseries 897} (2015) 717}
  [\href{https://arxiv.org/abs/1408.0278}{{\ttfamily 1408.0278}}].

\bibitem{Maltoni:2015ena}
F.~Maltoni, D.~Pagani and I.~Tsinikos, \emph{{Associated production of a
  top-quark pair with vector bosons at NLO in QCD: impact on $t\bar{t}H$
  searches at the LHC}},
  \href{https://doi.org/10.1007/JHEP02(2016)113}{\emph{JHEP} {\bfseries 02}
  (2016) 113} [\href{https://arxiv.org/abs/1507.05640}{{\ttfamily
  1507.05640}}].

\bibitem{vanDeurzen:2015cga}
H.~van Deurzen, R.~Frederix, V.~Hirschi, G.~Luisoni, P.~Mastrolia and
  G.~Ossola, \emph{{Spin Polarisation of $t\bar{t}\gamma\gamma$ production at
  NLO+PS with GoSam interfaced to MadGraph5\_aMC@NLO}},
  \href{https://doi.org/10.1140/epjc/s10052-016-4048-2}{\emph{Eur. Phys. J. C}
  {\bfseries 76} (2016) 221}
  [\href{https://arxiv.org/abs/1509.02077}{{\ttfamily 1509.02077}}].

\bibitem{Pagani:2021iwa}
D.~Pagani, H.-S. Shao, I.~Tsinikos and M.~Zaro, \emph{{Automated EW corrections
  with isolated photons: t$ \overline{t} $\ensuremath{\gamma}, t$ \overline{t}
  $\ensuremath{\gamma}\ensuremath{\gamma} and t\ensuremath{\gamma}j as case
  studies}}, \href{https://doi.org/10.1007/JHEP09(2021)155}{\emph{JHEP}
  {\bfseries 09} (2021) 155}
  [\href{https://arxiv.org/abs/2106.02059}{{\ttfamily 2106.02059}}].

\bibitem{Melnikov:2011ta}
K.~Melnikov, M.~Schulze and A.~Scharf, \emph{{QCD corrections to top quark pair
  production in association with a photon at hadron colliders}},
  \href{https://doi.org/10.1103/PhysRevD.83.074013}{\emph{Phys. Rev. D}
  {\bfseries 83} (2011) 074013}
  [\href{https://arxiv.org/abs/1102.1967}{{\ttfamily 1102.1967}}].

\bibitem{Bevilacqua:2018woc}
G.~Bevilacqua, H.~B. Hartanto, M.~Kraus, T.~Weber and M.~Worek, \emph{{Hard
  Photons in Hadroproduction of Top Quarks with Realistic Final States}},
  \href{https://doi.org/10.1007/JHEP10(2018)158}{\emph{JHEP} {\bfseries 10}
  (2018) 158} [\href{https://arxiv.org/abs/1803.09916}{{\ttfamily
  1803.09916}}].

\bibitem{Bevilacqua:2018dny}
G.~Bevilacqua, H.~B. Hartanto, M.~Kraus, T.~Weber and M.~Worek, \emph{{Precise
  predictions for $t\bar{t}\gamma/t\bar{t}$ cross section ratios at the LHC}},
  \href{https://doi.org/10.1007/JHEP01(2019)188}{\emph{JHEP} {\bfseries 01}
  (2019) 188} [\href{https://arxiv.org/abs/1809.08562}{{\ttfamily
  1809.08562}}].

\bibitem{Bergner:2018lgm}
J.~Bergner and M.~Schulze, \emph{{The top quark charge asymmetry in
  $t\bar{t}\gamma$ production at the LHC}},
  \href{https://doi.org/10.1140/epjc/s10052-019-6707-6}{\emph{Eur. Phys. J. C}
  {\bfseries 79} (2019) 189}
  [\href{https://arxiv.org/abs/1812.10535}{{\ttfamily 1812.10535}}].

\bibitem{Bevilacqua:2019quz}
G.~Bevilacqua, H.~B. Hartanto, M.~Kraus, T.~Weber and M.~Worek,
  \emph{{Off-shell vs on-shell modelling of top quarks in photon associated
  production}}, \href{https://doi.org/10.1007/JHEP03(2020)154}{\emph{JHEP}
  {\bfseries 03} (2020) 154}
  [\href{https://arxiv.org/abs/1912.09999}{{\ttfamily 1912.09999}}].

\bibitem{Denner:1999gp}
A.~Denner, S.~Dittmaier, M.~Roth and D.~Wackeroth, \emph{{Predictions for all
  processes $e^+ e^- \to$ 4 fermions $+$ $\gamma$}},
  \href{https://doi.org/10.1016/S0550-3213(99)00437-X}{\emph{Nucl. Phys. B}
  {\bfseries 560} (1999) 33}
  [\href{https://arxiv.org/abs/hep-ph/9904472}{{\ttfamily hep-ph/9904472}}].

\bibitem{Denner:2005fg}
A.~Denner, S.~Dittmaier, M.~Roth and L.~H. Wieders, \emph{{Electroweak
  corrections to charged-current e+ e- ---\ensuremath{>} 4 fermion processes:
  Technical details and further results}},
  \href{https://doi.org/10.1016/j.nuclphysb.2011.09.001}{\emph{Nucl. Phys. B}
  {\bfseries 724} (2005) 247}
  [\href{https://arxiv.org/abs/hep-ph/0505042}{{\ttfamily hep-ph/0505042}}].

\bibitem{Melnikov:2009dn}
K.~Melnikov and M.~Schulze, \emph{{NLO QCD corrections to top quark pair
  production and decay at hadron colliders}},
  \href{https://doi.org/10.1088/1126-6708/2009/08/049}{\emph{JHEP} {\bfseries
  08} (2009) 049} [\href{https://arxiv.org/abs/0907.3090}{{\ttfamily
  0907.3090}}].

\bibitem{Melnikov:2011qx}
K.~Melnikov, A.~Scharf and M.~Schulze, \emph{{Top quark pair production in
  association with a jet: QCD corrections and jet radiation in top quark
  decays}}, \href{https://doi.org/10.1103/PhysRevD.85.054002}{\emph{Phys. Rev.
  D} {\bfseries 85} (2012) 054002}
  [\href{https://arxiv.org/abs/1111.4991}{{\ttfamily 1111.4991}}].

\bibitem{Campbell:2012uf}
J.~M. Campbell and R.~K. Ellis, \emph{{Top-Quark Processes at NLO in Production
  and Decay}}, \href{https://doi.org/10.1088/0954-3899/42/1/015005}{\emph{J.
  Phys. G} {\bfseries 42} (2015) 015005}
  [\href{https://arxiv.org/abs/1204.1513}{{\ttfamily 1204.1513}}].

\bibitem{Behring:2019iiv}
A.~Behring, M.~Czakon, A.~Mitov, A.~S. Papanastasiou and R.~Poncelet,
  \emph{{Higher order corrections to spin correlations in top quark pair
  production at the LHC}},
  \href{https://doi.org/10.1103/PhysRevLett.123.082001}{\emph{Phys. Rev. Lett.}
  {\bfseries 123} (2019) 082001}
  [\href{https://arxiv.org/abs/1901.05407}{{\ttfamily 1901.05407}}].

\bibitem{Czakon:2020qbd}
M.~Czakon, A.~Mitov and R.~Poncelet, \emph{{NNLO QCD corrections to leptonic
  observables in top-quark pair production and decay}},
  \href{https://doi.org/10.1007/JHEP05(2021)212}{\emph{JHEP} {\bfseries 05}
  (2021) 212} [\href{https://arxiv.org/abs/2008.11133}{{\ttfamily
  2008.11133}}].

\bibitem{Harlander:2020cyh}
R.~V. Harlander, S.~Y. Klein and M.~Lipp, \emph{{FeynGame}},
  \href{https://doi.org/10.1016/j.cpc.2020.107465}{\emph{Comput. Phys. Commun.}
  {\bfseries 256} (2020) 107465}
  [\href{https://arxiv.org/abs/2003.00896}{{\ttfamily 2003.00896}}].

\bibitem{Bevilacqua:2022ozv}
G.~Bevilacqua, M.~Lupattelli, D.~Stremmer and M.~Worek, \emph{{Study of
  additional jet activity in top quark pair production and decay at the LHC}},
  \href{https://doi.org/10.1103/PhysRevD.107.114027}{\emph{Phys. Rev. D}
  {\bfseries 107} (2023) 114027}
  [\href{https://arxiv.org/abs/2212.04722}{{\ttfamily 2212.04722}}].

\bibitem{Actis:2016mpe}
S.~Actis, A.~Denner, L.~Hofer, J.-N. Lang, A.~Scharf and S.~Uccirati,
  \emph{{RECOLA: REcursive Computation of One-Loop Amplitudes}},
  \href{https://doi.org/10.1016/j.cpc.2017.01.004}{\emph{Comput. Phys. Commun.}
  {\bfseries 214} (2017) 140}
  [\href{https://arxiv.org/abs/1605.01090}{{\ttfamily 1605.01090}}].

\bibitem{Actis:2012qn}
S.~Actis, A.~Denner, L.~Hofer, A.~Scharf and S.~Uccirati, \emph{{Recursive
  generation of one-loop amplitudes in the Standard Model}},
  \href{https://doi.org/10.1007/JHEP04(2013)037}{\emph{JHEP} {\bfseries 04}
  (2013) 037} [\href{https://arxiv.org/abs/1211.6316}{{\ttfamily 1211.6316}}].

\bibitem{Denner:2016kdg}
A.~Denner, S.~Dittmaier and L.~Hofer, \emph{{Collier: a fortran-based Complex
  One-Loop LIbrary in Extended Regularizations}},
  \href{https://doi.org/10.1016/j.cpc.2016.10.013}{\emph{Comput. Phys. Commun.}
  {\bfseries 212} (2017) 220}
  [\href{https://arxiv.org/abs/1604.06792}{{\ttfamily 1604.06792}}].

\bibitem{Denner:2000bj}
A.~Denner, S.~Dittmaier, M.~Roth and D.~Wackeroth, \emph{{Electroweak radiative
  corrections to $e^+ e^- \to W^+ W^- \to 4$ fermions in double pole
  approximation: The RACOONWW approach}},
  \href{https://doi.org/10.1016/S0550-3213(00)00511-3}{\emph{Nucl. Phys. B}
  {\bfseries 587} (2000) 67}
  [\href{https://arxiv.org/abs/hep-ph/0006307}{{\ttfamily hep-ph/0006307}}].

\bibitem{Accomando:2004de}
E.~Accomando, A.~Denner and A.~Kaiser, \emph{{Logarithmic electroweak
  corrections to gauge-boson pair production at the LHC}},
  \href{https://doi.org/10.1016/j.nuclphysb.2004.11.019}{\emph{Nucl. Phys. B}
  {\bfseries 706} (2005) 325}
  [\href{https://arxiv.org/abs/hep-ph/0409247}{{\ttfamily hep-ph/0409247}}].

\bibitem{Denner:2016jyo}
A.~Denner and M.~Pellen, \emph{{NLO electroweak corrections to off-shell
  top-antitop production with leptonic decays at the LHC}},
  \href{https://doi.org/10.1007/JHEP08(2016)155}{\emph{JHEP} {\bfseries 08}
  (2016) 155} [\href{https://arxiv.org/abs/1607.05571}{{\ttfamily
  1607.05571}}].

\bibitem{Bevilacqua:2011xh}
G.~Bevilacqua, M.~Czakon, M.~V. Garzelli, A.~van Hameren, A.~Kardos, C.~G.
  Papadopoulos et~al., \emph{{HELAC-NLO}},
  \href{https://doi.org/10.1016/j.cpc.2012.10.033}{\emph{Comput. Phys. Commun.}
  {\bfseries 184} (2013) 986}
  [\href{https://arxiv.org/abs/1110.1499}{{\ttfamily 1110.1499}}].

\bibitem{Draggiotis:1998gr}
P.~Draggiotis, R.~H.~P. Kleiss and C.~G. Papadopoulos, \emph{{On the
  computation of multigluon amplitudes}},
  \href{https://doi.org/10.1016/S0370-2693(98)01015-6}{\emph{Phys. Lett. B}
  {\bfseries 439} (1998) 157}
  [\href{https://arxiv.org/abs/hep-ph/9807207}{{\ttfamily hep-ph/9807207}}].

\bibitem{Draggiotis:2002hm}
P.~D. Draggiotis, R.~H.~P. Kleiss and C.~G. Papadopoulos, \emph{{Multijet
  production in hadron collisions}},
  \href{https://doi.org/10.1007/s10052-002-0955-5}{\emph{Eur. Phys. J. C}
  {\bfseries 24} (2002) 447}
  [\href{https://arxiv.org/abs/hep-ph/0202201}{{\ttfamily hep-ph/0202201}}].

\bibitem{Bevilacqua:2013iha}
G.~Bevilacqua, M.~Czakon, M.~Kubocz and M.~Worek, \emph{{Complete Nagy-Soper
  subtraction for next-to-leading order calculations in QCD}},
  \href{https://doi.org/10.1007/JHEP10(2013)204}{\emph{JHEP} {\bfseries 10}
  (2013) 204} [\href{https://arxiv.org/abs/1308.5605}{{\ttfamily 1308.5605}}].

\bibitem{vanHameren:2009dr}
A.~van Hameren, C.~G. Papadopoulos and R.~Pittau, \emph{{Automated one-loop
  calculations: A Proof of concept}},
  \href{https://doi.org/10.1088/1126-6708/2009/09/106}{\emph{JHEP} {\bfseries
  09} (2009) 106} [\href{https://arxiv.org/abs/0903.4665}{{\ttfamily
  0903.4665}}].

\bibitem{Ossola:2006us}
G.~Ossola, C.~G. Papadopoulos and R.~Pittau, \emph{{Reducing full one-loop
  amplitudes to scalar integrals at the integrand level}},
  \href{https://doi.org/10.1016/j.nuclphysb.2006.11.012}{\emph{Nucl. Phys. B}
  {\bfseries 763} (2007) 147}
  [\href{https://arxiv.org/abs/hep-ph/0609007}{{\ttfamily hep-ph/0609007}}].

\bibitem{Ossola:2007ax}
G.~Ossola, C.~G. Papadopoulos and R.~Pittau, \emph{{CutTools: A Program
  implementing the OPP reduction method to compute one-loop amplitudes}},
  \href{https://doi.org/10.1088/1126-6708/2008/03/042}{\emph{JHEP} {\bfseries
  03} (2008) 042} [\href{https://arxiv.org/abs/0711.3596}{{\ttfamily
  0711.3596}}].

\bibitem{vanHameren:2010cp}
A.~van Hameren, \emph{{OneLOop: For the evaluation of one-loop scalar
  functions}}, \href{https://doi.org/10.1016/j.cpc.2011.06.011}{\emph{Comput.
  Phys. Commun.} {\bfseries 182} (2011) 2427}
  [\href{https://arxiv.org/abs/1007.4716}{{\ttfamily 1007.4716}}].

\bibitem{Catani:1996vz}
S.~Catani and M.~H. Seymour, \emph{{A General algorithm for calculating jet
  cross-sections in NLO QCD}},
  \href{https://doi.org/10.1016/S0550-3213(96)00589-5}{\emph{Nucl. Phys. B}
  {\bfseries 485} (1997) 291}
  [\href{https://arxiv.org/abs/hep-ph/9605323}{{\ttfamily hep-ph/9605323}}].

\bibitem{Catani:2002hc}
S.~Catani, S.~Dittmaier, M.~H. Seymour and Z.~Trocsanyi, \emph{{The Dipole
  formalism for next-to-leading order QCD calculations with massive partons}},
  \href{https://doi.org/10.1016/S0550-3213(02)00098-6}{\emph{Nucl. Phys. B}
  {\bfseries 627} (2002) 189}
  [\href{https://arxiv.org/abs/hep-ph/0201036}{{\ttfamily hep-ph/0201036}}].

\bibitem{Campbell:2004ch}
J.~M. Campbell, R.~K. Ellis and F.~Tramontano, \emph{{Single top production and
  decay at next-to-leading order}},
  \href{https://doi.org/10.1103/PhysRevD.70.094012}{\emph{Phys. Rev. D}
  {\bfseries 70} (2004) 094012}
  [\href{https://arxiv.org/abs/hep-ph/0408158}{{\ttfamily hep-ph/0408158}}].

\bibitem{Nagy:1998bb}
Z.~Nagy and Z.~Trocsanyi, \emph{{Next-to-leading order calculation of four jet
  observables in electron positron annihilation}},
  \href{https://doi.org/10.1103/PhysRevD.62.099902}{\emph{Phys. Rev. D}
  {\bfseries 59} (1999) 014020}
  [\href{https://arxiv.org/abs/hep-ph/9806317}{{\ttfamily hep-ph/9806317}}].

\bibitem{Nagy:2003tz}
Z.~Nagy, \emph{{Next-to-leading order calculation of three jet observables in
  hadron hadron collision}},
  \href{https://doi.org/10.1103/PhysRevD.68.094002}{\emph{Phys. Rev. D}
  {\bfseries 68} (2003) 094002}
  [\href{https://arxiv.org/abs/hep-ph/0307268}{{\ttfamily hep-ph/0307268}}].

\bibitem{Bevilacqua:2009zn}
G.~Bevilacqua, M.~Czakon, C.~G. Papadopoulos, R.~Pittau and M.~Worek,
  \emph{{Assault on the NLO Wishlist: pp ---\ensuremath{>} t anti-t b anti-b}},
  \href{https://doi.org/10.1088/1126-6708/2009/09/109}{\emph{JHEP} {\bfseries
  09} (2009) 109} [\href{https://arxiv.org/abs/0907.4723}{{\ttfamily
  0907.4723}}].

\bibitem{Czakon:2015cla}
M.~Czakon, H.~B. Hartanto, M.~Kraus and M.~Worek, \emph{{Matching the
  Nagy-Soper parton shower at next-to-leading order}},
  \href{https://doi.org/10.1007/JHEP06(2015)033}{\emph{JHEP} {\bfseries 06}
  (2015) 033} [\href{https://arxiv.org/abs/1502.00925}{{\ttfamily
  1502.00925}}].

\bibitem{Czakon:2009ss}
M.~Czakon, C.~G. Papadopoulos and M.~Worek, \emph{{Polarizing the Dipoles}},
  \href{https://doi.org/10.1088/1126-6708/2009/08/085}{\emph{JHEP} {\bfseries
  08} (2009) 085} [\href{https://arxiv.org/abs/0905.0883}{{\ttfamily
  0905.0883}}].

\bibitem{vanHameren:2007pt}
A.~van Hameren, \emph{{PARNI for importance sampling and density estimation}},
  {\emph{Acta Phys. Polon. B} {\bfseries 40} (2009) 259}
  [\href{https://arxiv.org/abs/0710.2448}{{\ttfamily 0710.2448}}].

\bibitem{vanHameren:2010gg}
A.~van Hameren, \emph{{Kaleu: A General-Purpose Parton-Level Phase Space
  Generator}},  \href{https://arxiv.org/abs/1003.4953}{{\ttfamily 1003.4953}}.

\bibitem{Bevilacqua:2010qb}
G.~Bevilacqua, M.~Czakon, A.~van Hameren, C.~G. Papadopoulos and M.~Worek,
  \emph{{Complete off-shell effects in top quark pair hadroproduction with
  leptonic decay at next-to-leading order}},
  \href{https://doi.org/10.1007/JHEP02(2011)083}{\emph{JHEP} {\bfseries 02}
  (2011) 083} [\href{https://arxiv.org/abs/1012.4230}{{\ttfamily 1012.4230}}].

\bibitem{Alwall:2006yp}
J.~Alwall et~al., \emph{{A Standard format for Les Houches event files}},
  \href{https://doi.org/10.1016/j.cpc.2006.11.010}{\emph{Comput. Phys. Commun.}
  {\bfseries 176} (2007) 300}
  [\href{https://arxiv.org/abs/hep-ph/0609017}{{\ttfamily hep-ph/0609017}}].

\bibitem{Bevilacqua:2016jfk}
G.~Bevilacqua, H.~B. Hartanto, M.~Kraus and M.~Worek, \emph{{Off-shell Top
  Quarks with One Jet at the LHC: A comprehensive analysis at NLO QCD}},
  \href{https://doi.org/10.1007/JHEP11(2016)098}{\emph{JHEP} {\bfseries 11}
  (2016) 098} [\href{https://arxiv.org/abs/1609.01659}{{\ttfamily
  1609.01659}}].

\bibitem{Bern:2013zja}
Z.~Bern, L.~J. Dixon, F.~Febres~Cordero, S.~H\"oche, H.~Ita, D.~A. Kosower
  et~al., \emph{{Ntuples for NLO Events at Hadron Colliders}},
  \href{https://doi.org/10.1016/j.cpc.2014.01.011}{\emph{Comput. Phys. Commun.}
  {\bfseries 185} (2014) 1443}
  [\href{https://arxiv.org/abs/1310.7439}{{\ttfamily 1310.7439}}].

\bibitem{PDF4LHCWorkingGroup:2022cjn}
{\scshape PDF4LHC Working Group} collaboration, \emph{{The PDF4LHC21
  combination of global PDF fits for the LHC Run III}},
  \href{https://doi.org/10.1088/1361-6471/ac7216}{\emph{J. Phys. G} {\bfseries
  49} (2022) 080501} [\href{https://arxiv.org/abs/2203.05506}{{\ttfamily
  2203.05506}}].

\bibitem{NNPDF:2017mvq}
{\scshape NNPDF} collaboration, \emph{{Parton distributions from high-precision
  collider data}},
  \href{https://doi.org/10.1140/epjc/s10052-017-5199-5}{\emph{Eur. Phys. J. C}
  {\bfseries 77} (2017) 663}
  [\href{https://arxiv.org/abs/1706.00428}{{\ttfamily 1706.00428}}].

\bibitem{Buckley:2014ana}
A.~Buckley, J.~Ferrando, S.~Lloyd, K.~Nordstr\"om, B.~Page, M.~R\"ufenacht
  et~al., \emph{{LHAPDF6: parton density access in the LHC precision era}},
  \href{https://doi.org/10.1140/epjc/s10052-015-3318-8}{\emph{Eur. Phys. J. C}
  {\bfseries 75} (2015) 132} [\href{https://arxiv.org/abs/1412.7420}{{\ttfamily
  1412.7420}}].

\bibitem{Bailey:2020ooq}
S.~Bailey, T.~Cridge, L.~A. Harland-Lang, A.~D. Martin and R.~S. Thorne,
  \emph{{Parton distributions from LHC, HERA, Tevatron and fixed target data:
  MSHT20 PDFs}},
  \href{https://doi.org/10.1140/epjc/s10052-021-09057-0}{\emph{Eur. Phys. J. C}
  {\bfseries 81} (2021) 341}
  [\href{https://arxiv.org/abs/2012.04684}{{\ttfamily 2012.04684}}].

\bibitem{Hou:2019efy}
T.-J. Hou et~al., \emph{{New CTEQ global analysis of quantum chromodynamics
  with high-precision data from the LHC}},
  \href{https://arxiv.org/abs/1912.10053}{{\ttfamily 1912.10053}}.

\bibitem{Workman:2022ynf}
{\scshape Particle Data Group} collaboration, \emph{{Review of Particle
  Physics}}, \href{https://doi.org/10.1093/ptep/ptac097}{\emph{PTEP} {\bfseries
  2022} (2022) 083C01}.

\bibitem{Denner:2019vbn}
A.~Denner and S.~Dittmaier, \emph{{Electroweak Radiative Corrections for
  Collider Physics}},
  \href{https://doi.org/10.1016/j.physrep.2020.04.001}{\emph{Phys. Rept.}
  {\bfseries 864} (2020) 1} [\href{https://arxiv.org/abs/1912.06823}{{\ttfamily
  1912.06823}}].

\bibitem{Jezabek:1988iv}
M.~Jezabek and J.~H. Kuhn, \emph{{QCD Corrections to Semileptonic Decays of
  Heavy Quarks}},
  \href{https://doi.org/10.1016/0550-3213(89)90108-9}{\emph{Nucl. Phys. B}
  {\bfseries 314} (1989) 1}.

\bibitem{Denner:2012yc}
A.~Denner, S.~Dittmaier, S.~Kallweit and S.~Pozzorini, \emph{{NLO QCD
  corrections to off-shell top-antitop production with leptonic decays at
  hadron colliders}},
  \href{https://doi.org/10.1007/JHEP10(2012)110}{\emph{JHEP} {\bfseries 10}
  (2012) 110} [\href{https://arxiv.org/abs/1207.5018}{{\ttfamily 1207.5018}}].

\bibitem{Cacciari:2008gp}
M.~Cacciari, G.~P. Salam and G.~Soyez, \emph{{The anti-$k_t$ jet clustering
  algorithm}}, \href{https://doi.org/10.1088/1126-6708/2008/04/063}{\emph{JHEP}
  {\bfseries 04} (2008) 063} [\href{https://arxiv.org/abs/0802.1189}{{\ttfamily
  0802.1189}}].

\bibitem{Frixione:1998jh}
S.~Frixione, \emph{{Isolated photons in perturbative QCD}},
  \href{https://doi.org/10.1016/S0370-2693(98)00454-7}{\emph{Phys. Lett. B}
  {\bfseries 429} (1998) 369}
  [\href{https://arxiv.org/abs/hep-ph/9801442}{{\ttfamily hep-ph/9801442}}].

\bibitem{Denner:2017kzu}
A.~Denner and M.~Pellen, \emph{{Off-shell production of top-antitop pairs in
  the lepton+jets channel at NLO QCD}},
  \href{https://doi.org/10.1007/JHEP02(2018)013}{\emph{JHEP} {\bfseries 02}
  (2018) 013} [\href{https://arxiv.org/abs/1711.10359}{{\ttfamily
  1711.10359}}].

\bibitem{Denner:2015yca}
A.~Denner and R.~Feger, \emph{{NLO QCD corrections to off-shell top-antitop
  production with leptonic decays in association with a Higgs boson at the
  LHC}}, \href{https://doi.org/10.1007/JHEP11(2015)209}{\emph{JHEP} {\bfseries
  11} (2015) 209} [\href{https://arxiv.org/abs/1506.07448}{{\ttfamily
  1506.07448}}].

\bibitem{Bevilacqua:2019cvp}
G.~Bevilacqua, H.~B. Hartanto, M.~Kraus, T.~Weber and M.~Worek, \emph{{Towards
  constraining Dark Matter at the LHC: Higher order QCD predictions for
  $t\bar{t}+Z(Z\to \nu_\ell \bar{\nu}_\ell)$}},
  \href{https://doi.org/10.1007/JHEP11(2019)001}{\emph{JHEP} {\bfseries 11}
  (2019) 001} [\href{https://arxiv.org/abs/1907.09359}{{\ttfamily
  1907.09359}}].

\bibitem{Stremmer:2021bnk}
D.~Stremmer and M.~Worek, \emph{{Production and decay of the Higgs boson in
  association with top quarks}},
  \href{https://doi.org/10.1007/JHEP02(2022)196}{\emph{JHEP} {\bfseries 02}
  (2022) 196} [\href{https://arxiv.org/abs/2111.01427}{{\ttfamily
  2111.01427}}].

\bibitem{Bern:2011pa}
Z.~Bern, G.~Diana, L.~J. Dixon, F.~Febres~Cordero, S.~Hoche, H.~Ita et~al.,
  \emph{{Driving Missing Data at Next-to-Leading Order}},
  \href{https://doi.org/10.1103/PhysRevD.84.114002}{\emph{Phys. Rev. D}
  {\bfseries 84} (2011) 114002}
  [\href{https://arxiv.org/abs/1106.1423}{{\ttfamily 1106.1423}}].

\bibitem{Campbell:2017dqk}
J.~M. Campbell, R.~K. Ellis and C.~Williams, \emph{{Driving missing data at the
  LHC: NNLO predictions for the ratio of $\gamma+j$ and $Z+j$}},
  \href{https://doi.org/10.1103/PhysRevD.96.014037}{\emph{Phys. Rev. D}
  {\bfseries 96} (2017) 014037}
  [\href{https://arxiv.org/abs/1703.10109}{{\ttfamily 1703.10109}}].

\bibitem{Chen:2019zmr}
X.~Chen, T.~Gehrmann, N.~Glover, M.~H\"ofer and A.~Huss, \emph{{Isolated photon
  and photon+jet production at NNLO QCD accuracy}},
  \href{https://doi.org/10.1007/JHEP04(2020)166}{\emph{JHEP} {\bfseries 04}
  (2020) 166} [\href{https://arxiv.org/abs/1904.01044}{{\ttfamily
  1904.01044}}].

\bibitem{Chawdhry:2019bji}
H.~A. Chawdhry, M.~Czakon, A.~Mitov and R.~Poncelet, \emph{{NNLO QCD
  corrections to three-photon production at the LHC}},
  \href{https://doi.org/10.1007/JHEP02(2020)057}{\emph{JHEP} {\bfseries 02}
  (2020) 057} [\href{https://arxiv.org/abs/1911.00479}{{\ttfamily
  1911.00479}}].

\bibitem{Gehrmann:2020oec}
T.~Gehrmann, N.~Glover, A.~Huss and J.~Whitehead, \emph{{Scale and isolation
  sensitivity of diphoton distributions at the LHC}},
  \href{https://doi.org/10.1007/JHEP01(2021)108}{\emph{JHEP} {\bfseries 01}
  (2021) 108} [\href{https://arxiv.org/abs/2009.11310}{{\ttfamily
  2009.11310}}].

\bibitem{Chawdhry:2021hkp}
H.~A. Chawdhry, M.~Czakon, A.~Mitov and R.~Poncelet, \emph{{NNLO QCD
  corrections to diphoton production with an additional jet at the LHC}},
  \href{https://doi.org/10.1007/JHEP09(2021)093}{\emph{JHEP} {\bfseries 09}
  (2021) 093} [\href{https://arxiv.org/abs/2105.06940}{{\ttfamily
  2105.06940}}].

\bibitem{Badger:2021ohm}
S.~Badger, T.~Gehrmann, M.~Marcoli and R.~Moodie, \emph{{Next-to-leading order
  QCD corrections to diphoton-plus-jet production through gluon fusion at the
  LHC}}, \href{https://doi.org/10.1016/j.physletb.2021.136802}{\emph{Phys.
  Lett. B} {\bfseries 824} (2022) 136802}
  [\href{https://arxiv.org/abs/2109.12003}{{\ttfamily 2109.12003}}].

\bibitem{Chen:2022gpk}
X.~Chen, T.~Gehrmann, E.~W.~N. Glover, M.~H\"ofer, A.~Huss and R.~Sch\"urmann,
  \emph{{Single photon production at hadron colliders at NNLO QCD with
  realistic photon isolation}},
  \href{https://doi.org/10.1007/JHEP08(2022)094}{\emph{JHEP} {\bfseries 08}
  (2022) 094} [\href{https://arxiv.org/abs/2205.01516}{{\ttfamily
  2205.01516}}].

\bibitem{Badger:2023mgf}
S.~Badger, M.~Czakon, H.~B. Hartanto, R.~Moodie, T.~Peraro, R.~Poncelet et~al.,
  \emph{{Isolated photon production in association with a jet pair through
  next-to-next-to-leading order in QCD}},
  \href{https://arxiv.org/abs/2304.06682}{{\ttfamily 2304.06682}}.

\end{thebibliography}



\providecommand{\href}[2]{#2}\begingroup\raggedright\endgroup

\end{document}